\documentclass[11pt, a4paper]{article} 

\usepackage[english]{babel} 
\usepackage{booktabs}
\usepackage{comment}
\usepackage[utf8]{inputenc}
\usepackage[T1]{fontenc} 

\usepackage{amsmath,amsfonts,amsthm} 
\numberwithin{equation}{section}
\usepackage{amssymb}
\usepackage{bm}

\usepackage{tikz} 
\usepackage{tikz-cd}
\usetikzlibrary{positioning,arrows} 
\usetikzlibrary{decorations.pathreplacing}
\usepackage{tikzsymbols}

\usetikzlibrary{shapes.geometric, arrows}
\tikzstyle{box} = [rectangle, rounded corners, minimum width=3cm, minimum height=1cm,text centered, draw=black, fill=blue!20, text width=3cm]
\tikzstyle{nobox} = [rectangle, rounded corners, minimum width=3cm, minimum height=1cm,text centered, draw=white, fill=white, text width=3cm]
\tikzstyle{arrow} = [thick,->,>=stealth]

\usepackage{geometry} 
\usepackage{setspace} 
\usepackage{fancyhdr}
\usepackage[numbers, compress, square]{natbib}
\usepackage{hyperref}

\usepackage{epigraph}
\usepackage{afterpage}
\usepackage{color}
\hypersetup{
     colorlinks   = true,
     linkcolor    = black,
     citecolor    = blue,
     urlcolor     = {blue!80!black}
}

\newcommand{\deriv}{\mathrm{d}}

\usepackage{textcomp, gensymb}
\usepackage{empheq}
\usepackage[makeroom]{cancel}
\usepackage{physics}

\newcommand*\widefbox[1]{\fbox{\hspace{1em}#1\hspace{1em}}}

\usepackage{caption}
\usepackage{subcaption}
\usepackage{epstopdf} 

\begin{document}

\begin{titlepage} 

\center \vfill
	
{\huge\bfseries \textsc{BMS symmetries of \\ gravitational scattering}}

\vfill

Xavier \textsc{Kervyn} \\ \href{mailto:xpmk2@cam.ac.uk}{xpmk2@cam.ac.uk} / \href{mailto:kervyn.xavier@gmail.com}{kervyn.xavier@gmail.com}

\vfill

\textsc{Department of Applied Mathematics and Theoretical Physics} \\ \textsc{University of Cambridge} \\
Centre for Mathematical Sciences, \\ Wilberforce Rd, Cambridge CB3 0WA, United Kingdom

\vfill

August 2023

\vfill

\begin{abstract}
    
    After motivating the relevance of the Bondi-Metzner-Sachs (BMS) group over the last decades, we review how concepts such as Penrose diagrams and the covariant phase space formalism can be used to understand the asymptotic structure of asymptotically flat spacetimes (AFS). We then explicitly construct the asymptotic symmetry group of AFS in $3+1$ dimensions, the BMS group. Next, we apply this knowledge to the usual far-field scattering problem in general relativity, which leads to the unravelling of the intrinsic features of gravity in the infrared. In particular, we work out the connections between asymptotic symmetries, soft theorems in quantum field theories and gravitational memory effects. We restrict to the study of this \textit{infrared triangle} through the lens of supertranslations here, but the analogous features that can be found in the case of superrotations or for other gauge theories are also motivated at the end of our discussion. We conclude with an overview of the implications of the infrared triangle of gravity for the formulation of an approach to quantum gravity through holography, as well as a brief discussion of its potential in tackling the black hole information paradox.
    \vspace{\baselineskip}

    This review article arose from an essay submitted for the partial fulfillment of the requirements for the degree of \textit{Master of Advanced Study in Applied Mathematics} (Part III of the Mathematical Tripos) at the University of Cambridge, set by Dr.~Prahar Mitra and submitted in May 2023. It is aimed at advanced undergraduate students or early postgraduate students willing to learn about the role of asymptotic symmetries in the context of flat space holography, with only basic knowledge of quantum field theory and general relativity assumed.
    
    \end{abstract}

\vfill

\end{titlepage}

\newpage

\thispagestyle{empty} 
\tableofcontents

\newpage

\setcounter{page}{1} 

\section{Introduction}  
    The current theoretical exploration in the search for further unification of the forces of Nature is very often based on the search for new symmetries of the theory. As Weinberg said,
    \begin{quote}
        ``Nothing in physics seems so hopeful to me as the idea that it is possible for a theory to have a high degree of symmetry was hidden from us in everyday life. The physicist's task is to find this deeper symmetry.'' -- \textit{The Forces of Nature. (1976)}
    \end{quote}
    In mathematics, this notion ties beautifully to the concepts of group theory, which in turn provides a general framework for the study of the laws of physics. Gravity is no exception. In this article, we review the insights provided by the study of asymptotic symmetries and their associated asymptotic symmetry group on general relativity, the \textit{BMS group}, restricting ourselves to $3+1$ dimensions.

    \subsection*{Historical overview} Since Einstein's publication of the theory of general relativity in 1915, group-theoretical methods have been instrumental not only in deriving new solutions of the associated Einstein equations and understanding their properties \cite{carmeli_group_2000, stephani_exact_2003},
    but also in probing the asymptotic and conformal structure of spacetime, notably through the work of Penrose and collaborators in the 1960s \cite{penrose_asymptotic_1963, penrose_conformal_1964, penrose_structure_1968, penrose_zero_1997}. At the same time, Bondi, Metzner and Sachs \cite{bondi_gravitational_1962, sachs_gravitational_1962, sachs_asymptotic_1962} first formulated a set of symmetries that describe the behavior of gravitational waves in asymptotically flat spacetimes. These were shown to be the building blocks of a so-called \textit{asymptotic symmetry group}, the \textit{BMS group}, of which McCarthy and collaborators investigated the representations and a number other properties in the 1970s \cite{mccarthy_structure_1972, mccarthy_representations_1972, 
    mccarthy_representations_1997,
    mccarthy_representations_1973}.
    However, while this work provided a foundational understanding, it was not immediately clear how to incorporate the BMS group into a broader theory of gravity.

    \smallskip
    In the subsequent decades, research in theoretical physics shifted towards other areas, such as quantum field theory, string theory, and the study of black holes. These fields garnered significant attention and resources, leading to a temporary diversion of focus away from BMS-related work. 
    In 1993, Christodoulou and Klainerman (CK) \cite{christodoulou_global_1993} proved the nonlinear gravitational stability of Minkowski space-time. In doing so, they also provided a prescription for matching data incoming from future and past null infinity, which is necessary to study scattering processes. At the time though, string theory was still in full bloom, retaining much of the attention of the scientific community, while the result did not seem of big importance to the general relativity community. It was only in the early 2010s that Barnich and Trossaert reignited interest in the BMS group \cite{barnich_aspects_2010}, opening new avenues of research along the lines of string theory and holography, where symmetries are key ingredients of the duality. Combining their results with the work of CK, Strominger, along with collaborators, then made significant contributions to our understanding of the BMS group \cite{strominger_bms_2014, kapec_semiclassical_2014, strominger_gravitational_2016, hawking_soft_2016, strominger_black_2017}, see \cite{strominger_lectures_2017} and references therein. Among others, their work connected the BMS symmetries to the holographic principle, which relates gravitational theories in higher dimensions to lower-dimensional quantum field theories, as well as to the black hole information paradox.  
    
    \smallskip
    Interestingly, these developments were concomitant with the successful detection of gravitational waves by the Laser Interferometer Gravitational-Wave Observatory (LIGO) in 2015, about 50 years after the publication of BMS's seminal papers. While no experimental data was needed for the aforementioned theoretical progress, the experimental breakthroughs highlighted the relevance of the BMS group and its potential implications for the fundamental nature of spacetime. We briefly mention some the key insights provided so far by this prism of study of the BMS group in the next paragraphs, in hindsight of and with references to our discussion in the rest of this review article.

    \subsection*{Key insights from the BMS group}

\begin{figure}[h!]
            \centering
            \includegraphics[width=0.4\linewidth]{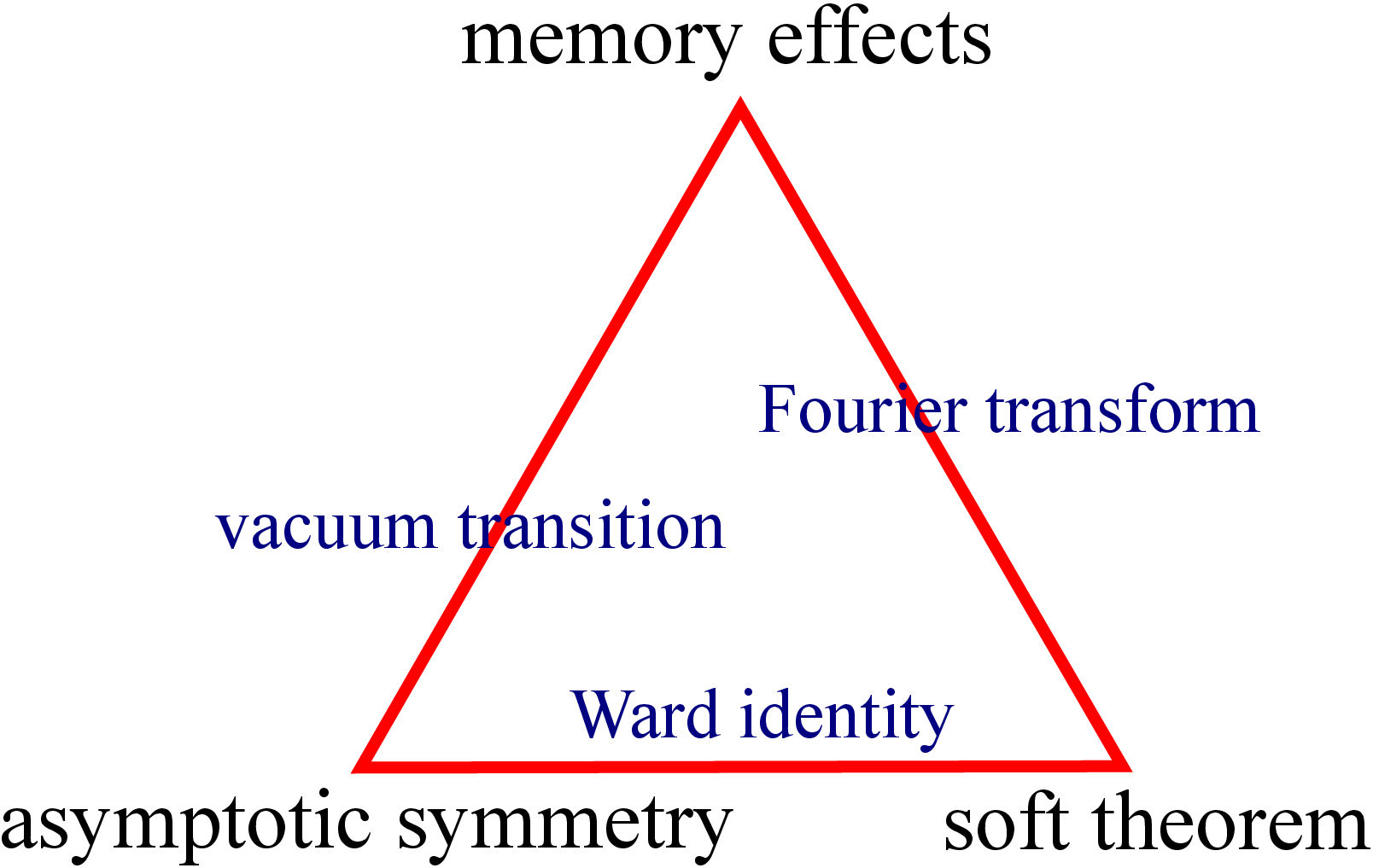}
            \caption{The infrared triangle \label{fig:IRtriangle}}
        \end{figure}

    \paragraph{Infrared structure of gauge theories} BMS were originally expecting gravity in asymptotically flat spacetimes to resemble special relativity as one recedes from an isolated gravitational system. Solving the Killing equation to obtain the vector fields preserving the form of their metric, they instead found that the BMS group is larger than the Poincaré group. We follow their steps and construct the generators of the BMS group in section~\ref{sec:gravity in AFS}.  Strominger shed a new light on the implications of this feature by showcasing a triangular equivalence relation between \textit{asymptotic symmetries}, \textit{soft theorems} and \textit{memory effects} as pictured in Fig.~\ref{fig:IRtriangle} \cite{strominger_bms_2014, he_bms_2015, strominger_gravitational_2016}. 
     In this diagram, asymptotic symmetries -- such as the BMS symmetries -- are only one of the three equivalent entries to a much richer dictionary, allowing for a better understanding of of each of the corners. The explicit derivation of this triangle in the case of BMS supertranslations will be the backbone of this review. We shall see that it has profound consequences for gravity, showing for instance that the vacuum in general relativity is not unique. Copies of the same triangle were subsequently constructed for electromagnetism \cite{kapec_new_2015, campiglia_asymptotic_2014}, BMS superrotations \cite{cachazo_evidence_2014}, and other gauge theories \cite{kapec_asymptotic_2017, strominger_asymptotic_2014, he_asymptotic_2023}, each involving various soft theorems and unravelling new symmetries or memory effects (work in progress).

    \paragraph{Gravitational wave astronomy}
    The upper corner of Fig.~\ref{fig:IRtriangle}, \textit{memory effects}, refers to long-lasting changes in the spacetime geometry caused by the passage of gravitational waves. These effects can lead to a permanent displacement of test masses or the acquisition of a new state of motion. Since BMS symmetries govern the asymptotic behavior of gravitational waves in asymptotically flat spacetimes, they play a crucial role in their propagation. Understanding them is thus essential for accurately analyzing and interpreting the signals received from gravitational wave detectors. As shall be made explicit in section~\ref{sec: memory effects} of this review, the BMS group provides insights into the radiation patterns, conservation laws, and energy-momentum properties of gravitational waves.

    \paragraph{Quantum gravity and holography} The holographic conjecture is a major advance in theoretical physics which relates gravitational theories in higher dimensions to lower-dimensional quantum field theories; the prime example of it being the AdS/CFT correspondence \cite{brown_central_1986, maldacena_large_1999}. This is still an active area of research and many results remain unknown, especially for (asymptotically) flat spacetimes -- in fact in any but asymptotically negatively curved backgrounds. Recent research suggested that the BMS group could be used to address this problem in asymptotically flat spacetimes, by allowing to recast scattering amplitudes of any four-dimensional theory with non-abelian gauge group as two-dimensional correlation functions on the asymptotic two-sphere at null infinity \cite{he_2d_2016}.  
    This led to the proposal that gravity in four-dimensional asymptotically flat spacetimes may be dual to a theory living on the “celestial sphere” at infinity \cite{pasterski_flat_2017}, a program known as \textit{celestial holography} \cite{raclariu_lectures_2021}. The BMS symmetries have been linked to the symmetries of the boundary theory in the holographic dual, offering potential insights into the nature of spacetime, information, and quantum gravity \cite{pasterski_celestial_2021}. Many complexities and challenges however still need to be carefully addressed.

    \paragraph{Black hole physics} BMS symmetries also led to crucial insights into the conservation laws of black holes. They govern the behavior of the waves they emit and thus provide a framework to compute conserved quantities such as energy, momentum, and angular momentum which are needed to characterize the dynamics and evolution of such objects.
    Second, the BMS group also has connections to the \textit{black hole information paradox} \cite{hawking_breakdown_1976}, which arises due to the conflict between the principles of quantum mechanics and classical general relativity, suggesting that information can be lost in the process of black hole evaporation. In particular, BMS supertranslations within the BMS group have been linked to changes in the geometry and dynamics of the event horizon through \textit{soft hairs} \cite{strominger_black_2017, hawking_superrotation_2017, hawking_soft_2016}. The links of the BMS group to quantum gravity and the holographic principle could help to understand the holographic nature of black holes and the role of the event horizon in encoding information about the interior \cite{kapec_area_2017}, see also \cite{strominger_lectures_2017} and references therein. This research direction thus has the potential to shed light on the fundamental nature of black holes and their relation to quantum gravity, though many questions remain to be addressed.

    \subsection*{Plan of the article} This review article is structured as follows. In section \ref{sec:section 1}, we introduce key concepts (Penrose diagram, asymptotic symmetries, covariant phase space formalism, Poincaré group) as well as coordinate conventions to be used throughout this article. Section \ref{sec:gravity in AFS} is then devoted to the construction of the BMS group. In section~\ref{sec: IR structure of gravity} we eventually build upon these newly constructed asymptotic symmetries of gravitational scattering to understand the intricate structure of gravity. We focus on the role of supertranslations and their connections to soft theorems and memory effects. Section~\ref{sec: conclusion} wraps up our discussion by presenting the infrared triangle of gravity, its analogue for BMS superrotations or other gauge theories, and then mentioning the role of the latter in formulating an approach to quantum gravity through flat space holography or tackling the black hole information paradox.

\newpage
\section{Asymptotic structure of Minkowski spacetime} \label{sec:section 1}

    In this section, we carry out the conformal compactification of Minkowski and introduce the different regions of interest for the scattering problem in gravity. Next, we explicit the scattering problem and set up the coordinate system that will predominantly be used throughout this article. Finally, we introduce the covariant phase space formalism as a set of methods allowing for the construction of a symplectic form on the phase space of a covariant field theory. We end with a lightning overview of the generators of the Poincaré group, the group of isometries of flat space in $3+1$ dimensions. 
    
    \subsection{Conformal compactification and scattering in Minkowski \label{sec: conformal compactification Minkowski}}
        The goal of conformal compactification is to extend the notion of infinity in a geometric space, such as a manifold or a metric space, in a way that makes it possible to study the properties of the space near infinity \cite{carroll_spacetime_2019}. In practice, it consists of a succession of clever coordinate transformations that allow to bring infinity "at a finite coordinate distance away". 

\begin{figure}[h!]
        \centering
        \begin{subfigure}[b]{0.495\linewidth}
            \begin{center}
            \includegraphics[width=\linewidth]{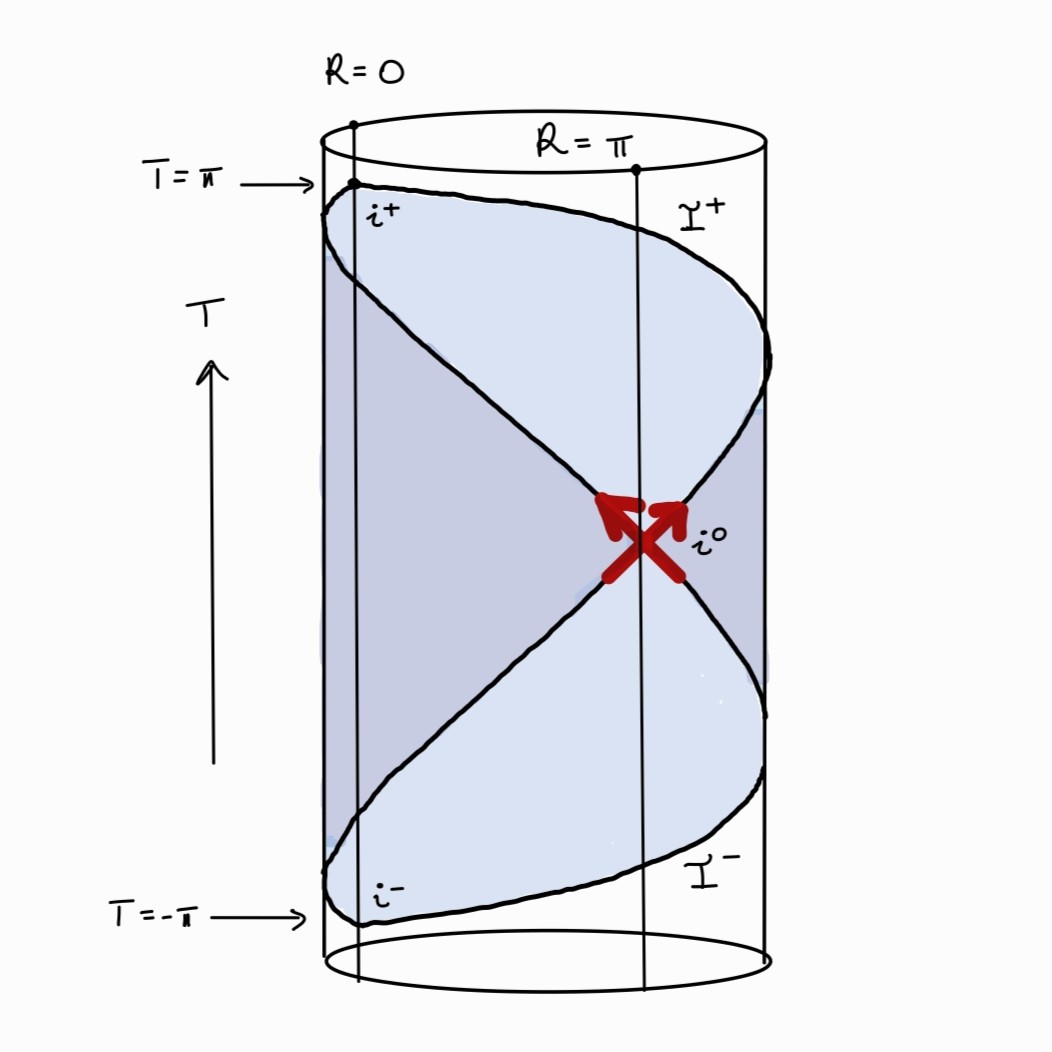}
            \caption{\label{fig:einstein static universe}}
            \end{center}
        \end{subfigure}
        \begin{subfigure}[b]{0.495\linewidth}
            \begin{center}
            \includegraphics[width=\linewidth]{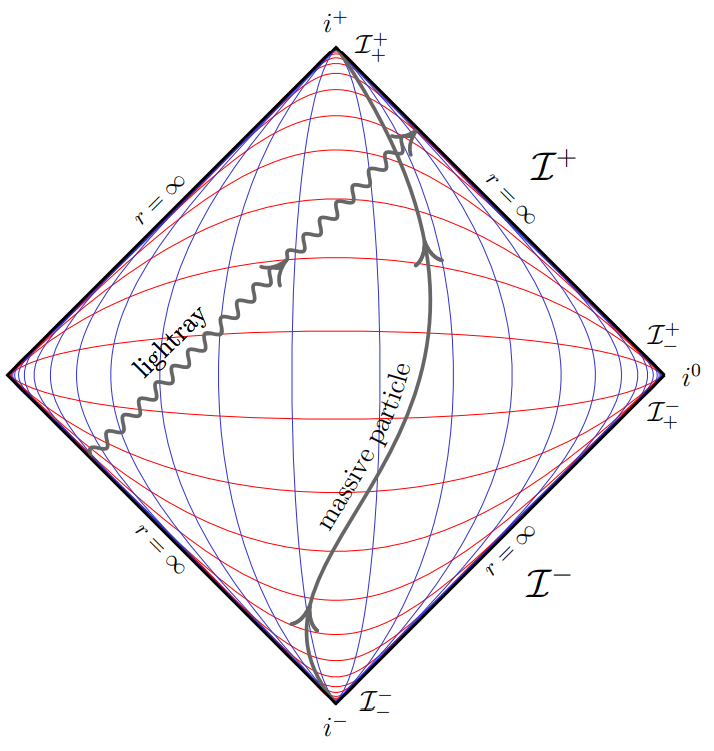}
            \caption{\label{fig:penrose minkowski}}
            \end{center}
        \end{subfigure}
        \caption{Conformal compactification of Minkowski space \cite{strominger_lectures_2017}}
        \end{figure}

        \smallskip
        Let us carry out this procedure for Minkowski, $\mathbb{R}^{1,3}$. The latter is usually described by Cartesian coordinates $(x^0, x^1, x^2, x^3) \in \mathbb{R}^{4}$ and the flat metric $\eta = \text{diag}(-1, +1, +1, +1)$, such that the invariant line element writes $\deriv s^2 = \eta_{\mu \nu} \deriv x^\mu \deriv x^\nu$. Our first action will be to transform to spherical coordinates $(x^1, x^2, x^3) \to (r, \theta, \phi)$, with $x^0 \equiv t$. In these coordinates, the line element writes 
        \begin{equation*}
            \deriv s^2 = - \deriv t^2 + \deriv r^2 + r^2 \deriv \Omega^2,\quad  \text{where } \deriv \Omega^2 = \deriv \theta^2 + \sin^2 \theta \deriv \phi^2.
        \end{equation*} We then switch to the \textit{retarded} and \textit{advanced} null coordinates
        \begin{equation}
            u = t - r, \quad v = t + r, \qquad - \infty < u \leq v < \infty, \label{eq:retarded advanced null coord}
        \end{equation} in terms of which we now have $\deriv s^2 = - \deriv u \deriv v + \frac{1}{4}(v - u)^2 \deriv \Omega^2$. Defining $U = \arctan u$ and $V = \arctan v$, we get
        \begin{equation*}
            \deriv s^2 = \frac{1}{4\cos^2 U \cos^2 V} [-4 \deriv U   \deriv V + \sin^2 (V-U) \deriv \Omega^2], \quad - \pi/2 < U \leq V < \pi/2.
        \end{equation*} Transforming back to timelike coordinate $T$ and radial coordinate $R$ via $T = V+U$, $R = V-U$, in the respective ranges $0 \leq R < \pi$, $\abs{T} + R < \pi$, we arrive at
        \begin{equation*}
            \deriv s^2 = \frac{1}{\omega^2(T,R)} \underbrace{\left( - \deriv T \otimes \deriv T + \deriv R \otimes \deriv R + \sin^2 R \deriv \Omega^2 \right)}_{=: \deriv \tilde s^2}, \quad \omega = 2 \cos U \cos V = \cos T + \cos R.
        \end{equation*} The original line element $\deriv s^2$ is \textit{conformally equivalent} to the unphysical line element $\deriv \tilde s^2$. The latter describes the manifold $\mathbb{R} \times S^3$, which is no other than a region of Einstein static universe, pictured in Fig.~\ref{fig:einstein static universe}.
         The shaded region can be unrolled to portray Minkowski space as a triangle. This is our \textit{conformal diagram}. Note that our manipulations did not affect the compact coordinates $\theta, \phi$. Hence, each point on the conformal diagram should be thought of a 2-sphere. It is convenient to mirror that triangle to obtain a symmetric picture of the conformal diagram as in Fig.~\ref{fig:penrose minkowski}. On this picture, each two-sphere of constant $(R > 0, T)$ is represented by two points, one on the left and one on the right, which are exchanged by the antipodal map.
        
        \paragraph{Asymptotic regions} The causality of Minkowski translates into several regions of interest in the conformal diagram, also pictured in Figure~\ref{fig:penrose minkowski}: first, (future/past) \textit{timelike infinity} $i^\pm$ is such that all timelike geodesics begin at $i^-$ and end at $i^+$. In the unphysical coordinates,  $i^+ = (T=\pi, R=0)$, $i^- = (T = - \pi, R=0)$. Next, (future/past) \textit{null infinity} $\cal I^\pm$ are such that all null geodesics begin at $\mathcal{I}^-$ and end at $\cal I^+$. In the unphysical coordinates, $\mathcal{I}^- = (T = -\pi + R, 0<R<\pi)$, $\mathcal{I}^+ =(T=\pi - R, 0 < R< \pi)$. These are null surfaces with topology of $\mathbb{R} \times \mathbb{S}^2$. Last but not least, \textit{spatial infinity} $i^0$ is defined such that spacelike geodesics both begin and end at $i^0$ ($i^0 = (T=0, R=\pi)$). We now build upon these concepts to define our scattering problem.

        \subsection{Gravitational scattering at null infinity: coordinate conventions \label{sec: gravitational scattering}} 
        When studying \textit{scattering} processes, one is usually interested in knowing how a given initial state of a system transforms into a final state. In quantum mechanics, this is associated to the amplitude
        \begin{equation*}
            \mathcal{A}_\text{out,in} = \langle out | in \rangle,
        \end{equation*} where the \textit{scattering matrix} ($\cal S$-matrix) relates the ingoing and outgoing states, $| out \rangle = \mathcal{S} | in \rangle$. In this work, we ultimately want to look at the scattering of massless particles (or wavepackets) in asymptotically flat spacetimes, such as gravitational waves propagating. To gain insight into this problem, we will need to specify $|in\rangle$, i.e.~provide initial data at $\cal I^-$ (since we do not consider any stable massive particles, hence need not provide initial data at $i^-$). Furthermore, one usually assumes that the theory is weakly interacting in the far past and future. Incoming wavepackets then evolve towards each other, interact and come out on $\cal I^+$ (again, we do not consider stable massive particles here, so nothing happens at $i^+$). From  Fig.~\ref{fig:penrose minkowski}, we immediately see that what goes on near $i^0$ will be important. In particular, we will need to understand how to relate fields at $\cal I^-_+$ to fields at $\cal I^+_-$ by means of suitably defined matching conditions. Both these points will be addressed in section~\ref{sec: IR structure of gravity}. We now introduce a convenient set of coordinates for describing null infinity.
        
        \paragraph{Retarded and advanced coordinates} Retarded and advanced null coordinates \eqref{eq:retarded advanced null coord} naturally parametrize $\cal I^\pm$ respectively. As a matter of fact, the Minkowski metric in these coordinates writes 
        \begin{equation}
                \deriv s^2 = - \deriv u^2 - 2\deriv u \deriv r + r^2 \gamma_{AB}(x) \deriv x^A \deriv x^B 
                = - \deriv v^2 + 2\deriv v \deriv r + r^2 \gamma_{AB}(\tilde x) \deriv \tilde x^A \deriv \tilde x^B. \label{eq: Minkowski generic}
        \end{equation} Here, $x^A$ and $\tilde x^A$ ($A\in \{2,3\}$) are antipodal coordinate systems on the asymptotic 2-sphere, while $\gamma_{AB}$ is the round metric on the 2-sphere. Future null infinity $\cal I^+$ is then best described in retarded coordinates $(u, r, x^A)$ as the boundary located at $r = \infty$, keeping $(u, x^A)$ fixed, whereas $\cal I^-$ is best described in advanced coordinates $(v, r, \tilde x^A)$ as the limit $r \to \infty$, keeping $(v, \tilde x^A)$ fixed. A specific choice for $x^A$ is preferred in the literature: stereographic coordinates.
        
\begin{figure}[h!]
            \centering
            \begin{subfigure}[b]{0.495\linewidth}
                \begin{center}
                \includegraphics[width=\linewidth]{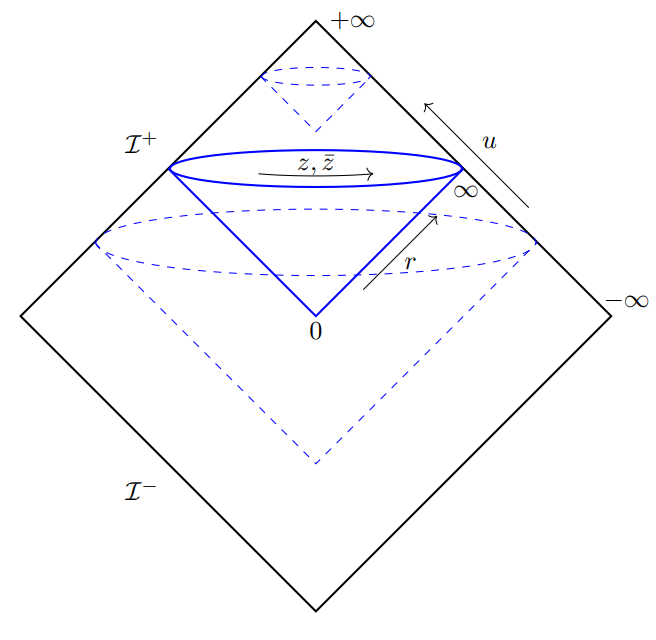}
                \caption{\label{fig:retarded minkowski}}
                \end{center}
            \end{subfigure}
            \begin{subfigure}[b]{0.495\linewidth}
                \begin{center}
                \includegraphics[width=\linewidth]{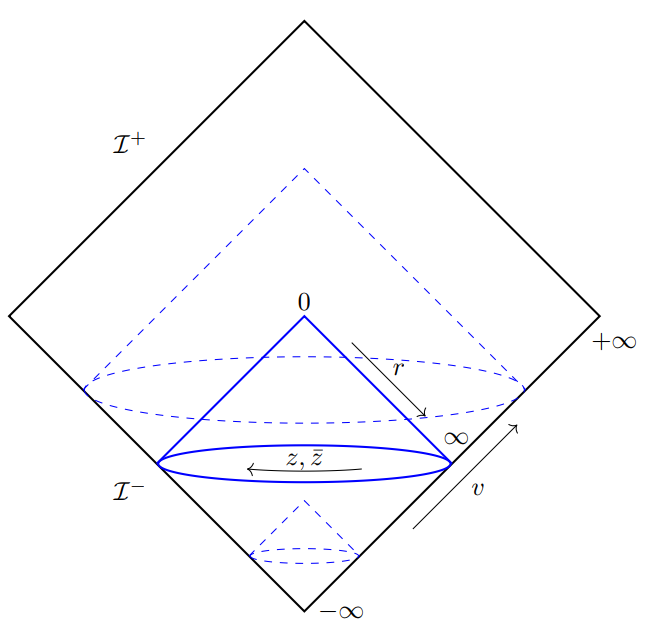}
                \caption{\label{fig:advanced minkowski}}
                \end{center}
            \end{subfigure}
            \caption{Retarded and advanced coordinates on Minkowski space \cite{strominger_lectures_2017}}
        \end{figure}

        \paragraph{Stereographic coordinates} It is useful to think of $\mathbb{S}^2$ as the Riemann sphere $\mathbb{C} \cup \infty$. Starting with standard spherical coordinates $(\theta, \phi) \in [0, \pi[ \times [0, 2 \pi[$, we use the \textit{stereographic projection} St to describe the sphere $\mathbb{S}^2$ as a complex 1-dimensional manifold \cite{prinz_lie_2022}:
        \begin{equation}
            \text{St}: \mathbb{S}^2 \backslash \{(0,0,1)\} \to \mathbb{C} \cong \mathbb{R}^2,\quad \text{St}(\theta, \phi) \mapsto z := e^{i \phi} \tan\frac{\theta}{2}, \label{eq:def stereographic coord}
        \end{equation} so that $\bar z = e^{-i\phi} \tan(\theta/2)$. This projection can be extended to a diffeomorphism $\kappa$ such that
        \begin{equation*}
            \kappa: \mathbb{S}^2 \to \mathbb{C} \cup \{\infty\}, \quad \kappa(\mathbf{z}) = \left\{ \begin{tabular}{ll}
                \text{St}($\mathbf{z}$) & $\mathbf{z} \in \mathbb{S}^2 \backslash \{(0,0,1)\}$ \\
                 $\infty$ & $\text{else}$
            \end{tabular} \right.
        \end{equation*} where we denoted the resulting complex coordinates by $\mathbf{z} = (z, \bar z)$. The sets of coordinates $(u, r, z, \bar z)$ and $(v, r, z, \bar z)$ are pictured in Figure~\ref{fig:retarded minkowski} and \ref{fig:advanced minkowski} respectively.
         The former is related to standard Cartesian coordinates by
        \begin{equation}
            u = t - \sqrt{x_i x^i}, \quad r = \sqrt{x_ix^i}, \quad z = \frac{x^1 + ix^2}{x^3 + \sqrt{x_i x^i}}, \quad \bar z = \frac{x^1 - ix^2}{x^3 + \sqrt{x_i x^i}}, \label{eq: stereographic in cartesian}
        \end{equation} and inversely
        \begin{equation}
            t = u + r, \quad x^1 + i x^2 = \frac{2 rz}{1+ z \bar z}, \quad x^3 = \frac{r(1-z \bar z)}{1+ z \bar z}. \label{eq: cartesian in stereographic}
        \end{equation}In these coordinates, the $\mathbb{S}^2$ line element takes the form $\deriv \Omega^2 = 2 \gamma_{z \bar z} \deriv z  \deriv \bar z$, with $\gamma_{z \bar z} = 2/(1 + z \bar z)^2$ and  $\gamma_{zz} = \gamma_{\bar z \bar z} = 0$. The Minkowski line element \eqref{eq: Minkowski generic} writes
        \begin{equation}
                \deriv s^2 = - \deriv u^2 - 2\deriv u  \deriv r + 2 r^2 \gamma_{z \bar z} \deriv z \deriv \bar z 
                = - \deriv v^2 + 2\deriv v \deriv r +2 r^2 \gamma_{z \bar z} \deriv z \deriv \bar z.
                \label{eq:Minkowski stereographic}
        \end{equation} From \eqref{eq:def stereographic coord} wee see that $z$ runs over entire complex plane, $z = 0$ is the north pole, $z = \infty$ is the south pole, $z \bar z = 1$ is the equator and the antipodal stereographic coordinates $(\tilde z, \tilde{\bar z})$ are related to $(z, \bar z)$ as $\tilde z = - 1/\bar z$, $\tilde{\bar z} = - 1/z$. The minus signs ensure that the $z$ in the advanced coordinates denotes the antipodal point on the sphere to the $z$ in the retarded coordinates.\footnote{Hence, for a light ray that crosses Minkowski space, the initial value of $z$ in advanced coordinates is equal to the final value of $z$ in retarded coordinates.}

    \subsection{Covariance, asymptotic symmetries and Covariant Phase Space formalism \label{sec: covariance and asymptotic symmetries}}
        In general relativity, the \textit{principle of general covariance} establishes the equivalence of all observers. This translates in gravity being a gauge theory under the gauge group of diffeomorphic automorphisms over the spacetime manifold. However, when analyzing the asymptotic behavior of the gravitational field around a boundary -- be it repelled to infinity -- the situation drastically changes. 
        
        \paragraph{Asymptotic symmetries} The presence of a boundary explicitly breaks general covariance by imposing a choice of a particular class of observers that all agree with the position of the boundary and the assorted set of boundary conditions for the dynamical fields under consideration \cite{fiorucci_leaky_2021, ciambelli_cornering_2023}. While a majority of transformations still describe a pure redundancy of the theory with zero charge, some of the residual gauge transformations preserving the structure around the boundary are promoted to physical symmetries of the theory. 
        
        One defines \textit{residual symmetries} as the symplectomorphisms preserving the dynamics in the bulk and the boundary conditions. Among these, \textit{trivial symmetries} still have vanishing Noether charges (and therefore are true redundancies of the system), while \textit{asymptotic symmetries}\footnote{The name is rather misleading, since these are not approximate but \textit{exact} symmetries of the theory in the asymptotic region (e.g.~infinity) of spacetime.} now acquire non-vanishing Noether charges. They are, therefore, physical transformations acting non-trivially on the field space, mapping the system into an inequivalent configuration. From the perspective of the scattering problem mentioned in the previous section, it will be useful to study the action of the \textit{asymptotic symmetry group} on the $\cal S$-matrix,
        \begin{equation*}
            \text{asymptotic symmetry group} = \frac{\text{residual gauge symmetries}}{\text{trivial gauge symmetries}}.
        \end{equation*} In fact, the restriction to asymptotic symmetries is needed if one wishes to deal with a well-defined Poisson bracket of physical charges -- see the notes of Ciambelli et al.~for a pedagogical review \cite{ciambelli_cornering_2023}. 
        
        \smallskip
        In the following, looking at radiative gravity in four dimensions will lead us to work in asymptotically flat spacetimes (AFS), for which boundary conditions are needed (section~\ref{sec:gravity in AFS}). There is a large freedom in the choice of falloffs and gauge conditions one can pick, provided these are weak enough to allow for all physically reasonable solutions, but strong enough to permit the construction of the charges of asymptotic symmetries. We will follow the analysis of Bondi, van der Burg and Metzner \cite{bondi_gravitational_1962} to derive the asymptotic symmetry group of asymptotically flat spacetimes at null infinity: the BMS group. Most of section~\ref{sec: IR structure of gravity} will then be devoted to investigating the effect of the relevant charges on the $\cal S$-matrix and their implications on the infrared structure of gravity. Before doing so, we now formulate a prescription for deriving the charges associated to asymptotic symmetries.

        \smallskip
        We have just fleshed out the implications of non-trivial conserved charges for gravitational scattering. While the scattering problem is usually most naturally discussed within the Hamiltonian formalism, in relativistic theories however it is difficult to use since the choice of a preferred set of time slices inevitably destroys manifest covariance. This problem is common to all relativistic covariant field theories, and usually avoided by restricting to Lagrangians (classical) and path integrals  (quantum) approaches, but there remain some applications, such as the initial value problem, for which the tools inherited from the Hamiltonian formalism are too convenient to dispense with.\footnote{For example it is only in the Hamiltonian formalism that one can do a proper accounting of the degrees of freedom in a system and define thermodynamic quantities such as energy and entropy \cite{harlow_covariant_2020}.}
        
\begin{figure}[!htbp]
            \vspace{0.3cm}
            \centering
            \begin{tikzpicture}[node distance=1.5cm]
                \node (start) [box] {Spacetime $(\mathcal{M},g)$};
                \node (in1) [box, right of=start, xshift=4cm] {Configuration space $\mathfrak{F}$};
                \node (in2) [nobox, below of=in1] {Lagrangian form $L = \epsilon \mathcal{L}$};
                \node (in3) [nobox, below of=in2] {$\delta L$, equations of motion};
                \node (in4) [box, left of=in3, xshift=-4cm] {Solution space $\mathfrak{S} \subset \mathfrak{F}$};
                \node (in5) [nobox, below of=in4] {define $\omega$};
                \node (in6) [nobox, below of=in5] {obtain $\tilde{\Omega}_\Sigma$ and $\tilde{\Theta}_\Sigma$};
                \node (in7) [nobox, right of=in6, xshift=4cm, text width=5cm] {identify equivalence class of degeneracy subspaces};
                \node (end) [box, right of=in7, xshift=4cm] {Phase space $\Gamma \equiv \mathfrak{S}/\sim$};
                \node (extra) [nobox, right of=in2, xshift=4cm] {Canonical transformations \& Poisson brackets};
                
                \draw [arrow] (start) -- node[midway, above=6pt, align=center, anchor=east] {1} (in1);
                \draw [arrow] (in1) -- node[anchor=south] {} (in2);
                \draw [arrow] (in2) -- node[anchor=south] {} (in3);
                \draw [arrow] (in3) -- node[midway, above=6pt, align=center, anchor=east] {2} (in4);
                \draw [arrow] (in4) -- node[midway, right=15pt, align=center, anchor=east] {3} (in5);
                \draw [arrow] (in5) -- node[midway, right=15pt, align=center, anchor=east] {4} (in6);
                \draw [arrow] (in6) -- node[midway, above=6pt, align=center, anchor=east] {5} (in7);
                \draw [arrow] (in7) -- node[midway, above=6pt, align=center, anchor=east] {6} (end);
                \draw [arrow] (end) -- node[midway, left=1cm, align=center, anchor=north, text width=2cm] {symplectic geometry} (extra);
            \end{tikzpicture}
            \caption{Covariant phase space formalism \label{fig:CPS formalism}}
        \end{figure}
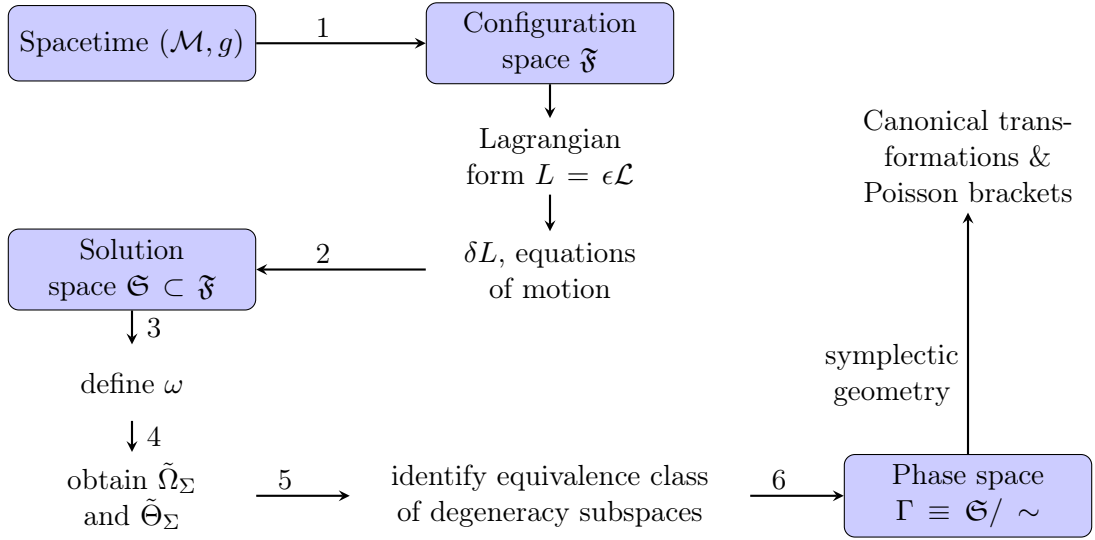

        \paragraph{CPS procedure} Crnkovic, Witten, Iyer, Lee, Wald and Zoupas \cite{crnkovic_covariant_1986, crnkovic_symplectic_1988, lee_local_1990, iyer_properties_1994, wald_general_2000} developed a formalism which incorporates the powerful features of phase space analyses without abandoning covariance: the \textit{covariant phase space} (CPS) formalism. It allows to build the phase space\footnote{What physicists refer to as a \textit{phase space} is also known as \textit{symplectic manifold} $(\Gamma, \Omega)$ by mathematicians, that is a smooth manifold $\Gamma$ equipped with a closed, degenerate two-form $\Omega$. Recall a $q$-form $\Omega \in \Omega^q(\Gamma)$ is an antisymmetric $(0,q)$ tensor field on $\Gamma$, which is said to be \textit{closed} if: $\deriv  \Omega = 0$, and \textit{non-degenerate} if: $i_X \Omega= 0 \Rightarrow X = 0$, $\forall X \in T \Gamma$.} of a covariant theory starting from its Lagrangian, without having to explicitly refer to the Hamiltonian. The construction of surface charges at the boundary then builds upon the variational principle\footnote{Another definition for the surface charges was developed in parallel by Barnich and Brandt \cite{barnich_covariant_2002}, but relying on the equations of motion rather than the variational principle \cite{fiorucci_leaky_2021}.} as well as symplectic methods. At the end of the day, the CPS formalism boils down to an algorithm involving six essential steps, schematically pictured in Fig.~\ref{fig:CPS formalism} and summarized in what follows.
        \begin{enumerate}
            \item Starting from a four-dimensional spacetime $(\mathcal{M},g)$, the configuration space $\mathfrak{F}$ is first obtained as the space of all allowed field configurations, defined by imposing boundary conditions on the field of the generally covariant theory. The Lagrangian density $\cal L$ governing the dynamics of the system can then be constructed. It generally depends both on the metric $g_{ab}$, the matter fields $\psi$ and a finite number of their derivatives, which we combine under the collective variable $\phi \equiv (g_{ab}, \psi)$ in the following. Under the variation $\phi \to \phi + \delta \phi$, $\cal L$ changes as $\mathcal{L} \to \mathcal{L} + \delta \mathcal{L}$, with
            \begin{equation}
                \delta \mathcal{L} = \delta \phi^i \frac{\partial \mathcal{L}}{\partial \phi^i} + \partial_\mu \delta \phi^i \frac{\partial \mathcal{L}}{\partial(\partial_\mu \phi^i)} + ... \equiv \delta \phi^i \mathcal{E}_i + \partial_\mu \theta^\mu [\phi, \delta \phi],
                \label{eq: variational principle}
            \end{equation} where
            \begin{equation*}
                \mathcal{E}_i \equiv \frac{\partial \mathcal{L}}{\partial \phi^i} - \partial_\mu \left( \frac{\partial \mathcal{L}}{\partial (\partial_\mu \phi^i)} \right) + \partial_\mu \partial_\nu \left( \frac{\partial \mathcal{L}}{\partial (\partial_\mu \partial_\nu \phi^i)} \right) + ...
            \end{equation*} are the \textit{Euler-Lagrange} terms and $\theta$ is the \textit{symplectic potential current density} obtained by successive application of the Leibniz rule to get rid of the terms of the form $\partial_\mu \delta \phi^i$, $\partial_\mu \partial_\nu \delta \phi^i$, etc.~and factor out the variation $\delta \phi^i$ in front of $\mathcal{E}_i$. From the perspective of the variational bicomplex, the \textit{Lagrangian form} $L \equiv \epsilon \mathcal{L}$ is a $4$-form on $\cal M$ and a function on $\mathfrak{F}$ \cite{he_covariant_2020}.
            In the language of differential forms, \eqref{eq: variational principle} writes
            \begin{equation}
                \delta L = \delta \phi^i \frac{\delta L}{\delta \phi^i} + \deriv \theta[\phi, \delta \phi]. \label{eq: variational principle form}
            \end{equation} $\theta$ is now the $3$-form on $\mathcal{M}$ associated with the symplectic current potential. By construction, it is defined only up to a closed form in spacetime $\kappa \in \Omega^{2}(\mathcal{M})$, $\theta \to \theta + \deriv \kappa$.
            \item Requiring that the variational principle hold amounts to the Euler-Lagrange equations  $\mathcal{E}_i = 0$. This takes us to the solution space $\mathfrak{S} \subset \mathfrak{F}$. Configurations in $\mathfrak{S}$ are said to be \textit{on-shell}.
            \item Next, we define the symplectic current density $\mathbf{\omega}$ as the exterior derivative of $\theta$ on $\mathfrak{S}$,
            \begin{equation*}
                \omega = \deriv \theta \in \Omega^{3}(\mathcal{M}) \times \Omega^2(\mathfrak{S}).
            \end{equation*} By construction, $\omega$ is closed in $\mathfrak{S}$ and $\cal M$ \cite{he_covariant_2020}. For two field variations $\delta_1 \phi$ and $\delta_2 \phi$, we have
            \begin{equation}
                \omega[\phi, \delta_1 \phi, \delta_2 \phi] = \delta_1 \theta[\phi, \delta_2 \phi] - \delta_2 \theta[\phi, \delta_1 \phi]. \label{eq: symplectic form current density}
            \end{equation}
            \item Our next task is to integrate $\omega$ and $\theta$ over a Cauchy slice\footnote{A Cauchy surface is a subset of the manifold which is intersected by every maximal causal curve exactly \textit{once}. Once the initial data is fixed on such a codimension 1 surface, the field equations lead to the evolution of the system in the entire spacetime \cite{compere_advanced_2019}.} $\Sigma$ of $\mathcal{M}$. This yields the \textit{presymplectic form}
            \begin{equation*}
                \tilde \Omega_\Sigma [\phi, \delta_1 \phi, \delta_2 \phi] \equiv \int_\Sigma \omega[\phi, \delta_1 \phi, \delta_2 \phi]
            \end{equation*} and \textit{presymplectic potential} $\tilde \Theta_\Sigma$ such that $\tilde \Omega_\Sigma = \deriv \tilde \Theta_\Sigma$. By construction the former is closed and thus a good candidate for the symplectic form of the phase space of our theory, but nothing guarantees that it is indeed non-degenerate (hence the name "presymplectic"). In particular, the freedom in defining $\theta$ up to a closed 2-form in \eqref{eq: variational principle form} implies that $\tilde \Omega_\Sigma$ is not uniquely defined.
            \item We thus need to construct the equivalence classes [$\sim$] of degeneracy subspaces of $\tilde \Omega_\Sigma$, and then quotient $\mathfrak{S}$ by the latter. This fixes $\Theta_\Sigma = \tilde{\Theta}_\Sigma|_\Gamma$ and $\Omega_\Sigma = \tilde{\Omega}_\Sigma|_\Gamma$. The latter is now nondegenerate and  independent of $\Sigma$ provided the variations $\delta_1 \phi$ and $\delta_2 \phi$ obey the equations of motion \cite{alessio_note_2019}.
            \item The phase space of our theory is $\Gamma = \mathfrak{S}/\sim$, with associated symplectic form $\Omega_\Sigma$ and symplectic potential $\Theta_\Sigma$. 
        \end{enumerate} 

        \paragraph{Canonical transformations} We sketched how one constructs the symplectic form for a theory on a given Cauchy surface. Let us now see how the conserved charges follow. Omitting the reference to $\Sigma$ for now, we have \cite{he_covariant_2020}
        \begin{equation*}
            \Omega = \deriv \Theta \text{ (up to closed one-form) } \Rightarrow \Omega(X, Y) = X(\Theta(Y)) - Y(\Theta(X)) - \Theta([X, Y]).
        \end{equation*} In particular, $\Omega \in \Omega^2(\Gamma)$ but can also be thought as $\Omega \in \Omega^1(\Gamma)$ by leaving one slot empty,
        \begin{equation}
            \Omega: T \Gamma \to \Omega^1(\Gamma), \quad \Omega(X) \equiv - \iota_X \Omega \quad \text{such that } \Omega(X)(Y) = - \Omega(Y)(X) = - \Omega(X, Y), \label{eq:symplectic one form}
        \end{equation} with $\iota_X$ denoting the \textit{interior derivative}\footnote{The interior product $\iota_X \Omega$ is the contraction of the differential form $\Omega$ with the vector field $X$, $\iota_X : \Omega^p(M) \to \Omega^{p-1}(M)$ with $(\iota_X \Omega)(X_1, ..., X_{p-1}) = \Omega(X, X_1, ..., X_{p-1})$.} on $\Gamma$ and $T\Gamma$ the tangent space to the symplectic manifold $\Gamma$. $\Omega$ being non-degenerate, one can find an inverse map $\Omega^{-1}: \Omega^1(\Gamma) \to T\Gamma$ that also acts as antisymmetric bilinear map acting on $\Omega^1(\Gamma) \times \Omega^1(\Gamma)$. Diffeomorphisms on $\Gamma$ that preserve the symplectic form $\Omega$ are called \textit{canonical transformations}. They are generated by \textit{Hamiltonian vector fields} $X_f$ obeying 
        \begin{equation}
            \mathcal{L}_{X_f} \Omega = 0, \label{eq: canonical transformation}
        \end{equation} with $\cal L$ the Lie derivative with respect to the metric on our spacetime. Using Cartan's homotopy formula $\mathcal{L}_X = \deriv \iota_X + \iota_X \deriv$ and the closedness of $\Omega$, one obtains
        \begin{equation*}
            \mathcal{L}_{X_f} \Omega = \deriv \iota_{X_f} \Omega + \iota_{X_f} \cancel{\deriv \Omega} \overset{!}{=} 0 \Rightarrow \iota_{X_f} \Omega = - \deriv f \in \Omega^1(\Gamma),        
        \end{equation*} for $f \in \mathcal{F}(\Gamma)$ the \textit{Hamiltonian charge}. $\deriv f$ can here be seen as the infinitesimal charge associated with the symmetry generated by the Hamiltonian vector field $X_f$. Using \eqref{eq:symplectic one form}, we can write an explicit relation between this infinitesimal charge and the symplectic (one-)form:
        \begin{equation}
            \Omega(X_f) = \deriv f \Leftrightarrow X_f = \Omega^{-1}(\deriv f). \label{eq: symplectic form and hamiltonian charge}
        \end{equation} Restoring the reference to the Cauchy slice of interest, we write
        \begin{equation}
            \cancel{\delta} Q_\xi [\phi, \delta \phi] = \int_{\Sigma} \omega[\phi, \delta \phi, \delta_\xi \phi] = \Omega_{\Sigma}[\phi, \delta \phi, \delta_\xi \phi] \label{eq: infinitesimal charge}
        \end{equation} for the infinitesimal charge associated to a symmetry generated by the vector field $\xi$. Here, we introduced the notation $\cancel{\delta}$ used by Alessio and Arzano \cite{alessio_note_2019} to emphasize that \eqref{eq: infinitesimal charge} might not be an exact differential, or \textit{integrable}. It is straightforward to build upon this symplectic structure and define the Hamiltonian charges associated to canonical transformations generated by the Lie bracket of Hamiltonian vector fields, which we call \textit{Poisson bracket}. Let $X_f$ and $X_g$ be two such vector fields corresponding to two Hamiltonian charges $f,g$ on phase space respectively. Then \eqref{eq: canonical transformation} $\Rightarrow$ \eqref{eq: symplectic form and hamiltonian charge} can be rewritten as
        \begin{equation*}
            \mathcal{L}_{[X_f, X_g]} \Omega = [ \mathcal{L}_{X_f}, \mathcal{L}_{X_g}] \Omega = 0 \Rightarrow \deriv h = \Omega(X_f, X_g) \equiv - \{f, g\} \quad \text{(Poisson bracket).}
        \end{equation*} With this definition, we recover the usual properties of the Poisson brackets in analytical mechanics.

    \subsection{Symmetries of flat space and Poincaré group}
        Before looking at gravity in asymptotically flat spacetimes, let us briefly identify the symmetries of flat space. In $3+1$ dimensional Minkowski, there are ten isometries which form the well-known \textit{Poincaré group}. The generators $\zeta$ of the latter solve the Killing equation
        \begin{equation*}
            \mathcal{L}_\zeta g_{ab} = 0 \Leftrightarrow \nabla_a \zeta_b + \nabla_b \zeta_a = 0.
        \end{equation*} This gives rise to three \textit{rotation} generators
        \begin{equation}
            \xi_{12} = x^1 \partial_2  - x^2 \partial_1 , \quad \xi_{13} = - x^1 \partial_3 + x^3 \partial_1, \quad \xi_{32} = x^2 \partial_3 - x^3 \partial_2 \label{eq: rotation generators}
        \end{equation} and three \textit{boost} generators
        \begin{equation}
            \xi_{01} = x^1 \partial_0 + x^0 \partial_1, \quad \xi_{02} = x^2 \partial_0 + x^0 \partial_2, \quad \xi_{03} = x^3 \partial_0 + x^0 \partial_3 \label{eq: boost generators}
        \end{equation} which together form the \textit{Lorentz group}. On top of this, one also gets an Abelian normal subgroup of four spacetime \textit{translations}
        \begin{equation}
            \xi_0 = \partial_0, \quad \xi_1 = \partial_1, \quad \xi_2 = \partial_2, \quad \xi_3 = \partial_3. \label{eq: translation generators}
        \end{equation} As a result, we have the group structure: $\text{Poincaré} = \text{Lorentz} \ltimes \text{Translations}.$ We do not prove these standard results here and refer the interested reader to e.g.~the work of Hirata \cite{hirata_lecture_2011} or Compère and Fiorucci \cite{compere_advanced_2019} for the construction of these generators and their associated charges. We now turn to the asymptotic symmetries of general relativity in asymptotically flat spacetimes. In particular, we will see that the asymptotic symmetry group of such spaces is larger than the Poincaré group, meaning that general relativity does not reduce to special relativity at large distances! 

\section{Asymptotic symmetries of asymptotically flat spacetimes \label{sec:gravity in AFS}}

    In section~\ref{sec: covariance and asymptotic symmetries}, we explained how boundary conditions give rise to asymptotic symmetries. We now apply these ideas to look at asymptotically flat spacetimes. We first define the latter and then introduce a metric treatment that will be convenient of the study of gravitational scattering: the Bondi-Sachs formalism. We later on make good use of this formalism to derive the asymptotic symmetry group of asymptotically flat spacetimes, the BMS group.

    \subsection{Asymptotically flat spacetimes}
        In general relativity, an \textit{asymptotically flat spacetime} (AFS) corresponds to the intuitive notion of an \textit{isolated system}. The formal definition of the latter is however not so straightforward, notably because the metric acts both as physical field and background. Consider a system alone in the universe, described by a spacetime $\cal M$. As one recedes from the system, we expect its influence to decrease, so we expect $\cal M$ to resemble flat Minkowski spacetime $\mathbb{R}^{1,3}$, with this approximation becoming even better the farther away we go. This is the intuitive picture. The formal definition of an AFS involves the concept of \textit{asymptotically simple} spacetime \cite{wald_general_1984, frauendiener_asymptotic_2006}. We omit the formal definition here for conciseness, only stating that an asymptotically simple spacetime provides all the necessary conditions to perform a conformal compactification in the spirit of what we did in section~\ref{sec: conformal compactification Minkowski}. To arrive at an AFS, one must make additional assumptions on behaviour of the curvature near $\cal I$.
        
        \paragraph{Asymptotic flatness} With an asymptotically simple spacetime at hand, there are two ways to define asymptotic flatness according to Compère and Fiorucci \cite{compere_advanced_2019}, either
        \begin{enumerate}
            \item using covariant objects but involving unphysical fields such as a conformal factor used to do a Penrose compactification of spacetime (one would then expect the causal structure of the AFS to resemble the one of Minkowski seen in section~\ref{sec: conformal compactification Minkowski}); or
            \item using an adapted coordinate system and specifying fall-off conditions.
        \end{enumerate}  We follow the second route in this article, which allows for an easier analysis of the details the asymptotic structure, even though it involves a choice of coordinates which makes unclear whether the definition is still covariant. Both approaches are of course expected to be equivalent, and the choices made by BMS can indeed be justified in terms of Penrose compactification of $\mathcal{I}^+$ \cite{tamburino_gravitational_1966, madler_bondi-sachs_2016}. 
        
    \subsection{Bondi-Sachs metric}

        The description of an AFS by means of suitable coordinates and falloffs is due to BMS \cite{bondi_gravitational_1962}, who described the propagation of gravitational waves in four-dimensional AFS endowed with an additional axial and reflexion symmetry ($\phi \to - \phi$). To do so, they introduced a convenient choice of metric satisfying the so-called \textit{Bondi gauge}.
    
        \paragraph{Bondi gauge} As motivated in section~\ref{sec: gravitational scattering}, it is convenient to use set of coordinates $(u, r, \theta, \phi)$, where $u$ is the retarded time encountered previously in \eqref{eq:retarded advanced null coord}, and $\theta, \phi$ the usual spherical coordinates. One can foliate the original spacetime in a family of null hypersurfaces $\Gamma_c = \{(u,r,x^A): u = c\}$ and define $r$ to be the future-pointing null radial coordinate along these 
        \begin{figure}[h!]
            \centering
            \includegraphics[width=0.65\textwidth]{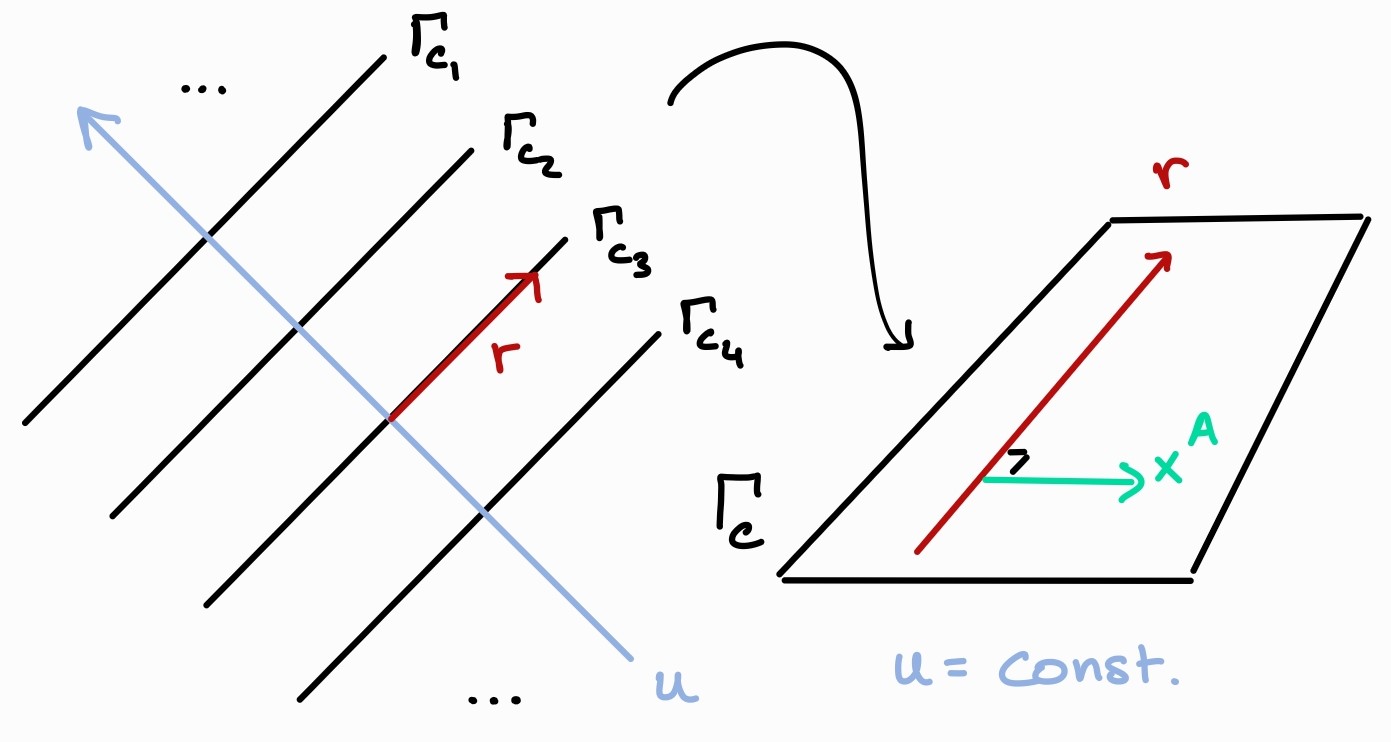}
            \caption{Bondi coordinates}
            \label{fig:Bondi coord}
        \end{figure}
        hypersurfaces, while $x^A$ ($A=2,3$) are two compact angular coordinates. This setup is schematically depicted in Figure~\ref{fig:Bondi coord}.
        Null hypersurfaces have null normal vector $k_\mu = - \partial_\mu u$, which is also tangent to the surface \cite{aretakis_lecture_2012}. In particular, 
        \begin{equation*}
            k^2 = k^\mu k_\mu  = g^{\mu \nu} (\partial_\mu u) (\partial_\nu u) = 0 \quad \Rightarrow \quad g^{uu} = 0.
        \end{equation*}Since $x^A$ and $u$ are orthogonal directions we also have 
        \begin{equation}
            k^\mu \partial_\mu x^A = - g^{\mu \nu} (\partial_\mu u)(\partial_\nu x^A) = 0 \quad \Rightarrow \quad g^{uA} = 0. \label{eq: guA}
        \end{equation} Hence for $g^{\mu \nu}$ to be non-degenerate we need $g^{ur} \neq 0$. Actually, we can be more precise. For $X$ and $Y$ two causal vectors belonging to the same connected component of the light cone, one must have that $g_{ab}X^aY^b \leq 0$ \cite{reall_part_2022} so in our case $g^{ur} < 0$. Schematically, 
        \begin{equation*}
            g^{\mu \nu} = \begin{pmatrix}
                0 & g^{ur} & 0 & 0 \\
                g^{ur} & g^{rr} & g^{r\theta} & g^{r\phi} \\
                0 & g^{r\theta} & g^{22} & g^{33} \\
                0 & g^{r\phi} & g^{23} & g^{33}
            \end{pmatrix}^{\mu \nu}
        \end{equation*} where we introduced the inverse $2 \times 2$ metric $g^{AB}$ for the angular components. Inverting $g^{\mu \nu}$, one can easily check that we get $g_{rr} = g_{rA} = 0$. Meanwhile, the radial coordinate $r$ is still unspecified. The choice of BMS was to define it as the \textit{radial luminosity distance} along a null geodesic and impose
        \begin{equation}
            \partial_r \text{det}(r^{-2} g_{AB}) = 0. \label{eq:determinant gauge condition}
        \end{equation} We have arrived at the \textit{Bondi gauge}
        \begin{empheq}[box=\widefbox]{equation}
            g_{rr} = g_{rA} = 0 \quad \text{and} \quad \partial_r \text{det}\left(\frac{g_{AB}}{r^2}\right) = 0. \label{eq: bondi gauge}
        \end{empheq} It is worth emphasizing that  our construction is possible for any metric, since in $D$ dimensions there are $D$ coordinate choices to be made, and the Bondi gauge amounts precisely to $1 + (D-2) + 1 = D$ conditions. Let us now see how to parametrize the metric in this gauge.

        \paragraph{Bondi-Sachs metric} We now construct the metric in the Bondi gauge. We saw that $g^{ur} < 0$. A natural parametrization is thus $g^{ur} = - e^{2\beta}$ with $\beta \equiv \beta(u,r,x^A)$. We can then define $g^{rr} \equiv e^{-4 \beta} U$ and $g^{rA} \equiv \frac{1}{2}e^{-2 \beta} U^A$ with $U,U^A$ all functions of $(u,r,x^A)$ to get
        \begin{equation}
            g^{\mu \nu} = 
            \begin{pmatrix}
                0 & - e^{-2\beta} & 0 & 0 \\
                - e ^{-2 \beta} & e^{-4 \beta} U & \frac{1}{2}e^{-2 \beta}U^2 & \frac{1}{2}e^{-2 \beta}U^3 \\
                0 & \frac{1}{2}e^{-2 \beta}U^2 & g^{22} & g^{23} \\
                0 & \frac{1}{2}e^{-2 \beta}U^3 & g^{23} & g^{33}
            \end{pmatrix}^{\mu \nu}. \label{eq: full covar matrix}
        \end{equation} The form of these coefficients is motivated by the work of BMS \cite{bondi_gravitational_1962}, albeit slightly different in order to match the notation of Strominger \cite{strominger_lectures_2017} and simplify calculations later on; the key point being that $g^{ur} < 0$. Inverting \eqref{eq: full covar matrix}, one finds for the covariant components 
        \begin{equation}
            g_{\mu \nu} = 
            \begin{pmatrix} 
            -U + \frac{1}{4} g_{AB} U^A U^B & - e^{2 \beta} & \frac{1}{2}g_{2A}U^A & \frac{1}{2}g_{3A} U^A \\
            -e^{2 \beta} & 0 & 0 & 0 \\
            \frac{1}{2}g_{2A}U^A & 0 & g_{22} & g_{23} \\
            \frac{1}{2}g_{3A} U^A & 0 & g_{23} & g_{33}
            \end{pmatrix}_{\mu \nu}. \label{eq: Bondi metric}
        \end{equation} Such a metric corresponds to the line element
        \begin{empheq}[box=\widefbox]{equation}
            \deriv s^2 = - U \deriv u^2 -  2e^{2 \beta} \deriv u \deriv r + g_{AB}\left(\deriv x^A+ \frac{1}{2}U^A \deriv u\right) \left(\deriv x^B + \frac{1}{2}U^B \deriv u\right). \label{eq: Bondi Sachs metric}
        \end{empheq} This is the \textit{retarded Bondi-Sachs metric}. Choosing the advanced time $v$ instead of $u$ in our parametrization similarly yields the \textit{advanced Bondi-Sachs metric}
        \begin{empheq}[box=\widefbox]{equation*}
            \deriv s^2 = - V \deriv v^2 + 2 e^{2 \beta^-}\deriv v \deriv r + g_{AB}^-\left(\deriv x^A - \frac{1}{2} V^A \deriv v\right) \left(\deriv x^B - \frac{1}{2}V^B \deriv v\right),
        \end{empheq} where we introduced $\beta^-$, $g_{AB}^-$, $V$ and $V^A$. Switching $x^A = (\theta, \phi) \to (z, \bar z)$ as in \eqref{eq:def stereographic coord}, we have
        \begin{multline}
            \deriv s^2 = - U \deriv u^2 - 2e^{2 \beta} \deriv u \deriv r + \left[g_{zz} \left(\deriv z + \frac{1}{2} U^z \deriv u\right)^2 +  g_{z\bar z}\left(\deriv z+ \frac{1}{2}U^z \deriv u\right) \left(\deriv \bar z + \frac{1}{2}U^{\bar z} \deriv u\right) + \text{c.c.}\right] \\
             = - U \deriv u^2 - 2e^{2\beta} \deriv u \deriv r + \left[g_{zz}\left(\deriv z^2 + U^z \deriv u \deriv z + \frac{1}{4}U^z U^z \deriv u^2 \right)\right.  \\ \left. + g_{z \bar z} \left(\deriv z \deriv \bar z+ \frac{1}{2}U^{\bar z} \deriv u \deriv z + \frac{1}{2}U^z \deriv u \deriv \bar z + \frac{1}{4} U^z U^{\bar z} \deriv u^2\right) + \text{c.c.}\right],
            \label{eq: Bondi Sachs metric stereographic}
        \end{multline} while for advanced coordinates one gets
        \begin{multline}
            \deriv s^2 = - V \deriv v^2 - 2e^{2\beta^-} \deriv v \deriv r + \left[g^-_{zz}\left(\deriv z^2 - V^z \deriv v \deriv z + \frac{1}{4}V^z V^z \deriv v^2 \right)\right.  \\ \left. + g^-_{z \bar z} \left(\deriv z \deriv \bar z- \frac{1}{2}V^{\bar z} \deriv v \deriv z - \frac{1}{2}V^z \deriv v \deriv \bar z + \frac{1}{4} V^z V^{\bar z} \deriv v^2\right) + \text{c.c.}\right]. \label{eq: advanced Bondi Sachs metric stereographic}
        \end{multline} Again, we stress that this construction is general, meaning that \textit{any} metric can be put in this form. 
        
        \paragraph{Asymptotics at infinity} Now, if we want to focus on AFS, we have to impose falloff conditions on the components of the metric. At $\cal I^+$, this amounts to performing an analytic expansion\footnote{It is assumed \cite{bondi_gravitational_1962} that for any choice of $u$ one can take the limit $r \to \infty$ along each ray. Newman and Unti \cite{newman_behavior_1962} replaced this condition with a weaker statement. As a matter of fact, Penrose \cite{penrose_asymptotic_1963, friedrich_peeling_2018} showed that the \textit{peeling theorem} is violated in four dimensions, and we cannot expect the expansion to be analytic in general. However, in turns out that the analycity assumption is sufficient in the case of BMS supertranslations, which is precisely within the scope of our considerations in this article.} of the coefficients $U, \beta$ and $U^A$ in powers of $1/r$ at large distances. There is no \textit{a priori} preferred method to determine these falloff conditions on the metric components. Following \cite{bondi_gravitational_1962,sachs_asymptotic_1962, flanagan_conserved_2017} we write:
        \begin{empheq}[box=\widefbox]{equation}
            \begin{aligned}
                U(u,r,x^A) &= 1 - \frac{2m(u,x^A)}{r} + \frac{U_2(u,x^A)}{r^2} + \mathcal{O}(r^{-3}), \\
                \beta(u,r,x^A) &= \frac{\beta_1(u,x^A)}{r} + \frac{\beta_2(u,x^A)}{r^2} + \frac{\beta_3(u,x^A)}{r^3} + \mathcal{O}(r^{-4}), \\
                U^A(u,r,x^B) &= \frac{U^{A}_{2}(u,x^B)}{r^2} + \frac{U^A_3(u,x^B)}{r^3} + \frac{U^A_4(u,x^B)}{r^4} +\mathcal{O}(r^{-5}), \\
                g_{AB}(u,r,x^C) &= r^2 \gamma_{AB}(x^C) + r C_{AB}(u,x^C) + D_{AB}(u,x^C) + \mathcal{O}(r^{-1}).
            \end{aligned} \label{eq: falloffs metric}
        \end{empheq} We adopt the convention that capital Roman indices $A, B$ etc.~are raised and lowered with the  round metric on the 2-sphere $\gamma_{AB}$ and its inverse $\gamma^{AB}$, and we denote by $D_A$ the covariant derivative associated with $\gamma_{AB}$. In particular, we have that the trace of $C_{AB}$ vanishes by \eqref{eq:determinant gauge condition}, since
        \begin{equation*}
            \text{det} \left( \frac{g_{AB}}{r^2} \right) = \text{det}\left( \gamma_{AB} + \frac{C_{AB}}{r} + \mathcal{O}(r^{-2}) \right) = \text{det}(\gamma) \left( 1 + \frac{C^A\,_A}{r} + \mathcal{O}(r^{-2}) \right)
        \end{equation*} and thus $\partial_r \text{det}(g_{AB}/r^2) = 0$ requires $\gamma^{AB}C_{AB} = 0$ to $\mathcal{O}(r^{-1})$. For $x^A = (z, \bar z)$ the stereographic coordinates \eqref{eq:def stereographic coord}, 
        \begin{equation*}
            \gamma^{AB} C_{AB} = 0 \Leftrightarrow 2 \frac{(1+ z \bar z)^2}{2} C_{z \bar z} = 0 \Rightarrow C_{z \bar z} = 0,
        \end{equation*} while $C_{zz}$ and $C_{\bar z \bar z}$ are left unspecified. From \eqref{eq: Bondi Sachs metric stereographic} we can read off the metric coefficients
        \begin{gather}
            g_{uu} = - U +  \frac{1}{4} g_{zz} U^z U^z + \frac{1}{4} g_{\bar z \bar z} U^{\bar z} U^{\bar z} + \frac{1}{2} g_{z \bar z} U^z U^{\bar z}, \quad \text{and} \quad g_{ur} = -e^{2 \beta} \\ 
            g_{uz} = \frac{1}{2} g_{zz} U^z + \frac{1}{2} g_{z \bar z} U^{\bar z}, \quad \text{and} \quad
            g_{u \bar z} = \frac{1}{2} g_{\bar z \bar z} U^{\bar z} + \frac{1}{2} g_{z \bar z} U^z. \label{eq: guz guzbar}
        \end{gather} The angular components of the metric write to $\mathcal{O}(r)$
        \begin{equation*}
            g_{zz} = r^2 \cancel{\gamma_{zz}} +  rC_{zz}, \quad
            g_{\bar z \bar z} = r^2 \cancel{\gamma_{\bar z \bar z}} +  rC_{\bar z \bar z},  \quad \text{and} \quad
            g_{z \bar z} = r^2 \gamma_{z \bar z} + r \cancel{C_{z \bar z}}.
        \end{equation*} These together with the falloffs \eqref{eq: falloffs metric} for $U$, $\beta$ and $U^A$  help rewrite
        \begin{equation}
           \begin{aligned}
               g_{uu} &= - \left(1 - \frac{2m}{r} \right) + \left[\frac{r C_{zz}}{4}\left( \frac{U^z_2}{r^2} + \frac{U^z_3}{r^3} \right)^2 + \text{c.c.} \right] + \frac{r^2 \gamma_{z \bar z}}{2}  \left( \frac{U^z_2}{r^2} + \frac{U^z_3}{r^3} \right) \left( \frac{U^{\bar z}_2}{r^2} + \frac{U^{\bar z}_3}{r^3} \right) \\ &= - 1 + \frac{2m}{r} + \mathcal{O}(r^{-2}).
           \end{aligned} \label{eq: guu asymptotic}
        \end{equation} Assuming $\beta_1/r \ll 1$, we also get
        \begin{equation*}
            g_{ur} = -\exp\left(\frac{2\beta_1}{r} + \mathcal{O}(r^{-2})\right) = - 1  - \frac{2\beta_1}{r} + \mathcal{O}(r^{-2}),
        \end{equation*} and finally
        \begin{equation*}
            \begin{aligned}
                g_{uz} &= \frac{r C_{zz}}{2} \left( \frac{U^z_2}{r^2} + \frac{U^z_3}{r^3} \right) + \frac{r^2 \gamma_{z \bar z}}{2} \left( \frac{U^{\bar z}_2}{r^2} + \frac{U^{\bar z}_3}{r^3} \right) = \frac{1}{2}U_{2z} + \frac{1}{r} \left( \frac{1}{2} C_{zz} U^z_2 + \frac{1}{2} U_{3z} \right) + \mathcal{O}(r^{-2}) \\ \text{and} \quad
                g_{u \bar z} &= \frac{1}{2}U_{2\bar z} + \frac{1}{r} \left( \frac{1}{2} C_{\bar z \bar z} U^{\bar z}_2 + \frac{1}{2} U_{3 \bar z} \right) + \mathcal{O}(r^{-2}) \qquad \Rightarrow g_{uA} = \frac{1}{2}\gamma_{AB}U^B_2 + \mathcal{O}(r^{-1}).
            \end{aligned}   
        \end{equation*}
         These coefficients correspond to a metric of the form
        \begin{equation}
            g_{\mu \nu} = \begin{pmatrix}
                -\left(1 - \frac{2m}{r} \right) & - 1  - \frac{2\beta_1}{r} & g_{uz} & g_{u \bar z} \\
                - 1  - \frac{2\beta_1}{r} & 0 & 0 & 0 \\
                g_{uz} & 0 & r C_{zz} & r^2 \gamma_{z \bar z} \\
                g_{u \bar z} & 0 & r^2 \gamma_{z \bar z} & rC_{\bar z \bar z}
            \end{pmatrix}_{\mu \nu} + \mathcal{O}(r^{-2}). \label{eq: asymptotic metric matrix}
        \end{equation} To determine the various coefficients in the expansion \eqref{eq: falloffs metric}, we require that \eqref{eq: Bondi metric} satisfy the \textit{Einstein equations}
        \begin{equation}
            G_{\mu \nu} \equiv R_{\mu \nu} - \frac{1}{2}R g_{\mu \nu} = 8 \pi G T_{\mu \nu}. \label{eq: einstein equations matter}
        \end{equation} We now need to specify the behaviour of $T_{\mu \nu}$ as $r \to \infty$ to make progress \cite{flanagan_conserved_2017}
        \begin{gather*}
            T_{uu} = \frac{1}{r^2} \hat T_{uu}(u, x^A) + \mathcal{O}(r^{-3}), \qquad
            T_{rr} = \frac{1}{r^4} \hat T_{rr}(u, x^A) + \frac{1}{r^5} \tilde T_{rr}(u, x^A) + \mathcal{O}(r^{-6}), \\
            T_{uA} = \frac{1}{r^2} \hat T_{uA}(u, x^A) + \mathcal{O}(r^{-3}), \qquad
            T_{rA} = \frac{1}{r^3} \hat T_{rA}(u, x^A) + \mathcal{O}(r^{-4}), \\
            T_{AB} = \frac{1}{r}\hat T(u, x^A) \gamma_{AB} + \mathcal{O}(r^{-2}), \qquad T_{ur} = \mathcal{O}(r^{-4}).
        \end{gather*} Using stress-energy conservation $\nabla^\beta T_{\alpha \beta} = 0$, one is then able to simplify $\hat T_{rA}$ and ultimately obtain and expression for the coefficient of our expansion by enforcing \eqref{eq: einstein equations matter}.  For simplicity, we will later on restrict to the case where $\hat T = 0$, as done by Strominger \cite{strominger_bms_2014}, Kapec et al.~\cite{kapec_semiclassical_2014}, He et al.~\cite{he_bms_2015} or Alessio and Arzano \cite{alessio_note_2019}. The detailed calculations of the expansion coefficients was carried out with the help of \verb!Mathematica! (Appendix \ref{app: coefficients metric expansion}). We summarize here the results. 
        The gauge condition \eqref{eq:determinant gauge condition} gives 
        \begin{equation*}
            D_{z\bar z} = \frac{1}{2} \gamma^{z \bar z} C_{\bar z \bar z} C_{zz}.
        \end{equation*}From the $rA$ piece of \eqref{eq: einstein equations matter}, we get
        \begin{equation}
            U_{2A} = D^B C_{BA} \quad \Rightarrow \quad g_{uA} = \frac{1}{2}D^B C_{BA} + \mathcal{O}(r^{-1}). \label{eq: U2A} 
        \end{equation} Finding an expression for $U_{3A}$ is more involving; one needs to keep additional terms in the expansion. It is convenient to write it as
        \begin{equation*}
            U_{3z} = \frac{4}{3}N_z + \frac{4}{3}u D_z m - \frac{1}{4} D_z C_{zz} C^{zz} - C_{zz} \gamma^{z \bar z} D^{\bar z} C_{\bar z \bar z},
        \end{equation*} implying
        \begin{equation*} 
            g_{uz} = \frac{1}{2}D^z C_{zz} + \frac{1}{6r} C_{zz} D_z C^{zz} + \frac{2}{3r} N_z + \mathcal{O}(r^{-2}).
        \end{equation*}Meanwhile, the $rr$ piece of \eqref{eq: einstein equations matter} yields
        \begin{equation}
            \beta_{1} =0 \text{ and } \beta_2 = - \frac{1}{32} C_{AB}C^{AB} \quad \Rightarrow \quad g_{ur} = - 1 + \mathcal{O}(r^{-2}).  \label{eq: fallof g_ur}
        \end{equation} The other terms in the expansion will not play a role in what we consider later on. We have arrived at our desired result: the class of allowed metrics in Bondi gauge whose falloffs ensure a description of an AFS at large radial distances, with line element given by
        \begin{empheq}[box=\widefbox]{multline}
            \deriv s^2 = -\deriv u^2 - 2 \deriv u \deriv r + 2 r^2 \gamma_{z \bar z} \deriv z \deriv \bar z + \frac{2m}{r} \deriv u^2 + r C_{zz} \deriv z^2 + r C_{\bar z \bar z} \deriv \bar z^2 \\ + D^z C_{zz} \deriv u \deriv z + D^{\bar z} C_{\bar z \bar z} \deriv u \deriv \bar z + \frac{1}{r} \left( \frac{4}{3}(N_z + u D_z m) - \frac{1}{4} D_z(C_{zz} C^{zz}) \right) \deriv u \deriv z + \text{c.c.} + ... \label{eq:asymptotic metric retarded}
        \end{empheq} If we readily identify the flat Minkowski metric \eqref{eq:Minkowski stereographic} in the first line, it is not obvious why the remainder of this expression takes this particular form, with factors of $4/3$ and all that. It is however common practice in the literature to write such expressions in hindsight of the calculations to come, which often helps simplifying expressions or making explicit the physical interpretation of some of the coefficients. Speaking of physical significance, let us thus pause for a moment to explicit the several physically relevant fields we encountered in our expansion of the metric coefficients:
        \begin{itemize}
            \item the \textit{Bondi mass aspect} $m(u, z, \bar z)$ gives the angular density of energy of the spacetime as measured from a point at $\cal I^+$ labeled by $u$ and in the direction pointed out by the angles $z, \bar z$. The \textit{Bondi mass} is obtained after performing an integration of $m$ on the sphere: 
            \begin{equation*}
                M_B = \oint_{\mathbb{S}^2_\infty} \deriv^2 \Omega \, m(u, z, \bar z).
            \end{equation*} One can show that $\partial_u M_B(u) \leq 0$ for pure gravity or gravity coupled to matter obeying the null energy condition. Physically, the radiation carried by gravitational waves or null matter such as electromagnetic fields escapes through $\cal I^+$ and lowers the energy of spacetime when the retarded time $u$ evolves: this is the \textit{mass loss} \cite{bondi_gravitational_1962}. At $u \to \infty$, the Bondi mass equates the \textit{ADM energy}, or total energy of a Cauchy slice of spacetime \cite{compere_advanced_2019}.
            \item the traceless and symmetric field $C_{AB}$. $C_{zz}$ and $C_{\bar z \bar z}$ can be seen as two polarization modes for the gravitational waves. Later on, we will see that they also encode the helicity modes of gravitons in the context of soft theorems. The \textit{(retarded) Bondi news tensor} is defined as
            \begin{equation}
                N_{AB} = \partial_u C_{AB}. \label{eq: bondi news tensor}
            \end{equation} Its square is proportional to the energy flux across $\cal I^+$.
            \item finally, we also introduced the \textit{angular momentum aspect} $N^A(u, z, \bar z)$. It is closely related to the angular density of angular momentum with respect to the origin ($r = 0$).
        \end{itemize} Great. We now have an expression for the asymptotic behaviour of the metric of an AFS at $\cal I^+$. However, we have not yet imposed all of Einstein's equations. Looking at the $uu $ and $uA$ parts of \eqref{eq: einstein equations matter} gives us additional constraints (Appendix \ref{app: coefficients metric expansion}),
        \begin{equation}
            \begin{aligned}
                \partial_u m &= - 4 \pi G \hat T_{uu} - \frac{1}{4}N_{zz}N^{zz} + \frac{1}{4} [D^{z2} N_{zz} + D^{\bar z2}N_{\bar z \bar z}] \\
                &= - T_{uu} + \frac{1}{4} [D^{z2} N_{zz} + D^{\bar z2}N_{\bar z \bar z}] = - T_{uu} + \frac{1}{4}\partial_u [D^{z} U_{2z} + D^{\bar z}U_{2\bar z}],
            \end{aligned} \label{eq:constraint mass aspect}
        \end{equation} and
        \begin{multline}
            \partial_u N_z  = \frac{1}{4} D_z [D_z^2 C^{zz} - D_{\bar z}^2 C^{\bar z \bar z}] \ - u D_z \partial_u m + \frac{1}{4} D_z (C_{zz} N^{zz}) + \frac{1}{2} C_{zz} D_z N^{zz} - 8 \pi G \hat T_{uz}. \label{eq:constraint angular momentum aspect}
        \end{multline}These are the \textit{constraint equations}, which tell us the time evolution of $m$ and $N_z$. We learn that to first and second subleading order in the $1/r$ expansion, $C_{zz}$ and $C_{\bar z \bar z}$ is the only "free data" that we need to assign, since all the other components of the metric are determined through \eqref{eq:constraint mass aspect} and \eqref{eq:constraint angular momentum aspect} once initial conditions for $m$, $N^A$ and $C_{AB}$ have been provided. 

        \smallskip
        We can repeat the same steps when looking at the behaviour of the metric at $\cal I^-$. Starting from \eqref{eq: advanced Bondi Sachs metric stereographic}, we find that the metric in advanced Bondi coordinates $(v, r, z, \bar z)$ has the large $r$ expansion
        \begin{empheq}[box=\widefbox]{equation*}
            \deriv s^2 = - \deriv v^2 + 2 \deriv v \deriv r + 2r^2 \gamma_{z \bar z} \deriv z \deriv \bar z + \frac{2m^-}{r} \deriv v^2 + r D_{zz} \deriv z^2 + r D_{\bar z \bar z} \deriv \bar z^2 + V_{2z} \deriv v \deriv z + V_{2\bar z} \deriv v \deriv \bar z + ...
        \end{empheq} with $m^-$ the advanced Bondi mass, $V_{2z} = -D^zD_{zz}$, $V_{2\bar z} = - D^{\bar z}D_{\bar z \bar z}$ and the constraint
        \begin{equation}
            \partial_v m^- = - T_{vv} -  \frac{1}{4}\partial_v [D^z V_{2z} + D^{\bar z} V_{2\bar z}], \label{eq:constraint mass aspect scri-}
        \end{equation} where $T_{vv}$ is the total incoming radiation flux at $\cal I^-$. We assume $g_{AB}^- = r^2 \gamma_{AB} + r D_{AB} + \mathcal{O}(1)$ and also denote by $M_{AB} = \partial_v D_{AB}$ the \textit{advanced Bondi news tensor}. In the following, we will mostly focus on what happens at $\cal I^+$. The analogous statements for $\cal I^-$ can be read off from $U^A \to - V^A$.

    \subsection{Bondi-Metzner-Sachs group in General Relativity \label{subsec: bms group}} 

        We now have a description of the behaviour of the Bondi metric of an AFS at our disposal. Our next task will be to find the form of the most general diffeomorphism $\xi$ which preserves the Bondi gauge conditions and the falloffs we required for an AFS: the generator of asymptotic symmetries. This will lead us to constructing the BMS algebra in four dimensions, $\mathfrak{bms}_4$.

        \subsubsection{\texorpdfstring{$\mathfrak{bms}_4$}{bms4} algebra} 
        Recall that the variation $g \to g + \delta g$ of the metric under a diffeomorphism generated by the vector field $\xi$ is given in a coordinate basis by \cite{reall_part_2022}
        \begin{equation}
            \delta g_{\mu \nu} \equiv \mathcal{L}_\xi g_{\mu \nu} = \xi^\rho \partial_\rho g_{\mu \nu} + g_{\mu \rho} \partial_\nu \xi^\rho + g_{\nu \rho} \partial_\mu \xi^\rho. \label{eq: Lie derivative metric}
        \end{equation} Hence, preservation of the Bondi gauge conditions \eqref{eq: bondi gauge} under $\xi$ writes
        \begin{equation}
            \boxed{\mathcal{L}_\xi g_{rr} = 0, \quad \mathcal{L}_\xi g_{rA} = 0, \quad \text{and}\quad \mathcal{L}_\xi \partial_r\text{det}\frac{g_{AB}}{r^2} = 0 \Rightarrow g^{AB} \mathcal{L}_\xi g_{AB} = 0,} \label{eq:gauge preservation condition}
        \end{equation} where the last implication stems from the fact that
        \begin{equation*} \begin{aligned}
                \mathcal{L}_\xi (\text{det}g_{AB}) &= \lim_{\epsilon \to 0} \frac{\text{det}(g'_{AB}) - \text{det}(g_{AB})}{\epsilon} = \lim_{\epsilon \to 0} \frac{\exp \text{Tr}[\log (g_{AB} + \mathcal{L}_\xi g_{AB}\epsilon)] - \text{det}(g_{AB})}{\epsilon}.
        \end{aligned} \end{equation*} Here we expanded $g_{AB}' = g_{AB} + \mathcal{L}_\xi g_{AB} \epsilon + \mathcal{O}(\epsilon^2)$ and made use of the relation $\text{det}(A) = \exp \text{Tr}(\ln A)$ for any square matrix $A$. Then a Taylor series expansion of $\exp$ and $\log$ gives
        \begin{equation*}
            \mathcal{L}_\xi (\text{det}g_{AB}) \approx \lim_{\epsilon \to 0} \frac{\text{det}(g_{AB}) + g^{AB} \mathcal{L}_\xi g_{AB} \epsilon + ... - \det(g_{AB})}{\epsilon} = g^{AB} \mathcal{L}_\xi g_{AB} \overset{!}{=} 0.
        \end{equation*}This last constraint amounts to asking that the angular metric $g_{AB}$ does not undergo any conformal rescaling under the transformation \cite{alessio_structure_2018}. Requiring that the asymptotic behaviour of \eqref{eq:asymptotic metric retarded} hold true\footnote{Horn \cite{horn_asymptotic_2022} investigated asymptotic symmetries which preserve the Bondi gauge condition but do not preserve the asymptotic falloff conditions for the metric near null boundary.} under $\xi$ amounts to enforcing:
        \begin{equation}
            \boxed{\mathcal{L}_\xi g_{uu} = \mathcal{O}(r^{-1}), \quad 
            \mathcal{L}_\xi g_{ur} = \mathcal{O}(r^{-2}), \quad
            \mathcal{L}_\xi g_{uA} = \mathcal{O}(1), \quad 
            \mathcal{L}_\xi g_{AB} = \mathcal{O}(r).} \label{eq: falloff preservation condition}
        \end{equation} The procedure for determining $\xi$ is as follows: one first solves the constraints \eqref{eq:gauge preservation condition} exactly, which allows to express the 4 components of $\xi^\mu$ in terms of 4 functions of $u, x^A$. The falloff preservation condition \eqref{eq: falloff preservation condition} can then be solved to reduce these 4 functions to only 3 functions on the 2-sphere, namely $T(x^A)$ and $R^B(x^A)$. These calculations are carried out in detail in Appendix \ref{app: determination BMS generators} and yield
        \begin{empheq}[box=\widefbox]{equation}
            \left.\xi_{T,R}\right|_{\mathcal{I}^+} = \left[T(x^C) + \frac{u}{2} D_A R^A(x^C) \right]\partial_u 
            + \left[R^A(x^C) \right] \partial_A. \label{eq: BMS generators scri+}
        \end{empheq} They correspond to the exact Killing vector fields at $\cal I^+$. In this expression $T(x^A)$ is unconstrained, while $R^A(x^B)$ obey the conformal Killing equation on the 2-sphere (Appendix \ref{app: determination BMS generators}),
        \begin{equation}
            D_A R_B + D_B R_A = \gamma_{AB} D_C R^C .\label{eq:conformal Killing equation}
        \end{equation} In stereographic coordinates $(z, \bar z)$, \eqref{eq:conformal Killing equation} implies that $R^z$ is holomorphic, $R^z \equiv R^z(z)$ and $R^{\bar z}$ antiholomorphic, $R^{\bar z} \equiv R^{\bar z}(\bar z)$; more on this in section~\ref{sec: subsec superrotations}. Extending \eqref{eq: BMS generators scri+} from future null infinity into the interior of the spacetime while still asking that \eqref{eq:gauge preservation condition} and \eqref{eq: falloff preservation condition} hold, we get \textit{asymptotic} Killing vectors (Appendix~\ref{app: determination BMS generators})
        \begin{empheq}[box=\widefbox]{multline}
            \xi_{T,R} = f \partial_u + \left[R^A - \frac{D^A f}{r} + \frac{(D_B f) C^{AB}}{2r^2} + \mathcal{O}(r^{-3}) \right]\partial_A \\ + \left[- \frac{rD_A R^A}{2} + \frac{D^2 f}{2r} - \frac{2 (D_A C^{AB}) D_B f + C^{AB} D_A D_B f}{4r} + \mathcal{O}(r^{-2}) \right] \partial_r \label{eq: BMS generators}
        \end{empheq} with
        \begin{equation}
            f(u, x^A) = T(x^A) + \frac{u}{2} D_B R^B(x^A). \label{eq: definition f from T and R}
        \end{equation} The vectors \eqref{eq: BMS generators} are known as the (asymptotic) \textit{BMS generators}. The most general diffeomorphisms generating a variation of our metric compatible with the Bondi gauge and the required falloffs for an AFS are obtained from $\xi_{T,R}$. They generate the asymptotic $\mathfrak{bms}_4$ algebra, which appears to be larger than the Poincaré algebra. This is further studied in the remainder of this section.
        
        \paragraph{$\mathfrak{bms}_4$ brackets} Our notation involving $T$ and $R$ suggests that there are two types of transformation generators of $\mathfrak{bms}_4$. Let us make this more explicit by computing the brackets of this algebra. First and foremost, note that there are two equivalent ways to go around this. One can start with \eqref{eq: BMS generators scri+} and compute the usual \textit{Lie bracket} $[\xi_1, \xi_2]  \equiv [\xi_{T_1,R_1}, \xi_{T_2,R_2}]$ (omitting the subscript $\cal I^+$). The vectors \eqref{eq: BMS generators scri+} are however only defined at $\cal I^+$ so one then has to check using \eqref{eq: BMS generators} that the commutation relations hold even away from the boundary (see \cite{strominger_lectures_2017} for an explicit computation). The alternative is to use a \textit{modified Lie bracket} 
        \begin{equation*}
            [\xi_1, \xi_2]_M = [\xi_2, \xi_2] - \delta^g_{\xi_1} \xi_2 + \delta^g_{\xi_2} \xi_1,
        \end{equation*} where $\delta^g_{\xi_1}\xi_2$ denotes the variation in $\xi_2$ under the variation of the metric induced by $\xi_1$. This accounts for the dependence of the asymptotic vector fields on the background metric, notably through $C_{AB}$ in \eqref{eq: BMS generators}. Barnich and Troessaert \cite{barnich_aspects_2010} showed that the latter vectors provide a faithful representation of $\mathfrak{bms}_4$ when equipped with $[\cdot,\cdot]_M$ and for $R^A$ a conformal Killing vector of the 2-sphere. In either way, denoting $\xi_{T,0} \equiv \xi_T$ and $\xi_{0,R} \equiv \xi_R$, we get
        \begin{equation}
            \boxed{[\xi_{T_1}, \xi_{T_2}] = 0, \quad \text{and} \quad [\xi_{R_1}, \xi_{R_2}] = \hat R^A \partial_A + \frac{u}{2} D_A \hat R^A \partial_u \equiv \xi_{\hat R}} \label{eq: bracket xi TT RR}
        \end{equation} with $\hat R^A = R_1^C \partial_C R_2^A - R_2^C \partial_C R_1^A$, as well as
        \begin{equation}
            \boxed{[\xi_R, \xi_T] = \left(R^A \partial_A T - \frac{T}{2}D_B R^B\right)\partial_u \equiv \hat T \partial_u \equiv \xi_{\hat T}.} \label{eq: bracket xi RT}
        \end{equation} These relations define the $\mathfrak{bms}_4$ algebra at $\cal I^+$, of which trivial boundary diffeomorphisms with $T = R^A \equiv 0$ form an ideal. Taking the quotient by this ideal, we are left with the asymptotic algebra of asymptotically flat spacetimes compatible with the Bondi-Sachs boundary conditions at $\cal I^+$. We will see later on that this algebra exponentiates to the group called BMS$^+$, and that replicating the exact same analysis for $\cal I^-$ starting this time from \eqref{eq: advanced Bondi Sachs metric stereographic} yields a second copy of the BMS group, BMS$^-$, acting on ingoing data at past null infinity. Before doing so, let us investigate the action of $\xi_T$ and $\xi_R$. In the next section, we show that the former vectors generate the so-called \textit{supertranslations}, while the latter generate the Lorentz transformations encountered in section~\ref{sec:section 1}.
    
        \subsubsection{Supertranslations}
        From the first of \eqref{eq: bracket xi TT RR}, we learn that the generators of supertranslations $\xi_T = T(x^A) \partial_u$ at $\cal I^+$ form an Abelian ideal of the $\mathfrak{bms}_4$ algebra. Inside the spacetime we have for the asymptotic Killing vector \eqref{eq: BMS generators}, working in the stereographic $x^A = (z, \bar z)$ coordinates
        \begin{equation}
            \xi_\text{T}^+ \approx f \partial_u +\left[- \frac{1}{r}(D^z f \partial_z + D^{\bar z} f \partial_{\bar z} ) + \frac{1}{2r^2}(C^{zz}D_z f\partial_z + C^{\bar z \bar z} D_z f\partial_{\bar z})\right] + D^zD_z f \partial_r. \label{eq:asymp supertranslations scri+}
        \end{equation} Here, we have renamed $T(x^A) \equiv f(x^A)$ since $R^B \equiv 0$ in \eqref{eq: definition f from T and R} and used $\frac{1}{2}D^2 f= D^zD_z f$ to match the notation of Strominger \cite{strominger_lectures_2017}. Similarly, one obtains in advanced Bondi coordinates
        \begin{equation}
            \xi_\text{T}^- \approx f^- \partial_v + \left[ \frac{1}{r}(D^{\bar z} f^- \partial_{\bar z} + D^z f^- \partial_z ) + \frac{1}{2r^2}(C^{zz}D_zf^- \partial_z + C^{\bar z \bar z}D_{\bar z}f^- \partial_{\bar z}) \right] - D^z D_z f^{-} \partial r. \label{eq:asymp supertranslations scri-}
        \end{equation} Since $f$ and $f^- \equiv f^-(z, \bar z)$ are arbitrary scalar fields on $\mathbb{S}^2$, the exponentiation of these vectors gives rise to an Abelian subgroup $S$ of BMS$_4$, which is infinite-dimensional. In fact, $S$ admits one unique normal finite subgroup that reproduces the Poincaré translations. This is discussed in what follows.

        \paragraph{Spacetime translations} We said that supertranslations are parametrized by an arbitrary function $f(z, \bar z)$ on $\mathbb{S}^2$. As such, it is natural to think of the latter as a linear superposition of spherical harmonics $Y^l_m$,
        \begin{equation}
            f(z, \bar z) = \sum_{l=0}^\infty \sum_{m=-l}^l f_{l,m}Y^l_m(z, \bar z), \quad f_{l,-m} = (-1)^m \bar f_{l,m}. \label{eq: f harmonic decomposition}
        \end{equation}For $l=0$ and $l=1$, we have for $\xi_T^+(f)$ given by \eqref{eq:asymp supertranslations scri+}
        \begin{equation}
            \xi_T^+(Y^0_0) = Y^0_0 \partial_u, \quad \text{and} \quad \xi_T^+(Y_1^m) = Y_1^m \partial_u - \frac{\gamma^{AB} \partial_B Y_1^m}{r} \partial_A + \frac{1}{2}D^2 Y_1^m \partial_r. \label{eq: xi harmonic}
        \end{equation} Using \eqref{eq:def stereographic coord} to write
    \begin{equation*}
        \frac{z + \bar z}{2 \sqrt{z \bar z}} = - \sin \phi, \quad \frac{z \bar z -1}{1 + z \bar z} =  - \cos \theta \quad \text{and} \quad \frac{2 \sqrt{z \bar z}}{1 + z \bar z} = \sin \theta,
    \end{equation*} the first few spherical harmonics can be recast in the $(u,r,z,\bar z)$ system as
    \begin{equation*}
             Y^0_0 = 1, \quad Y^0_1 = \frac{1- z \bar z}{1+z \bar z}, \quad Y_1^1 = \frac{z}{1 + z \bar z}, \quad Y^{-1}_1 = \frac{\bar z}{1 + z \bar z}.
        \end{equation*} Inserting this in \eqref{eq: xi harmonic} gives
        \begin{gather*} 
            \xi_T^+(Y^0_0) = \partial_u, \quad 
            \xi_T^+(Y^0_1) = \frac{1- z \bar z}{1 + z \bar z}(\partial_u -\partial_r) + \frac{z}{r} \partial_z + \frac{\bar z}{r} \partial_{\bar z} \\
            \xi_T^+(Y^1_1) = \frac{z}{1 + z \bar z} (\partial_u - \partial_r) + \frac{z^2}{2r}\partial_z - \frac{1}{2r} \partial_{\bar z}, \quad
            \xi_T^+(Y^{-1}_1) =  \frac{\bar z}{1 + z \bar z}(\partial_u - \partial_r) - \frac{1}{2r}\partial_z + \frac{\bar z ^2}{2r} \partial_{\bar z},
        \end{gather*} where we used that $\frac{1}{2}D^2 f(z, \bar z) = \frac{1}{2}(D^z D_z + D^{\bar z} D_{\bar z}) f(z, \bar z) = \gamma^{z \bar z} \partial_z \partial_{\bar z} f(z, \bar z).$ Besides, global spacetime translations \eqref{eq: translation generators} write in Bondi coordinates \cite{strominger_lectures_2017}
        \begin{equation*}\begin{aligned}
            \xi_0 &= \partial_u \\
            \xi_1 &= - \frac{z + \bar z}{1 + z \bar z} \partial_u + \frac{z + \bar z}{1 + z \bar z} \partial_r + \frac{1 - z^2}{2r} \partial_z + \frac{1 - \bar z^2}{2r}\partial_{\bar z}, \\
            \xi_2 &= \frac{i(z - \bar z)}{1 + z \bar z}\partial_u - \frac{i(z - \bar z)}{1 + z \bar z} \partial_r + \frac{i(1 + z^2)}{2r} \partial_z - \frac{i(1+\bar z^2)}{2r} \partial_{\bar z}, \\
            \xi_3 &= - \frac{1 - z \bar z}{1 + z \bar z} \partial_u + \frac{1 - z \bar z}{1 + z \bar z}\partial_r - \frac{z}{r}\partial_z - \frac{\bar z}{r}\partial_{\bar z}.
        \end{aligned}\end{equation*} Therefore, we identify
        \begin{gather*}
            \xi_T^+(Y_0^0) = \xi_0, \quad \xi_T^+(Y_1^0)= -\xi_3, \quad \xi_T^+(Y_1^1) = - \frac{\xi_1 + i \xi_2}{2}, \quad \xi_T^+(Y^{-1}_1) = -\frac{\xi_1 - i \xi_2}{2}.
        \end{gather*} Thus the usual spacetime translations are contained in the BMS supertranslations. Time translation $\xi_0$ implies energy conservation, while the three spatial translations $\xi_i$ imply ADM momentum conservation \cite{zheng_liang_bms_2017}.
        We have shown so far that $S$ is Abelian and contains the usual translations of Minkowski. Showing that it is in fact a \textit{normal} subgroup of BMS requires a more careful study of the structure of BMS transformations and their interplay \cite{alessio_asymptotic_2019}. In the next paragraph, we look at the action of supertranslations on the data we specified for our metric \eqref{eq:asymptotic metric retarded}.

        \paragraph{Effect of a supertranslation} The supertranslations at $\cal I^+$ act to shift individual light rays of null infinity forwards or backwards in retarded time. In practice, writing $\mathcal{L}_f \equiv \mathcal{L}_{\xi_f^+}$ with $f \equiv T$ for notational simplicity, we obtain at $\cal I^+$ (Appendix~\ref{app: effect supertranslation})
        \begin{align}
            \mathcal{L}_f C_{AB} &= f \partial_u C_{AB} + \gamma_{AB} D^2 f - 2 D_A D_B f, \label{eq: effect supertranslation Czz} \\
            \mathcal{L}_f N_{AB} &= f \partial_u N_{AB}, \label{eq: effect supertranslation Nzz} \\
            \mathcal{L}_f m &= f \partial_u m + \frac{1}{4} \left( N^{AB} D_A D_B f + 2 D_A f D_B N^{AB}\right), \label{eq: effect supertranslation m} \\
            \mathcal{L}_f U_{2z} &= f \partial_u U_{2z} + D^z f \partial_u C_{zz} - 2 D^z D^2_z f \label{eq: effect supertranslation Uz}
        \end{align} while at $\cal I^-$ we have, writing $f \equiv f^-$ again to simplify notation
        \begin{align}
            \mathcal{L}_{f} D_{zz} &= f \partial_v D_{zz} + 2 D^2_z f, \label{eq: supertrans cal- Czz}\\
            \mathcal{L}_f M_{zz} &= f \partial_v M_{zz}.\label{eq: supertrans cal- Nzz} \\
            \mathcal{L}_f V_{2z} &= - D^z f \partial_v D_{zz} + f \partial_v V_{2z} - 2 D^z D^2_z f. \label{eq: effect supertranslation Vz}
        \end{align}
    
        \subsubsection{Lorentz transformations and superrotations \label{sec: subsec superrotations}}
            We now focus on the second set of generators of $\mathfrak{bms}_4$ algebra, $\xi_{0,R} \equiv \xi_R$. Note that $f$ is again given here by \eqref{eq: definition f from T and R} with $T = 0$, and thus a function of $u$ and $x^A$. The vectors $\xi_R$ from \eqref{eq: BMS generators} are the generators of \textit{superrotations} and write
            \begin{equation*}
                \xi_R = \frac{u}{2} D_A R^A \partial_u + \left[R^A - \frac{u}{2r}D^A D_B R^B + \mathcal{O}(r^{-2})\right] \partial_A + \left[- \frac{r}{2} D_A R^A + \frac{u}{4}D^2 D_B R^B +  \mathcal{O}(r^{-1})\right] \partial_r
            \end{equation*} with $R^A(x^A)$ satisfying the conformal Killing equation \eqref{eq:conformal Killing equation}. In stereographic coordinates
            \begin{align*}
                \xi^{+u}_R &\approx \frac{u}{2} (D_z R^z + D_{\bar z} R^{\bar z}), \\
                \xi^{+z}_R &\approx R^z - \frac{u}{2r}D^z(D_z R^z + D_{\bar z} R^{\bar z}) = R^z + \frac{u}{2r}(R^z - D^z D_{\bar z} R^{\bar z}) \\ 
                \xi^{+\bar z}_R &\approx R^{\bar z} - \frac{u}{2r}D^{\bar z}(D_z R^z + D_{\bar z} R^{\bar z}) = R^{\bar z} + \frac{u}{2r}(R^{\bar z} - D^{\bar z} D_{z} R^{z})
            \end{align*} having used that $D^z D_z R^z = - R^z$ in the last two lines, and
            \begin{equation*}\begin{aligned}
                \xi_R^{+r} &\approx - \frac{r}{2}(D_z R^z + D_{\bar z}R^{\bar z}) + \frac{u}{4}(D^zD_z + D^{\bar z}D_{\bar z})(D_z R^z + D_{\bar z} R^{\bar z}) = - \frac{r+u}{2}(D_z R^z + D_{\bar z}R^{\bar z}).
            \end{aligned}\end{equation*} Therefore, in the $(u,r,z,\bar z)$ system, our asymptotic Killing vector $\xi_R^+$ writes
            \begin{multline}
                \xi_R^+ \approx \frac{u}{2} (D_z R^z + D_{\bar z} R^{\bar z})\partial_u - \frac{r+u}{2}(D_z R^z + D_{\bar z}R^{\bar z})\partial_r \\ + \left[R^z + \frac{u}{2r}(R^z - D^z D_{\bar z} R^{\bar z})\right]\partial_z + \left[R^{\bar z} + \frac{u}{2r}(R^{\bar z} - D^{\bar z} D_{z} R^{z})\right]\partial_{\bar z}. \label{eq: asymp superrotation scri+}
            \end{multline}

            \paragraph{Lorentz transformations} So far, we have not imposed any restrictions on $R^A$ apart from the fact that they are CKVs on the unit two-sphere. We have also mentioned that \eqref{eq:conformal Killing equation} implies that $R^z$ and $R^{\bar z}$ are repectively holomorphic and antiholomorphic, $\partial_z R^{\bar z} = \partial_{\bar z} R^z = 0$. Indeed, from \eqref{eq:conformal Killing equation} we learn that $D_z R^{\bar z} = D_{\bar z} R^z = 0$, which, using the Christoffel symbols for $\gamma$ in stereographic coordinates, imply
            \begin{equation*}
                D_z R^{\bar z} = \partial_z R^{\bar z} + \cancel{\Gamma^{\bar z}_{z z}} R^z + \cancel{\Gamma^{\bar z}_{\bar z z}} R^{\bar z} \overset{!}{=} 0 \Rightarrow \partial_z R^{\bar z} = 0,
            \end{equation*} and similarly $D_{\bar z} R^z \overset{!}{=} 0 \Rightarrow \partial_{\bar z} R^z = 0$. As such, both $R^z$ and $R^{\bar z}$ appear as a sum of monomial terms $R^z = z^k$, $k \in \mathbb{Z}$ when expanded in Laurent series. This restricts the possible choices for $R^z$. Indeed, considering $v_k = z^k \partial_z$, we have that when $k < 0$, $z^k$ is singular at the origin $z=0$ of the Riemann sphere, while for $k>2$ $z^k$ is singular\footnote{To see this, it is helpful to consider the transformation $z \to w = z^{-1}$, such that $v^k \to - w^{2-k} \partial_w$.} at the point at infinity $z = \infty$. Therefore $R^z$ is only well-defined for $k=0,1,2$, and similarly for $R^{\bar z}$, giving six valid asymptotic Killing vectors $\xi_R^+$ in total. These are parametrized by
            \begin{equation*}
                R^z \in \{1,z,z^2,i, iz, iz^2\}, \quad R^{\bar z} = \overline{R^z}.
            \end{equation*}
            Here comes the punchline: these six vector fields generate exactly the Lorentz transformations we know and love! Put differently, $R^{z}$ and $R^{\bar z}$ are well-defined globally on $\mathbb{S}^2$ if and only if the superrotations are in fact Lorentz transformations (LT). To see this, note that the LT Killing vectors \eqref{eq: rotation generators} and \eqref{eq: boost generators} write in the $(u,r,z, \bar z)$ coordinates \cite{strominger_lectures_2017}
            \begin{align*}
                \xi_{01} &= - \frac{u(z + \bar z)}{1 + z \bar z} \partial_u + \frac{(r+u)(z + \bar z)}{1 + z \bar z}\partial_r + \frac{(1 - z^2)(r+u)}{2r} \partial_z + \frac{(1 - \bar z^2)(r + u)}{2r} \partial_{\bar z}\\
                \xi_{02} &= \frac{iu(z - \bar z)}{1 + z \bar z}\partial_u - \frac{i(r+u)(z - \bar z)}{1 + z \bar z} \partial_r + \frac{i(1+z^2)(r+u)}{2r} \partial_z - \frac{i(1 + \bar z^2)(r+u)}{2r}\partial_{\bar z} \\
                \xi_{03} &= - \frac{u(1 - z \bar z)}{1+z \bar z} \partial_u + \frac{(r+u)(1- z \bar z)}{1 + z \bar z} \partial_r - \frac{z(r+u)}{r}\partial_z - \frac{\bar z(r+u)}{r}\partial_{\bar z}
            \end{align*}  for the boosts and \begin{equation*}
                \xi_{21} = iz \partial_z - i \bar z \partial_{\bar z}, \quad
                \xi_{32} = \frac{i(z^2 -1)}{2} \partial_z - \frac{i(\bar z^2 -1)}{2}\partial_{\bar z},
                \quad \xi_{13} = \frac{(z^2+1)}{2}\partial_z + \frac{(\bar z^2 +1)}{2}\partial_{\bar z}.
            \end{equation*} for the spacetime rotations. Following \cite{strominger_lectures_2017}, we write $\xi_{R^z} \equiv \zeta(a, b , c)$ for $a, b,c \in \mathbb{C}$
            \begin{align*}
                \zeta^u &= \frac{u}{2} \frac{2z(c - a^\ast) + 2 \bar z (c^\ast - a) + (b + b^\ast)(1-z \bar z)}{1 + z \bar z}, \\
                \zeta^r &=- \frac{u+r}{2} \frac{2z(c - a^\ast) + 2 \bar z (c^\ast - a) + (b + b^\ast)(1-z \bar z)}{1 + z \bar z}, \\
                \zeta^z &= a + b z + c z^2 + \frac{u}{2r}((a - c^\ast) + (b+ b^\ast)z + (c- a^\ast)z^2), \quad
                \zeta^{\bar z} = (\zeta^z)^\ast
            \end{align*} and thus identify
            \begin{gather*}
                \xi_{12} = \zeta(0, i, 0), \quad \xi_{23} = \zeta(- i/2, 0, i/2), \quad \xi_{31} = \zeta(1/2,0,1/2) \\
                \xi_{01}  =  \zeta(1/2,0,-1/2), \quad \xi_{02} = \zeta(i/2, 0, i/2), \quad \xi_{03} = \zeta(0,-1,0).
            \end{gather*} Here, we used
            \begin{equation*}
                D_z R^z = \partial_z R^z  - \frac{2 \bar z}{1 + z \bar z} R^{z} \quad \text{and} \quad D_{\bar z} R^{\bar z}  = \partial_{\bar z} R^{\bar z} - \frac{2z}{1 + z \bar z} R^{\bar z}.
            \end{equation*}This shows that BMS vector fields obtained from a linear combination of $\{1,z,z^2,i, iz, iz^2\}$ indeed correspond to the usual Lorentz generators, so the globally defined $\xi_R^+$ are simply the asymptotic Lorentz transformations. This is the historic form of the BMS group, sometimes also called \textit{global} BMS. Allowing for more general (and hence singular) superrotations gives rise to extensions of the BMS group, some of which will be introduced in the next section.
            
            \paragraph{Effect of a BMS superrotation} The determination of the action of a superrotation on the data at $\cal I^+$ follows the exact same procedure as for supertranslations. One computes the change in the metric components under the diffeomorphism using \eqref{eq: Killing contravariant}, and then looks at the relevant order in $1/r$ to identify e.g.~$\mathcal{L}_\xi m$, etc. We do not reproduce these calculations here for the sake of brevity, and instead state the results. We have at $\cal I^+$ \cite{alessio_note_2019}
            \begin{align}
                \mathcal{L}_\xi C_{zz} &= \frac{u}{2} D \cdot R N_{zz} + \mathcal{L}_R C_{zz} - \frac{1}{2} D \cdot R C_{zz} - u D^3_z R^z, \label{eq: effect BMS LT Czz}\\
                \mathcal{L}_\xi N_{zz} &= \frac{u}{2} D \cdot R \partial_u N_{zz} + \mathcal{L}_R N_{zz} - D^3_z R^z \label{eq: effect BMS LT Nzz}
            \end{align} where $\mathcal{L}_R C_{zz} = R \cdot D C_{zz} + 2 D_z R^z C_{zz}$ and similarly on $N_{zz}$.

            \subsubsection{Structure of the BMS group \label{sec: structure BMS}} 
            Our previous study of the BMS generators taught us that the standard BMS group is composed of an infinite-dimensional subgroup of supertranslations, $S$, containing the usual spacetime translations of flat space, as well as superrotations. If we discard the singular generators for the latter, we fall back on the well-known Lorentz group. Hence, we eventually get after exponentiation\footnote{The problem of the existence of exponentials for the BMS group is discussed by Prinz and Schmeding \cite{prinz_lie_2022}.}
            \begin{empheq}[box=\widefbox]{equation*}
                \text{BMS}_4 = \text{SO}(3,1) \ltimes S = \text{Lorentz} \ltimes \text{Supertranslations}
            \end{empheq} This is the historical form of the BMS group, which reproduces the semi-direct structure of the Poincaré group: the Lorentz group acts non-trivially on $S$ as it does on the usual global translations, which could be seen from the bracket \eqref{eq: bracket xi RT}. Since Translations $\subset$ Supertranslations, we have
            \begin{equation*}
                \text{Poincaré} = \text{Lorentz} \ltimes \text{Translations} \subset \text{BMS}_4.
            \end{equation*}The only difference between Poincaré and global BMS is thus that the translational part is enhanced in the latter. We shall see that this implies the degeneracy of the gravitational Poincaré vacua \cite{arcioni_exploring_2003}, a phenomenon at which we will look from the viewpoint of memory effects in section~\ref{sec: memory effects}. General relativity does indeed \textit{not} reduce to special relativity at large distances.

            \paragraph{Variants of the BMS group} The restriction of $R^z(z)$ to be a globally well-defined CKV on the two-sphere has been debated in the literature.\footnote{As a motivation for relaxing the well-definedness of $R$, note that crucial insights have been obtained in conformal field theories from singular conformal transformations on the sphere.} Two variants of the BMS group arise from the original BMS group discussed above by replacing the Lorentz group by a larger symmetry group:
            \begin{itemize}
                \item the extended BMS group eBMS, obtained by using two copies of the Bott-Virasoro group as superrotations  \cite{barnich_aspects_2010, barnich_symmetries_2010, barnich_supertranslations_2012}. This group emerged from the context of flat space holography, with the AdS/CFT correspondence in mind \cite{zheng_liang_bms_2017}.
                \item the generalized BMS group gBMS, obtained by allowing the index $k$ in the Laurent series of $R^z$ to run over all of $\mathbb{Z}$. As a result, superrotations comprise the entire group of diffeomorphisms on the 2-sphere, Diff$(\mathbb{S}^2)$. This extension first stem from the study of gravitational scattering and the equivalence of Ward identities with soft theorems \cite{campiglia_asymptotic_2014, campiglia_new_2015}. It is nowadays believed that gBMS is the correct asymptotic symmetry group of the theory as it is the only one to admit a generalization to higher spacetime dimensions.
            \end{itemize}
            
        \noindent 
        This concludes our construction of the asymptotic symmetry group of four-dimensional AFS. Next, we study the rich implications of this structure for the gravitational scattering problem.

\section{Infrared structure of gravity through supertranslations \label{sec: IR structure of gravity}}

    We are now able to exploit the BMS symmetries constructed above to probe the behaviour of gravitational scattering in the infrared (i.e.~in AFS, at large distances from an isolated system). In particular, we derive equivalence relations between the conservation laws obtained and so-called soft theorems. 

    \noindent \smallskip
    Note beforehand that two types of conservation laws can be singled out from the BMS group \cite{flanagan_conserved_2017}:  
    \begin{enumerate}
        \item those that relate quantities at one cross-section of $\cal I^+$ to another, and
        \item those that relate quantities at $\cal I^-$ to quantities at $\cal I^+$.
    \end{enumerate}  In the context of gravitational scattering where one seeks a $\mathcal{S}$-matrix relating $| out \rangle = \mathcal{S} | in \rangle$, it is natural to focus on the second type and ask whether we can find a relation of the form
    \begin{equation}
        B^+ \mathcal{S} - \mathcal{S}B^- = 0, \label{eq: symmetry gravitational scattering}
    \end{equation} for $B^\pm$ the infinitesimal generators (charges) of BMS$^\pm$. However, there is a subtlety we need to address before we further develop this approach: \eqref{eq: symmetry gravitational scattering} implicitly relates what happens at past- and future null infinity since we scatter massless particles. This means that what goes on near spatial infinity $i^0$ will be important. Specifically, we need to understand how to match on final data at $\mathcal{I}^+_-$ given initial data at $\mathcal{I}^-_+$ for it to make sense. Such a prescription was provided by Christodoulou and Klainerman \cite{christodoulou_global_1993} and is precisely the topic of the next subsection.

    \subsection{Christodoulou-Klainerman spaces \label{sec: CK spaces}}
    
    The particles considered in our scattering process are assumed to be weakly interacting in the far past and future. One could thus be tempted to naively equate the data at $\cal I^-$ with the data at $\cal I^+$ at spacelike infinity $i^0$. Things are however not so straightforward... Indeed, $i^0$ is a singular point in the conformal compactification of general AFS, preventing a fully general canonical identification between $\cal I^+$ and $\cal I^-$. There is a way out though, provided the metric under study lies in a suitable neighbourhood of the Minkowski metric. As argued by Strominger \cite{strominger_bms_2014}, we can assume that the configurations we analyze correspond to weakly interacting Christodoulou-Klainerman (CK) geometries. CK  showed \cite{christodoulou_global_1993} that there exists a class of initial data which decays sufficiently fast as spatial infinity, such that the "mapping" from $\cal I^-$ to $\cal I^+$ corresponds to a smooth geodesically complete solution. In such configurations, the Bondi news tensor falls off as
    \begin{equation}
        N_{zz}(u) \to |u|^{-3/2} \quad \text{or faster as } u \to \pm \infty, \label{eq: falloff bondi news}
    \end{equation} while $m$ and $N_A$ remain finite in the two limits. This will help us impose matching conditions for initial and late data. Recall that we seek asymptotically flat solutions to the Einstein equations which revert to the vacuum in the far past and future. Taking care of the Bondi news trivially determines $C_{zz}$, $C_{\bar z \bar z}$ up to an integration function. Meanwhile, from \eqref{eq: effect supertranslation Czz} we have
    \begin{equation*}
        \mathcal{L}_f C_{zz} = f \partial_u C_{zz} - 2 D^2_z f = f N_{zz} - 2 D^2_z f.
    \end{equation*} Hence vanishing of $N_{zz}$ as $u \to \pm \infty$ in CK spaces implies
    \begin{equation}
        \lim_{u \to \pm \infty} \mathcal{L}_f C_{zz} = - 2 D^2_z f|_{\mathcal{I}^+_\pm} 
        \Leftrightarrow \left. C_{zz} \right|_{\mathcal{I^+_\pm}} = \left. -  2D^2_{z} C \right|_{\mathcal{I}^+_\pm}, \quad \text{with $C(z, \bar z)$ such that } \mathcal{L}_fC = f, \label{eq: Czz CK spaces}
    \end{equation} having used that $C(z, \bar z)|_{\mathcal{I}^+_\pm} = C(z, \bar z)|_{\mathcal{I}^-_\mp}$ from the antipodal map relating stereographic coordinates. Similarly, we deduce at past null infinity $\cal I^-$ from \eqref{eq: supertrans cal- Czz} that (it is also assumed that the advanced Bondi news $M_{zz}$ vanishes as $v \to \pm \infty$ in CK spaces)
    \begin{equation*}
        D_{zz} |_{\mathcal{I}^-_\pm} = + 2 D^2_z D|_{\mathcal{I}^-_\pm}, \quad \text{for } D(z, \bar z) \text{ such that } \mathcal{L}_{f^-} D = f^-.
    \end{equation*} We now have a prescription on how to relate initial and late data, given by the continuity condition
    \begin{equation}
        \boxed{C_{zz}|_{\mathcal{I}^+_-} = - D_{zz} |_{\mathcal{I}^-_+} \Leftrightarrow C(z, \bar z) = - D(z, \bar z).} \label{eq: continuity condition}
    \end{equation}Let us turn to the study of the effect of BMS supertranslation generators (charges) on the $\cal S$-matrix.
        
    \subsection{BMS supertranslations and leading soft theorem \label{sec: supertranslations soft theorems}} 

        In the following, we make use of the covariant phase space formalism introduced in section~\ref{sec: covariance and asymptotic symmetries} to discuss the construction of the conserved charges associated to supertranslations. We then show that these charges yield a symmetry of the $\cal S$-matrix, implying energy conservation and a Ward identity. To conclude, we relate this identity to Weinberg's soft graviton theorem.

        \subsubsection{Supertranslation charges \label{sec: supertranslation charges}}
        To get the conserved charges, we first need to construct the symplectic form of the phase space of our theory. This is where the algorithm we fleshed out in section~\ref{sec: covariance and asymptotic symmetries} comes in handy.  The Lagrangian form considered here is no other than the Einstein-Hilbert Lagrangian
        \begin{equation}
            L = \frac{1}{16\pi G} \sqrt{-g} g^{\mu \nu} R_{\mu \nu} ( \deriv^4 x). \label{eq: Einstein Hilbert Lagrangian}
        \end{equation} Steps 1 to 4 of Figure~\ref{fig:CPS formalism} were carried out by Alessio and Arzano \cite{alessio_note_2019}, who assumed that
        \begin{equation*}
            \lim_{u \to \infty} C_{zz}(u, z, \bar z) = \phi^+_{zz}(z, \bar z) = D^2_z C^+, \quad \lim_{u \to - \infty} C_{zz}(u, z, \bar z) = \phi^-_{zz}(z, \bar z) = D^2_z C^-  \label{eq: boundary conditions Czz}
        \end{equation*} where $\phi^\pm_{zz}(z, \bar z)$ and $C^\pm(z, \bar z) = - 2 C|_{\mathcal{I}^+_\pm}$ are smooth, non-vanishing functions on $\mathbb{S}^2$. Integrating the Bondi news tensor \eqref{eq: bondi news tensor} along retarded time, one gets
        \begin{equation*}
            C_{zz}(u, z, \bar z) - \phi^-_{zz}(z, \bar z) = \int^u_{-\infty} \deriv u' N_{zz}(u', z, \bar z), \quad \phi^+_{zz}(z, \bar z) - C_{zz}(u, z, \bar z) = \int^\infty_u \deriv u' N_{zz}(u', z, \bar z).
        \end{equation*} Adding the two,
        \begin{equation}
            \int_{-\infty}^\infty \deriv u' N_{zz}(u', z, \bar z) = \phi^+_{zz} - \phi^-_{zz}  \equiv \Delta \phi_{zz} = D^2_z N, \label{eq: boundary field N}
        \end{equation} for $N(z, \bar z)$ a real boundary field. Meanwhile, substracting yields
        \begin{equation*}
            C_{zz}(u, z, \bar z) \equiv \frac{1}{2} \Delta \phi_{zz}(z, \bar z) + \phi^-_{zz}(z, \bar z) + \hat C_{zz}(u, z, \bar z) \label{eq: decomposition Czz}.
        \end{equation*} where 
        \begin{equation*}
            \hat C_{zz}(u, z, \bar z) \equiv \frac{1}{2} \left[ \int^u_{-\infty} \deriv u' N_{zz}(u', z, \bar z) - \int_u^\infty \deriv u' N_{zz}(u', z, \bar z) \right],
        \end{equation*} is the bulk contribution and the remaining terms are boundary terms. In particular, this means 
        \begin{equation*}
            N_{zz} = \partial_u C_{zz} = \partial_u \hat C_{zz}.
        \end{equation*}
        Choosing $\Sigma = \cal I^+$ as the Cauchy slice, one arrives at the presymplectic form
         \begin{equation}
            \tilde{\Omega}_{\mathcal{I}^+} = \frac{1}{16 \pi G} \int_{\mathcal{I}^+} \gamma_{z \bar z} \deriv^2 z \deriv u \delta \hat C_{zz} \wedge \delta N^{zz} + \frac{1}{32 \pi G} \int \gamma_{z \bar z} \deriv^2 z (\delta \phi^-_{zz} \wedge \delta \Delta \phi^{zz} + \delta \phi^-_{\bar z \bar z} \wedge \delta \Delta \phi^{\bar z \bar z}). \label{eq: presymplectic form simplified}
        \end{equation} To get the symplectic form, we need to carry out steps 5 and 6 of Fig.~\ref{fig:CPS formalism} and define our phase space. 
        
        \paragraph{Phase space} As mentioned in Section~\ref{sec: CK spaces}, we focus on CK spaces where the falloff \eqref{eq: falloff bondi news} of the Bondi news as $u \to \pm \infty$ guarantees that the first integral in \eqref{eq: presymplectic form simplified} converges. We argued that in such spaces, the only data we need specify is $C_{zz}$, which is further constrained by the condition \eqref{eq: continuity condition}. As a result, we choose for our phase space $\Gamma^+$ the set of $C_{zz}$ such that these hold,
        \begin{equation*}
            \Gamma^+ := \left\{ C_{zz} : C_{zz}|_{\mathcal{I}^+_\pm} = D^2_z C^\pm + \mathcal{O}(u^{-\epsilon}), \epsilon > 0\right\},
        \end{equation*} and similarly for $\Gamma^-$ at $\cal I^-$
        \begin{equation*}
             \Gamma^- := \left\{ D_{zz} : D_{zz}|_{\mathcal{I}^-_\pm} = D^2_z D^\pm + \mathcal{O}(u^{-\epsilon}), \epsilon > 0\right\}.
        \end{equation*}We also demand the variations $\delta C_{zz}$ and $\delta N_{zz}$ be also CK, that is to say
        \begin{align*}
            &\delta C_{zz} |_{\mathcal{I}^+_\pm} = C'^\pm_{zz}|_{\mathcal{I}^+_-} - C^\pm_{zz}|_{\mathcal{I}^+_-} = D^2_z C'^{\pm} - D_z^2C^\pm \overset{!}{=} D_z^2 \delta C^\pm \\ \text{and } &\delta N_{zz} \overset{u \to \pm \infty}{\longrightarrow} 0 \text{ as } u^{1-\epsilon}, \quad \epsilon > 0, \quad \text{and similarly at $\cal I^-$}.
        \end{align*} 
        
        \paragraph{Symplectic form} These considerations allow to go from \eqref{eq: presymplectic form simplified} to the symplectic form \cite{alessio_note_2019}
        \begin{equation}
            \Omega_{\mathcal{I}^+} = \frac{1}{16 \pi G} \int_{\mathcal{I}^+} \gamma_{z \bar z} \deriv^2 z \deriv u \delta \hat C_{zz} \wedge \delta N^{zz} + \frac{1}{16 \pi G} \int \gamma_{z \bar z} \deriv^2 z D^2_z \delta C \wedge D^{2z} \delta N, \label{eq: final symplectic form}
        \end{equation} where one denotes $C^- \equiv C$ for notational simplicity. This now converges on $\Gamma$ and we are in a position to apply the tools of symplectic geometry and derive Poisson brackets.

        \paragraph{Charges} From \eqref{eq: final symplectic form}, we can read off the non-vanishing Poisson brackets:
        \begin{align*}
           \text{bulk-bulk:}& \quad \{N_{\bar z \bar z}(u, z, \bar z), \hat C_{ww}(u', w, \bar w) \} = 16 \pi G \delta^2(z - w) \delta (u - u') \gamma_{z \bar z} \\
            \text{boundary-boundary:}& \quad \{ D^2_{\bar z} N(z, \bar z), D^2_wC(w, \bar w) \} = 16 \pi G \delta^2(z - w) \gamma_{z \bar z}.
        \end{align*} Now, recall that in symplectic geometry, the symplectic form $\Omega_{\mathcal{I}^+}$ and the infinitesimal charge $\cancel{\delta} Q_\xi$ associated with the symmetry generated by a given vector field $\xi$ are related by \eqref{eq: infinitesimal charge}. The finite charge can then be obtained by integrating $\cancel{\delta}Q_\xi$ along a path in the field space.\footnote{Such charge is said to be integrable if the integral does not depend on the particular path chosen, i.e.~if there exists a functional $Q_\xi$ such that $\cancel{\delta}Q_\xi = \delta(Q_\xi)$. $Q_\xi$ is conserved on shell \cite{alessio_note_2019}.} Eq.\eqref{eq: final symplectic form} yields $\cancel{\delta} Q_f = \delta Q_f$ where \cite{alessio_note_2019}
        \begin{equation}
            Q_f = -\frac{1}{16 \pi G} \int_{\mathcal{I}^+} \gamma_{z \bar z} \deriv^2 z \deriv u f N_{zz} N^{zz} + \frac{1}{8 \pi G} \int \gamma_{z \bar z} \deriv^2 z D^2_z f D^{2z} N \equiv Q^\mathcal{H}_f + Q^\mathcal{S}_f \label{eq: supertrans charge alessio}
        \end{equation} is our conserved charge under supertranslations. Note that it naturally decomposes into a \textit{soft} (linear in the fields) and \textit{hard} (quadratic in the fields) part, respectively denoted $Q^\mathcal{S}_f$ and $Q^\mathcal{H}_f$. We also introduced the boundary field $N(z, \bar z)$ \cite{alessio_note_2019} such that
         \begin{equation*}
             D^2 _z N(z, \bar z) = \int \deriv u' N_{zz}(u',z, \bar z).
         \end{equation*}Denoting $Q_f \equiv T^+$ to match the results of \cite{strominger_bms_2014} we have
        \begin{equation}
                T^+(f) = -\frac{1}{16 \pi G} \int_{\mathcal{I}^+} \gamma_{z \bar z} \deriv^2 z \deriv u f N_{zz} N^{zz} + \frac{1}{8 \pi G} \int \gamma_{z \bar z} \deriv^2 z D^2_z f D^{2z} N
                = \frac{1}{4 \pi G} \int_{\mathcal{I}^+_-} \deriv^2 z \gamma_{z \bar z} f m. \label{eq:supertranslation charges}
        \end{equation} Here, we performed an integration by parts on the last term in the first equality and used the constraint equation for the Bondi mass \eqref{eq:constraint mass aspect}. From the brackets, one finds \cite{strominger_bms_2014, alessio_note_2019}
        \begin{gather*}
            \{ T^+(f), N_{zz} \} = f \partial_u N_{zz}, \quad
            \{T^+(f), C_{zz}\} = f \partial_u C_{zz} - 2 D_z^2 f, \\
            \{T^+(f), N\} = 0, \quad \text{and} \quad
            \{T^+(f), C\} = - 2 f.
        \end{gather*} Comparing this with \eqref{eq: effect supertranslation Czz} and \eqref{eq: effect supertranslation Nzz} confirms that $T^+(f)$ is the generator of supertranslations at $\cal I^+$. Carrying out the same analysis at $\cal I^-$, one arrives at the supertranslation  charge
        \begin{equation}\begin{aligned}
            T^-(f^-) &= -\frac{1}{16\pi G} \int \deriv v \deriv^2 z f^- \gamma_{z \bar z} M_{zz}M^{zz} - \frac{1}{8 \pi G} \int \deriv^2 z \gamma_{z \bar z} f^- D_z^2 D_{\bar z}^2 M,  \\
            &= \frac{1}{4\pi G} \int_{\mathcal{I}^-_+} \deriv^2 z \gamma_{z \bar z} f^- m^- \quad \text{having used that } D_z^2 M = \int \deriv u' M_{zz}(u', z, \bar z). \label{eq:supertranslation charges scri-}
        \end{aligned}\end{equation} This charge has brackets
        \begin{gather*}
            \{ T^-(f^-), M_{zz} \} = f^- \partial_v M_{zz}, \quad
            \{T^-(f^-), D_{zz}\} = f^- \partial_v D_{zz} + 2 D_z^2 f^-, \\
            \{T^-(f^-), M\} = 0, \quad \text{and} \quad
            \{T^-(f^-), D\} = 2 f^-.
        \end{gather*} Inspecting \eqref{eq: supertrans cal- Czz} and \eqref{eq: supertrans cal- Nzz} shows again that $T^-(f^-)$ indeed generates supertranslations at $\cal I^-$.
    
        \subsubsection{Supertranslation invariance, soft graviton and Ward identity}
            Having obtained the supertranslation charges, we want to show that some supertranslations leave the $\cal S$-matrix invariant, i.e.~that $B^\pm$ commute with $\cal S$ for the generators $\xi_{T}$ in a subgroup BMS$^0$ of $\text{BMS}^+ \times \text{BMS}^-$. Let us determine this subgroup. From \eqref{eq: effect supertranslation Uz} and \eqref{eq: effect supertranslation Vz}, we learn that the supertranslations that are compatible with \eqref{eq: continuity condition} are those satisfying
            \begin{equation}
                V_{2z}|_{\mathcal{I}^-_+} = U_{2z}|_{\mathcal{I}^+_-} \Rightarrow f^- (z, \bar z) = f(z , \bar z). \label{eq: BMS0 supertranslation condition}
            \end{equation} To match the charges \eqref{eq:supertranslation charges} and \eqref{eq:supertranslation charges scri-}, we also need to impose the matching of the Bondi mass aspect
            \begin{equation}
                m(z, \bar z) |_{\mathcal{I}^+_-} = m^-(z, \bar z) |_{\mathcal{I}^-_+}. \label{eq: bondi mass matching condition}
            \end{equation} Note that in these expressions, the coordinates $(z, \bar z)$ on both sides of the equality are related by the antipodal map. The vector fields satisfying these two conditions constitute the supertranslations of BMS$^0$, which is usually referred to as the \textit{diagonal} BMS group.\footnote{The superrotations of BMS$^0$ will be discussed in section~\ref{sec:conclu superrot}.} 

            \smallskip
            We have not shown that the $\cal S$-matrix is invariant under $\xi_{T} \in \text{BMS}^0$ yet. To see this, note from \eqref{eq:supertranslation charges} that $T^+(1) = M$. Put differently, $T^+(1)$ is the ADM Hamiltonian (times $G$). Moreover, the charges at $\cal I^+$ obey \cite{strominger_bms_2014}
            \begin{equation}
                \{ T^+(f), T^+(f')\} = 0. \label{eq: bracket supertranslation charges}
            \end{equation} From our above matching conditions \eqref{eq: BMS0 supertranslation condition} and \eqref{eq: bondi mass matching condition}, we can finally write for $\xi_T \in \text{BMS}^0$
            \begin{equation}
                T(f) \equiv T^+(f) = T^-(f). \label{eq: equal T charges}
            \end{equation} Since $T(1)$ is the Hamiltonian, all $T(f)$ Poisson-commute with $T(1)$ \eqref{eq: bracket supertranslation charges} and $\cal S$ is constructed from exponentials of the Hamiltonian, we have $[T(f), \mathcal{S}] = 0$. This, combined with \eqref{eq: equal T charges}, is exactly what we needed to argue the supertranslation invariance of the $\cal S$-matrix under BMS$^0$,
            \begin{equation}
                \boxed{T^+(f) \mathcal{S} - \mathcal{S} T^-(f) = 0\quad \text{for $f$ and $m$ satisfying \eqref{eq: BMS0 supertranslation condition} and \eqref{eq: bondi mass matching condition}}.}\label{eq:supertranslation invariance}
            \end{equation} So far, all our considerations were purely classical. By writing \eqref{eq:supertranslation invariance} we have now just quantized our theory by promoting the charges to operators acting on Hilbert spaces and on the $\cal S$-matrix. 
            
            \paragraph{Energy conservation} There is another nice interpretation to the statement \eqref{eq:supertranslation invariance}. Using the constraints \eqref{eq:constraint mass aspect} and \eqref{eq:constraint mass aspect scri-} (assuming non vanishing stress tensor this time), we can rewrite the supertranslation charges at $\cal I^+$ \eqref{eq:supertranslation charges} and $\cal I^-$ \eqref{eq:supertranslation charges scri-} as
            \begin{equation*}
                T^+(f) = \frac{1}{4\pi G} \int \deriv u \deriv^2 z f \gamma_{z \bar z} \left[ -T_{uu} + \frac{1}{4} (D^2_z N^{zz} + D^2_{\bar z} N^{\bar z \bar z})\right]
            \end{equation*} and 
            \begin{equation*}
                T^-(f) = \frac{1}{4\pi G} \int \deriv v \deriv^2 z f \gamma_{z \bar z} \left[-T_{vv} - \frac{1}{4} (D^2_z M^{zz} + D^2_{\bar z} M^{\bar z \bar z} )\right].
            \end{equation*} We see that the local energy at a point includes not only a contribution from the stress tensor but also a term which is linear in the Bondi news and a total $u$ derivative. Then in particular for $f = \delta^2(z- w)$ our equality \eqref{eq: equal T charges} writes
            \begin{equation}
                \int_{\mathcal{I}^+} \deriv u  \gamma_{ z \bar z} \left[ -T_{uu} + \frac{1}{4} (D^2_z N^{zz} + D^2_{\bar z} N^{\bar z \bar z})\right] = \int_{\mathcal{I}^-} \deriv v \gamma_{z \bar z} \left[-T_{vv} - \frac{1}{4} (D^2_z M^{zz} + D^2_{\bar z} M^{\bar z \bar z} )\right]. \label{eq: energy conservation angle}
            \end{equation} This has a nice consequence: the energy flux at a point $w$ on $\cal I^+$ is equal to the integrated energy flux at the antipodal point $w$ on $\cal I^-$. In other words, the total accumulated energy incoming from every angle $(z, \bar z)$ on $\cal I^-$ equals the total accumulated energy emerging at the angle $(z, \bar z)$ on $\cal I^+$ \cite{strominger_bms_2014}. 

            \paragraph{Soft graviton} We now seek to relate the supertranslation invariance of the $\cal S$-matrix to a quantum Ward identity involving so called \textit{soft gravitons}. In quantum field theory, \textit{Ward identities} are fundamental relationships expressing the symmetries of a physical system by relating correlation functions of fields to each other. In the same way as fluctuations of quantum fields give rise to particles in quantum field theory, \textit{gravitons} denote the particles corresponding to small fluctuations of the spacetime metric around a flat background (Minkowski metric)
            \begin{equation*}
                g_{\mu \nu} = \eta_{\mu \nu} + h_{\mu \nu}.
            \end{equation*} The tensor $h_{\mu \nu}$ is then referred to as the \textit{graviton field}. Looking back at the form \eqref{eq:asymptotic metric retarded} of the asymptotic metric for an AFS in retarded coordinates,
            \begin{align*}
                \deriv s^2 &= -\deriv u^2 - 2 \deriv u \deriv r + 2 r^2 \gamma_{z \bar z} \deriv z \deriv \bar z + \frac{2m}{r} \deriv u^2 + r C_{zz} \deriv z^2 + r C_{\bar z \bar z} \deriv \bar z^2 + ... \\ &= \eta_{\mu \nu} \deriv x^\mu \deriv x^\nu + r C_{zz} \deriv z^2 + r C_{\bar z \bar z} \deriv \bar z^2 + ...
            \end{align*} we deduce that the graviton field in the $(u,r,z,\bar z)$ system is encoded through $C_{AB}$ since we can easily read off (up to normalization) \cite{he_bms_2015}
            \begin{equation}
                C_{zz}(u,z, \bar z) = \kappa \lim_{r \to \infty} \frac{1}{r} h^{out}_{zz}(r,u,z,\bar z), \quad \kappa^2 = 32\pi G. \label{eq: Czz graviton}
            \end{equation} Thus $C_{zz}$ and $C_{\bar z \bar z}$ can indeed be seen as the two helicity modes of the particle, as was already pointed out around \eqref{eq: bondi news tensor}. Since we are dealing with outgoing radiation at $\cal I^+$, $h_{z z}^{out}$ in fact denotes the free graviton field. It is common practice in QFT to expand the fields in momentum space
            \begin{equation}
                h_{\mu \nu}^{out}(x) = \sum_{\alpha = \pm} \int \frac{\deriv^3 q}{(2\pi)^3} \frac{1}{2 \omega_q}  \left[ \varepsilon^{\alpha \ast}_{\mu \nu}(\Vec{q}) a_\alpha^{out}(\Vec{q})e^{iq\cdot x} + \varepsilon^\alpha_{\mu \nu} (\Vec{q}) a_{\alpha}^{out}(\Vec{q})^\dagger e^{- i q \cdot x} \right]. \label{eq: mode expansion graviton}
            \end{equation} Here $q^0 \equiv \omega_q = \abs{\Vec{q}}$ since the graviton is massless, $\alpha = \pm$ are the two helicities of the particle and $\varepsilon_{\mu \nu}^\alpha$ is its polarization tensor. The quantization comes with a set of commutation relations for the graviton creation and annihilation operators $a_\alpha^{out\dagger}$ and $a_\alpha^{out}$
            \begin{equation*}
                [a_\alpha^{out}(\Vec{q}), a_\beta^{out}(\Vec{q}')^\dagger] = \delta_{\alpha \beta} (2 \omega_q) (2\pi)^3 \delta^{(3)}(\Vec{q}- \Vec{q}'), \quad \text{all the others vanish.}
            \end{equation*} To proceed, it is useful to introduce a careful parametrization\footnote{This is motivated by the fact that for $t,r \to \infty$, the wave packet for a massless particle with spatial momentum centered around $\Vec{p}$ localizes on the conformal sphere near 
            \begin{equation*}
                \Vec{p} = \omega \hat x \equiv \omega \frac{\Vec{x}}{r} = \frac{\omega}{1 + z \bar z}(z + \bar z, -i(z - \bar z), 1 - z \bar z), \quad \text{where $\Vec{p} \cdot \Vec{p} = \omega^2$ \cite{he_bms_2015}.}
            \end{equation*} } of the graviton four-momentum $q^\mu$,
            \begin{equation*}
                q^\mu = \frac{\omega_q}{1 + w \bar w} ( 1+ w \bar w, w + \bar w, - i(w - \bar w), 1 - w \bar w),
            \end{equation*} as well as of the polarization tensor $\varepsilon^{\pm \mu \nu} = \varepsilon^{\pm \mu}\varepsilon^{\pm \nu}$,
            \begin{equation*}
                \varepsilon^{+\mu}(\Vec{q}) = \frac{1}{\sqrt{2}}(\bar w, 1, -i, - \bar w), \quad \varepsilon^{-\mu}(\Vec{q}) = \frac{1}{\sqrt{2}}(w, 1, i, - w), \quad \text{such that } \varepsilon^{\pm\mu \nu}q_\nu = \varepsilon^{\pm \mu}\,_\mu = 0.
            \end{equation*} Now applying the chain rule $h_{zz} = \partial_z x^\mu \partial_z x^\nu h_{\mu \nu}$ to our mode expansion \eqref{eq: mode expansion graviton} for the graviton field
            \begin{equation*}
                h_{zz} = \partial_z x^\mu \partial_z x^\nu \sum_{\alpha = \pm} \int \frac{\deriv^3 q}{(2\pi)^3} \frac{1}{2 \omega_q}  \left[ \varepsilon^{\alpha \ast}_{\mu \nu}(\Vec{q}) a_\alpha^{out}(\Vec{q})e^{iq\cdot x} + \varepsilon^\alpha_{\mu \nu} (\Vec{q}) a_{\alpha}^{out}(\Vec{q})^\dagger e^{- i q \cdot x} \right]
            \end{equation*} and using the fact that with the above parametrization choices
            \begin{equation*}
                \partial_z x^\mu \varepsilon^{+}_\mu(\Vec{q}) = \frac{\sqrt{2}r \bar z(\bar w - \bar z)}{(1+z \bar z)^2}, \quad \partial_z x^\mu \varepsilon^{-}_\mu(\Vec{q}) = \frac{\sqrt{2}r(1+w \bar z)}{(1+ z \bar z)^2}
            \end{equation*} along with \eqref{eq: Czz graviton}, we can recast $C_{zz}$ directly in terms of the graviton mode operators \cite{he_bms_2015}
            \begin{equation*}
                C_{zz} = \kappa \lim_{r \to \infty} \frac{1}{r} \partial_z x^\mu \partial_z x^\nu \sum_{\alpha = \pm} \int \frac{\deriv^3 q}{(2\pi)^3} \frac{1}{\omega_q} \left[\varepsilon^{\alpha \ast}_{\mu \nu}(\Vec{q}) a_\alpha^{out}(\Vec{q})e^{- i \omega_q u - i \omega_q r (1 - \cos \theta)} + \text{h.c.} \right].
            \end{equation*} Here $\theta$ is the angle between $\Vec{x}$ and $\Vec{q}$ and we used the expressions \eqref{eq: cartesian in stereographic} for $x^\mu$ in terms of $(u,r,z, \bar z)$. Taking $r \to \infty$, this integral is can be evaluated by resorting to the \textit{stationary phase approximation}, i.e.~we approximate the oscillatory integrand by its dominant contributions, which occur when the phase of the integrand is stationary ($\theta = 0,\pi$). The contributions from other regions of the integrand that rapidly oscillate will cancel out or be negligible due to their oscillatory nature.  As it turns out, the contribution from $\theta = \pi$ also vanishes as $r \to \infty$, leaving us with
            \begin{equation}
                C_{zz} = - \frac{i \kappa}{4 \pi^2(1 + z \bar z)^2} \int_0^\infty \deriv \omega_q \left[ a_+^{out}(\omega_q \hat x) e^{- i \omega_q u} - a_-^{out}(\omega_q \hat x)^\dagger e^{i \omega_q u} \right]. \label{eq: expression Czz operators}
            \end{equation} Hence $C_{zz}$ is nothing but a Fourier transform of the momentum-space creation and annihilation operators. Acting with the inverse transform and defining the Fourier modes of the Bondi news 
            \begin{equation}
                N^\omega_{zz}(z, \bar z) \equiv \int_{- \infty}^\infty \deriv u e^{i \omega u} \partial_u C_{zz}, \label{eq: Fourier mode Bondi news}
            \end{equation} our expression \eqref{eq: expression Czz operators} becomes
            \begin{equation*}
                N_{zz}^\omega(z, \bar z) = - \frac{\kappa}{2 \pi(1+ z \bar z)^2} \int_0^\infty \deriv \omega_q \omega_q \left[ a_+^{out}(\omega_q \hat{x}) \delta(\omega_q - \omega) + a_-^{out}(\omega_q \hat{x})^\dagger \delta(\omega_q + \omega) \right].
            \end{equation*} The sign of $\omega$ will affect which of the terms in the integrand contributes. For $\omega = 0$, we average on both contributions and define the (hermitian) zero-mode $N^0_{zz}$ of the Bondi news as
            \begin{equation}
                N^0_{zz}(z, \bar z) \equiv \lim_{\omega \to 0^+} \frac{1}{2}(N_{zz}^\omega + N_{zz}^{-\omega})  = - \frac{\kappa}{4 \pi (1+z \bar z)^2} \lim_{\omega \to 0^+} \left[ \omega a_+^{out}(\omega \hat x) + \omega a_-^{out}(\omega \hat x)^\dagger \right]. \label{eq: N0 modes}
            \end{equation} Carrying out the same steps at $\cal I^-$ yields \cite{he_bms_2015}
            \begin{equation}
                M^0_{zz}  = - \frac{\kappa}{4\pi(1+z \bar z)^2} \lim_{\omega \to 0^+} \left[\omega a_+^{in}(\omega \hat x) + \omega a_-^{in} (\omega \hat x)^\dagger \right]. \label{eq: M0 modes}
            \end{equation} This achieves to show how $C_{zz}$ relates to gravitons. Both formuli will play a key role later on. 

            \smallskip
            Now, what do we mean by \textit{soft} gravitons? In scattering amplitudes, \textit{soft} contributions refer to the contributions from low-energy, low-momentum particles. In quantum field theory, we see that these contributions are \textit{linear} in the fields because they correspond to the first-order terms in a perturbative expansion, where each order is proportional to a power of the coupling constant. In contrast, \textit{hard} contributions correspond to high-energy, high-momentum particles or radiation which cannot be treated as small fluctuations and thus require a non-perturbative treatment. The non-linear terms in the perturbative expansion become important at higher orders, where they contribute to the hard contributions to the scattering amplitudes. This motivates our splitting of the charge \eqref{eq: supertrans charge alessio} since we now see that we only need to consider the soft parts of the charges to first order in perturbation theory. Incidentally, these are also related to the zero-modes of the advanced and retarded Bondi news we just derived through $C_{zz}$, etc. Great! We now have all the tools we need to reformulate \eqref{eq:supertranslation invariance} as a Ward identity, viewing it first as a statement for a soft graviton current.

            \paragraph{Soft graviton current} Time has finally come to look at scattering. Let us denote a in-state with energies $E_k^{in}$ incoming at points $z_k^{in}$ on the conformal $\mathbb{S}^2$ by $|z_1^{in}, z_2^{in},... \rangle$. Then the supertranslation generator $T^-$ acts as \cite{strominger_bms_2014}
            \begin{equation}
                T^-(f) |z_1^{in}, z_2^{in},... \rangle = F^-|z_1^{in}, z_2^{in},... \rangle + \sum_k E_k^{in} f(z_k^{in}) |z_1^{in}, z_2^{in},... \rangle, \label{eq: T in}
            \end{equation} where $F^-$ denotes the soft part of \eqref{eq:supertranslation charges scri-}
            \begin{equation*}
                F^- = -\frac{1}{8\pi G} \int_{\mathcal{I}^-} \deriv v \deriv^2 z D^2_{\bar z} f M^{\bar z}\,_z, \quad M_{zz} = \partial_v D_{zz}.
            \end{equation*} Labelling an outgoing state with energies $E_k^{out}$ by $|z_1^{out}, z_2^{out}, ... \rangle$, we see that $T^+$ similarly acts as
            \begin{equation}
                \langle z_1^{out}, z_2^{out},... | T^+(f) = \langle z_1^{out}, z_2^{out},... | F^+ + \sum_k E_k^{out} f(z_k^{out}) \langle z_1^{out}, z_2^{out},... |, \label{eq: T out}
            \end{equation} with $F^+$ the soft part of \eqref{eq:supertranslation charges}
            \begin{equation*}
                F^+ = \frac{1}{8 \pi G} \int_{\mathcal{I}^+} \deriv u \deriv^2 z D_{\bar z}^2 f N^{\bar z}\,_z.
            \end{equation*} Let us also define
            \begin{equation}
                F \equiv F^+ - F^- = \frac{1}{8 \pi G} \int \deriv^2z \gamma^{z \bar z} D_{\bar z}^2 f \left[ \int_{\mathcal{I}^-} \deriv v M_{zz} + \int_{\mathcal{I}^+} \deriv u N_{zz} \right]. \label{eq: definition F}
            \end{equation} The supertranslation Ward identity for the time ordered product $:F \mathcal{S}: = F^+ \mathcal{S} - \mathcal{S} F^-$ is then straightforwardly deduced from \eqref{eq:supertranslation invariance}, \eqref{eq: T in} and \eqref{eq: T out}
            \begin{multline}
                \langle z^{out}_1, z_2^{out}, ... | : F \mathcal{S}: | z^{in}_1, z_2^{in}, ... \rangle = \langle z^{out}_1, z_2^{out}, ... | F^+ \mathcal{S} - \mathcal{S}F^- | z^{in}_1, z_2^{in}, ... \rangle \\
                = \sum_k (E_k^{in}f(z_k^{in}) - E_k^{out} f(z_k^{out})) \langle z^{out}_1, z_2^{out}, ... | \mathcal{S} | z^{in}_1, z_2^{in}, ... \rangle. \label{eq:supertranslation ward identity}
            \end{multline} This relates $\cal S$-matrix elements with and without the insertion of $F$. In light of our expressions \eqref{eq: N0 modes} and \eqref{eq: M0 modes} for the modes of $N_{zz}$ and $M_{zz}$ in terms of graviton annihilation and creation operators, it is natural to think of $F$ as the operator corresponding to the insertion of a soft graviton to the scattering process. If there are $n$ incoming and $m$ outgoing particles, then total energy conservation requires
            \begin{equation*}
                \sum_{k=1}^m E_k^{out} = \sum_{k=1}^n E_k^{in}.
            \end{equation*} Choosing $f(w, \bar w) = (z-w)^{-1}$ in \eqref{eq: definition F}, we rewrite $F$ as the \textit{soft graviton current} $P_z$
            \begin{equation}
                F \left( \frac{1}{z-w} \right) \equiv P_z =  \frac{1}{2G} \left( \left. U_{2z} \right|_{\mathcal{I}^+_-}^{\mathcal{I}^+_+} - \left. V_{2z} \right|_{\mathcal{I}^-_-}^{\mathcal{I}^-_+} \right)=\frac{1}{2G} \left(\int_{- \infty}^\infty \deriv v \partial_v V_{2z} - \int_{-\infty}^\infty \deriv u \partial_u U_{2z} \right). \label{eq: def soft graviton current Pz}
            \end{equation} Using \eqref{eq: BMS0 supertranslation condition}, the matching \eqref{eq: energy conservation angle} of the charges at $\cal I^\pm$ can be rewritten as
            \begin{equation*}\begin{aligned}
                \int_{\mathcal{I}^+} \deriv u \left[ -\gamma_{z \bar z}T_{uu} + \frac{1}{4} \partial_u (\partial_z U_{2\bar z} + \partial_{\bar z}U_{2z}) \right] &= \int_{\mathcal{I}^-} \deriv v \left[ -\gamma_{z \bar z}T_{vv} + \frac{1}{4} \deriv v (\partial_z V_{2\bar z} + \partial_{\bar z} V_{2z} )\right] \\
                \Leftrightarrow  \gamma_{z \bar z} \left( \int_{\mathcal{I}^+} \deriv u T_{uu}  - \int_{\mathcal{I}^-}  \deriv v T_{vv} \right) &= \partial_{\bar z} \left[ \left. U_{2z} \right|_{\mathcal{I}^+_-}^{\mathcal{I}^+_+} - \left. V_{2z} \right|_{\mathcal{I}^-_-}^{\mathcal{I}^-_+} \right] \overset{\eqref{eq: def soft graviton current Pz}}{=} 4 G \partial_{\bar z} P_z.
            \end{aligned}\end{equation*} We can solve for $P_z$ in this special case by means of a Green function for $\partial_{\bar z}$,
            \begin{equation}
                P_z \equiv \frac{1}{4\pi G} \int \deriv^2 w \frac{\gamma_{w \bar w}}{z - w} \left( \int \deriv u T_{uu} - \int \deriv v T_{vv} \right). \label{eq: soft graviton current}
            \end{equation}
            
            \paragraph{Ward identity} We now construct the Ward identity for the soft graviton current. Consider the simple case where
            \begin{equation*}
                G \sum_k E_k^{out} = G \sum_k E_k^{in}
            \end{equation*} and the incoming and outgoing particles are localized at $(v_k, z_k^{in})$ and $(u_k, z_k^{out})$ at $\cal I^-$ and $\cal I^+$ respectively. Then $T_{uu}$ and $T_{vv}$ take the simple form
            \begin{equation*}
                T_{uu} = 4 \pi G \sum_k E_k^{out} \delta(u -u_k) \frac{\delta^2(z - z_k^{out})}{\gamma_{z \bar z}} \quad \text{and} \quad T_{vv} = 4 \pi G \sum_k E_k^{in} \delta(v - v_k) \frac{\delta^2(z - z_k^{in})}{\gamma_{z \bar z}}
            \end{equation*}while \eqref{eq: soft graviton current} simplifies to 
            \begin{equation*}
                P_z =  \sum_{k=1}^m \frac{E_k^{out}}{z - z_k^{out}} - \sum_{k=1}^n \frac{E_k^{in}}{z - z_k^{in}},
            \end{equation*} such that the matrix element \eqref{eq:supertranslation ward identity} reads
            \begin{equation}
                \boxed{\langle z^{out}_1, z_2^{out}, ... | : P_z \mathcal{S}: | z^{in}_1, z_2^{in}, ... \rangle = \left[ \sum_{k=1}^m \frac{E_k^{out}}{z - z_k^{out}} - \sum_{k=1}^n \frac{E_k^{in}}{z - z_k^{in}}\right] \langle z^{out}_1, z_2^{out}, ... | \mathcal{S} | z^{in}_1, z_2^{in}, ... \rangle.}\label{eq:supertranslation ward current}
            \end{equation} Thus \eqref{eq:supertranslation ward current} relates $\cal S$-matrix elements with and without insertions of the soft graviton current $P_z$. We now show that this is equivalent to Weinberg's leading soft graviton theorem.
    
            \subsubsection{Leading soft theorem} 

            As mentioned before, there are certain limits in which the scattering matrix of a particular process simplifies. In particular, if the energy of one or more of the massless particles involved in the collision is taken to be small in comparison to the energy or masses of the other particles in the process,  universal properties of Feynman diagrams and scattering amplitudes emerge. This is the \textit{soft} limit, which gives rise to \textit{soft theorems}. In what follows, we first state the results of Weinberg \cite{weinberg_infrared_1965}, who formulated these concepts for gravity. We then show how his results can be related to the Ward identity \eqref{eq:supertranslation ward current} derived previously from the BMS supertranslation invariance of the $\cal S$-matrix.
            
            \paragraph{Weinberg's soft graviton theorem} We focus here on the case of a free massless scalar field, as done by He et al.~\cite{he_bms_2015}. We are interested in computing the on-shell amplitude $\mathcal{M}(p_1', ..., p_m', p_1, ..., p_n)$ involving $n$ incoming (momenta $p_1, ..., p_n$) and $m$ outgoing (momenta $p_1', ..., p_m'$) massless scalars. Now, consider the exact same amplitude but with an additional outgoing soft graviton of momentum $q$ and polarization $\epsilon_{\mu \nu}(q)$ satisfying the gauge condition $q^\mu \epsilon_{\mu \nu} = \frac{1}{2} q_\nu \epsilon^\mu\,_\mu$, $\mathcal{M}_{\mu \nu}(q, p_1', ..., p'_m, p_1, ...,p_n)$. Weinberg's soft graviton theorem relates the two as
            \begin{empheq}[box=\widefbox]{equation}
                \mathcal{M}_{\mu \nu}(q, p_1', ..., p'_m, p_1, ...,p_n) = \frac{\kappa}{2} \left[ \sum_{k=1}^m \frac{p'_{k \mu} p'_{k \nu}}{p'_k \cdot q} - \sum_{k=1}^n \frac{p_{k\mu}p_{k\nu}}{p_k \cdot q}\right] \mathcal{M}(p_1', ..., p_m', p_1, ..., p_n), \label{eq: weinberg soft graviton theorem}
            \end{empheq} with $\kappa^2 = 32 \pi G$ as before and where the terms in brackets is the \textit{soft factor}, $S_{\mu \nu}^{m-n}$. The latter is universal and gauge invariant \cite{he_bms_2015}, meaning that the formula does not depend on any of the quantum numbers of the asymptotic particles involved in the $\mathcal{S}$-matrix element.
    
            \paragraph{Equivalence} We now want to show that \eqref{eq:supertranslation ward current} and \eqref{eq: weinberg soft graviton theorem} are equivalent statements. To this end, we need to recast the soft graviton current \eqref{eq: def soft graviton current Pz} in terms of standard momentum space creation and annihilation operators. From \eqref{eq: Fourier mode Bondi news} and \eqref{eq: boundary field N}, we see that
            \begin{equation*}
                N^0_{zz}(z, \bar z) = \int_{- \infty}^\infty \deriv u  \partial_u C_{zz} = D_z^2 N.
            \end{equation*} Similar considerations at $\cal I^-$ yield $M^0_{zz}(z,\bar z) = D_z^2 M$, for $M$ also a boundary field. Then \eqref{eq: def soft graviton current Pz}, together with $U_{2z} = D^z C_{zz}$ as in \eqref{eq: U2A} and $V_{2z} = - D^z D_{zz}$ means that we can write
            \begin{equation}
                P_z = -\frac{1}{2G} \left(\int_{- \infty}^\infty \deriv v \partial_v D^z D_{zz} + \int_{-\infty}^\infty \deriv u \partial_u D^z C_{zz} \right)
                = \frac{1}{2G} D^z \left(M^0_{zz} + N^0_{zz} \right)
                = \frac{1}{4G} \gamma^{z \bar z} \partial_{\bar z} \mathcal{O}_{zz}
                \label{eq: Pz Ozz}
            \end{equation} with $\mathcal{O}_{zz} = N_{zz}^0+M_{zz}^0$ the sum of the zero Fourier modes of the Bondi news \eqref{eq: N0 modes} and \eqref{eq: M0 modes}. Recall that in these formuli, $a^{in}_\pm$ ($a^{out}_\pm$) and $a_\pm^{in\dagger}$ ($a_\pm^{out\dagger}$) annihilate and create incoming (outgoing) gravitons on $\cal I^-$ ($\cal I^+$) with positive ($+$) or negative ($-$) helicity. Now, consider the matrix element
            \begin{equation*}
                \langle z_1^{out}, ... | : \mathcal{O}_{zz} \mathcal{S}: | z_1^{in}, ... \rangle.
            \end{equation*} On the one hand, using \eqref{eq: N0 modes} and \eqref{eq: M0 modes} we have
            \begin{equation*}\begin{aligned}
                \langle z_1^{out}, ... | : \mathcal{O}_{zz} \mathcal{S}: | z_1^{in}, ... \rangle 
                &= - \frac{\kappa}{4 \pi (1+ z \bar z)^2} \lim_{\omega \to 0^+} \langle z_1^{out}, ... | \left[\omega a_+^{out}(\omega \hat x) + \omega \cancel{a^{out}_- (\omega \hat x)^\dagger}\right] \mathcal{S}| z_1^{in}, ... \rangle \\
                &\qquad - \frac{\kappa}{4 \pi (1+ z \bar z)^2} \lim_{\omega \to 0^+} \langle z_1^{out}, ... | \mathcal{S} \left[ \omega \cancel{a_+^{in}(\omega \hat x)} + \omega a_-^{in} (\omega \hat x)^\dagger \right] | z_1^{in}, ... \rangle \\
                &= - \frac{\kappa}{2 \pi (1+z \bar z)^2} \lim_{\omega \to 0^+} \omega \langle z_1^{out}, ...|a_+^{out}(\omega \hat x) \mathcal{S} | z_1^{in}, ... \rangle.
            \end{aligned}\end{equation*} Here we used that $\langle z_1^{out}, ...|a_+^{out}(\omega \hat x) \mathcal{S}| z_1^{in}, ... \rangle = \langle z_1^{out}, ...| \mathcal{S} a_-^{in}(\omega \hat x)^\dagger| z_1^{in}, ... \rangle$ in the last line. On the other hand, \eqref{eq: weinberg soft graviton theorem} with a positive helicity outgoing graviton ($a_+^{out}(\Vec{q})$) allows to write~\cite{he_bms_2015}
            \begin{multline*}
                \langle z_1^{out}, ...|:\mathcal{O}_{zz} \mathcal{S}: |z_1^{in},... \rangle = \frac{8G}{(1+z \bar z)}\langle z_1^{out}, ...|\mathcal{S}| z_1^{in}, ... \rangle \\
                \times \left[\sum_{k=1}^m \frac{E_k^{out}(\bar z - \bar z_k^{out})}{(z-z_k^{out})(1+z_k^{out} \bar z_k^{out})} - \sum_{k=1}^n \frac{E_k^{in}(\bar z - \bar z_k^{in})}{(z - z_k^{in})(1+z_k^{in} \bar z_k^{in})} \right].
            \end{multline*} Then \eqref{eq: Pz Ozz} gives us
            \begin{equation*}\begin{aligned}
                \langle z_1^{out}, ...|:P_z \mathcal{S}:|z_1^{in}, ... \rangle &= \frac{1}{4G} \gamma^{z\bar z} \partial_{\bar z} \langle z_1^{out},... | : \mathcal{O}_{zz} \mathcal{S}:|z_1^{in}, ... \rangle \\
                &= \langle z_1^{out}, ...| \mathcal{S} | z_1^{in}, ... \rangle \left[ \sum_{k=1}^m \frac{E_k^{out}}{z - z_k^{out}} - \sum_{k=1}^n \frac{E^{in}_k}{z - z_k^{in}}\right] \\
                &\qquad + \langle z_1^{out}, ...| \mathcal{S} | z_1^{in}, ... \rangle \cancel{\left[ \sum_{k=1}^m \frac{E_k^{out} \bar z_k^{out}}{1+ z_k^{out} \bar z_k^{out}} - \sum_{k=1}^n \frac{E^{in}_k \bar z_k^{in}}{1+ z_k^{in}\bar z_k^{in}}\right]}.
            \end{aligned}\end{equation*} The last term vanishes because of total momentum conservation, meaning that we recover exactly \eqref{eq:supertranslation ward current}; the Ward identity for supertranslation invariance is equivalent to Weinberg's leading soft graviton theorem for the soft graviton current. Note that \eqref{eq: weinberg soft graviton theorem} naturally diverges as $q \to 0$. It is only thanks to the factor of $\omega$ in \eqref{eq: N0 modes} and \eqref{eq: M0 modes} that we are able to cancel this divergence in the soft limit and thereby pick up the residue of the soft factor. 

            \smallskip \noindent
            We now turn to the study of another aspect of the infrared structure of gravitational scattering that is highlighted by supertranslations: memory effects.
            
        \subsection{Gravitational memory effects \label{sec: memory effects}} 

            In this subsection we relate the BMS supertranslations to physical phenomena known as \textit{gravitational memory effects}. We first describe how the latter arise in general relativity, prior to describing how they relate to BMS supertranslations via so-called \textit{vacuum transitions}. In a second time, we show that memory effects can also be linked to soft theorems, in the spirit of what we saw in section~\ref{sec: supertranslations soft theorems}.
        
            \subsubsection{Displacement memory effect} 
\begin{figure}[h!]
                \centering
                \includegraphics[width=0.5\textwidth]{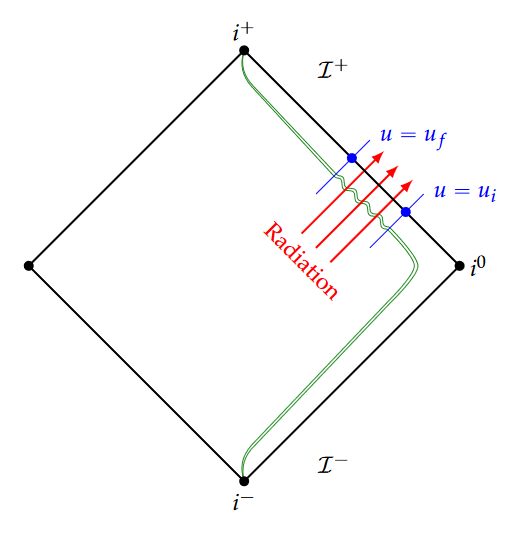}
                \caption{Penrose diagram for the displacement memory effect \cite{compere_advanced_2019}}
                \label{fig:displacement memory effect}
            \end{figure}

Let us introduce memory effects in gravity. Consider a pair of inertial observers ("detectors") travelling near future infinity, and suppose that a gravitational wave travels between the two during the retarded time interval $\Delta u \equiv u_f - u_i \ll r$. We assume that for any $u \notin [u_i, u_f]$, $N_{AB} = T_{\mu\nu} = 0$. This situation is depicted in Figure~\ref{fig:displacement memory effect}.
             It has been known since the work of Zeldovich \cite{zeldovich_radiation_1974}, Christodoulou \cite{christodoulou_nonlinear_1991}, Braginsky and Thorne \cite{braginsky_gravitational-wave_1987} and others that the passage of gravitational radiation in such a setup induces a permanent shift in the relative separation $s^\mu$ of the detectors. 
            
\begin{figure}
                \centering
                \includegraphics[width=\textwidth]{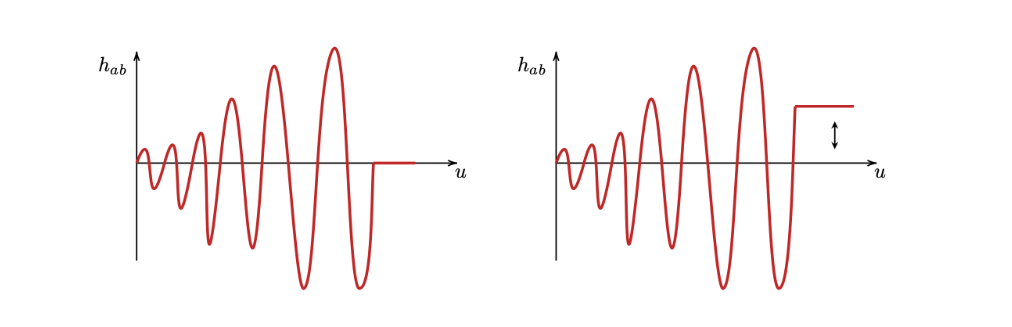}
                \caption{Sketch of the metric perturbation over time \cite{ciambelli_cornering_2023}}
                \label{fig:memory}
            \end{figure}

            \smallskip
            To see this, assume that the two inertial observers move along geodesics with 4-velocity $v^\mu$. Since both are located in the vicinity of $\cal I^+$, we can posit that $v^\mu \partial_\mu = \partial_u$ to leading order. We want to see how the passage of a gravitational wave affects our observers. We thus resort to the \textit{equation of geodesic deviation}, which quantifies the effect of tidal gravitational forces on neighboring free-falling objects. This writes \cite{strominger_lectures_2017, compere_advanced_2019, reall_part_2022}
            \begin{equation*}
                \nabla_v \nabla_v s^\mu = R^\mu\,_{\alpha \beta \gamma} v^\alpha v^\beta s^\gamma,
            \end{equation*} where $\nabla_v = v^\mu \nabla_\mu$ is the directional derivative along $v^\mu$. Both detectors being located on the same celestial sphere (i.e.~at the same value of $r$), we have $s^r = 0$. Expanding, one gets
            \begin{equation*}
                r^2 \gamma_{AB} \partial_u^2 s^B = R_{uAuB}s^B.
            \end{equation*} Noting that in Bondi gauge $R_{uAuB} = - \frac{r}{2} \partial_u^2 C_{AB} + \mathcal{O}(1)$ \cite{compere_advanced_2019}, this simplifies to
            \begin{equation*}
                \gamma_{AB}  \partial_u^2 s^B = \frac{1}{2r} \partial_u^2 C_{AB} s^B.
            \end{equation*} To proceed, let us expand $s^B = s^B_i + s^B_{sub}/r$ and integrate the above equation over $u$ to find
            \begin{equation*}
                \gamma_{AB} \Delta s_{sub}^B = \frac{1}{2r} \Delta C_{AB} s^B_i + \mathcal{O}(r^{-2}), \quad \Delta f(u) \equiv f(u_f) - f(u_i).
            \end{equation*} The case $\Delta s_{sub}^{B} \neq 0$ leads to the so-called \textit{displacement effect}.  To find the causes of this phenomenon, one must look at what could possibly lead to a change in $C_{AB}$. In their lecture notes, Compère and Fiorucci \cite{compere_advanced_2019} showed, using the constraint \eqref{eq:constraint mass aspect} on the Bondi mass aspect, that this either happens when $m$ varies between $u_i$ and $u_f$, when null matter reaches $\cal I^+$ between $u_i$ and $u_f$ (ex.~electromagnetic radiation), or when gravitational waves pass through $\cal I^+$ in the same interval. We have indeed argued above that the difference $\Delta C_{zz}$ need not vanish in our case since $C_{zz}$ (and $C_{\bar z \bar z}$) is left unconstrained in AFS, meaning that a shift in $s^\mu$ is possible: this is sometimes referred to as the \textit{Christodoulou effect}. See Fig.~\ref{fig:memory} for an illustration of the phenomenon, where the passage of the gravitational wave induces a permanent shift in the metric deformation.
             We now show that this can be seen as a consequence of BMS symmetry.

            \subsubsection{Vacuum transitions}  Recall how supertranslations act on $C_{zz}$ \eqref{eq: effect supertranslation Czz}, $N_{zz}$ \eqref{eq: effect supertranslation Nzz} and $m$ \eqref{eq: effect supertranslation m}. Therefore, starting with $C_{zz} = 0$, we have that after a \textit{pure}\footnote{By "pure", we mean one that is not an ordinary spacetime translation so that $D_z^2 f \neq 0$.} supertranslation $C_{zz}' = C_{zz} + \mathcal{L}_f C_{zz} \neq 0$. However, as $C_{zz}$ characterizes the vacuum ($N_{zz} = 0$), it must be that $C_{zz}'$ and $C_{zz}$ effectively describe physically inequivalent configurations. This shows that BMS supertranslations are the source of an infinite-dimensional degeneracy of vacua. 
            
            \smallskip
            Let us show that a vacuum transition under a supertranslation is equivalent to the displacement memory effect discussed above. Starting with the same setup as Fig.~\ref{fig:displacement memory effect}, we consider spacetimes for which
            \begin{gather*}
                m = M_i = \text{constant, and } C_{zz} = 0, N_{zz} = 0 \quad \text{for } u < u_i, \\
                m = M_f = \text{constant, and } C_{zz} \neq 0, N_{zz} = 0 \quad \text{for } u > u_f, 
            \end{gather*} while for $u_i \leq u \leq u_f$ $N_{zz}$ and/or $T_{uu}$ are nonzero on $\cal I^+$. We saw in \eqref{eq: Czz CK spaces} that in CK spaces we can find $C(z, \bar z)$ such that $C_{zz} = - 2 D^2_z C(z, \bar z)$ with $\mathcal{L}_f C = f$. In light of our discussion of vacuum transitions, we can now reinterpret this result as the statement that different vacua are related by supertranslations under which $C \to C + f$. Let us construct the supertranslation associated with a displacement memory effect. Integrating once again the constraint \eqref{eq:constraint mass aspect} on $m$  between $u_i$ and $u_f$ we find
            \begin{align*}
                \partial_u m &= - T_{uu} + \frac{1}{4} \left[ D^{2z} N_{zz}+ D^{\bar z2} N_{\bar z \bar z}\right] = - T_{uu} + \frac{1}{2} D_z^2 \partial_u C^{zz} \\
                \Leftrightarrow \quad D^2_z \Delta C ^{zz} &= 2 \int^{u_f}_{u_i} \deriv u T_{uu} + 2 \Delta m.
            \end{align*} The supertranslation $\Delta C \equiv f$ corresponding to such a change can be found by means of the Green function for $D_z^2 D_{\bar z}^2$:
            \begin{equation*}
                G(z, \bar z; z', \bar z') =- \frac{1}{\pi} \sin^2 \frac{\Theta}{2} \log \sin^2 \frac{\Theta}{2}, \quad \sin^2 \frac{\Theta(z, z')}{2} \equiv \frac{\abs{z - z'}^2}{(1+ z' \bar z')(1+ z \bar z)}
            \end{equation*}such that $D^2_z D^2_{\bar z} G(z,z';z', \bar z') = - \gamma_{z \bar z} \delta^2(z - z') + ...$, which yields
            \begin{equation}
                \Delta C(z, \bar z) = 2 \int \deriv^2 z' \gamma_{z' \bar z'} G(z,\bar z; z', \bar z') \left( \int_{u_i}^{u_f} \deriv u T_{uu}(z', \bar z') + \Delta m \right). \label{eq: explicit supertranslation}
            \end{equation} Indeed, a tedious calculation \cite{strominger_gravitational_2016} shows that
            \begin{equation*}
                \Delta C(z, \bar z) = - \frac{2}{\pi} \int \deriv^2 z' \gamma_{z' \bar z'}  \frac{\abs{z - z'}^2}{(1+ z' \bar z')(1+ z \bar z)} \log \frac{\abs{z - z'}^2}{(1+ z' \bar z')(1+ z \bar z)} \left( \int_{u_i}^{u_f} \deriv u T_{uu}(z', \bar z') + \Delta m \right) 
            \end{equation*} is such that
            \begin{equation*}
                D_z^2 D_{\bar z}^2 \Delta C = 2 \left( \int_{u_i}^{u_f} \deriv u T_{uu}(z', \bar z') + \Delta m
                \right)
            \end{equation*} and $C_{zz} = - 2 D^2_z C(z, \bar z)$ then gives us our result \eqref{eq: explicit supertranslation}. This is an explicit expression for the supertranslation induced by waves passing through $\cal I^+$, showing that the displacement effect is indeed equivalent to BMS supertranslations through vacuum transitions. Let us now wrap up this discussion by looking at how memory effects are related to the soft theorems of section~\ref{sec: supertranslations soft theorems}.
            
            \subsubsection{Memory and soft theorems} Finding a connection between soft theorems and memory effects is relatively straightforward in the case where both are associated to BMS supertranslations. On the one hand, Weinberg's soft graviton theorem \eqref{eq: weinberg soft graviton theorem} can be rewritten as
            \begin{equation*}
            \begin{aligned}
                \mathcal{M}_{\mu \nu}(q, p_1', ..., p'_m, p_1, ...,p_n) \epsilon^{\mu \nu} &= \frac{\kappa}{2} \left[ \sum_{k=1}^m \frac{p'_{k \mu} p'_{k \nu}}{p'_k \cdot q} - \sum_{k=1}^n \frac{p_{k\mu}p_{k\nu}}{p_k \cdot q}\right] \epsilon^{\mu \nu} \mathcal{M}(p_1', ..., p_m', p_1, ..., p_n) \\
                &= \frac{\kappa}{2} \left[ \sum_{k=1}^m \frac{p'_{k \mu} p'_{k \nu}}{p'_k \cdot \omega k} - \sum_{k=1}^n \frac{p_{k\mu}p_{k\nu}}{p_k \cdot \omega k}\right]^{TT} \epsilon^{\mu \nu} \mathcal{M}(p_1', ..., p_m', p_1, ..., p_n)
            \end{aligned}
            \end{equation*} by making an explicit reference to the transverse polarization $\epsilon_{\mu \nu}$ and the graviton $4$-momentum by  $q = (\omega, \omega \Vec{k})$ ($\omega \to 0$). Here, the superscript $TT$ denotes the transverse traceless part of the brackets. To go from the first to second line, we used that the graviton always couples to transverse and traceless polarizations, which is singled out from contraction of the terms in square brackets with $\epsilon^{\mu \nu}$. Meanwhile, Braginski and Thorne \cite{braginsky_gravitational-wave_1987} showed in previous work that the shift in this $TT$ part of the asymptotic metric at $\cal I^+$ resulting from the collision of large massive objects is \cite{strominger_gravitational_2016}
            \begin{equation*}
                \Delta h_{\mu \nu}^{TT}(\omega, \Vec{k}) = \frac{1}{r_0} \sqrt{\frac{G}{2\pi}} \left( \sum_{j=1}^n \frac{p'_{j \mu} p'_{j \nu}}{\omega k \cdot p'_j} - \sum_{j=1}^m \frac{p_{j\mu} p_{j \nu}}{\omega k \cdot p_j} \right)^{TT}
            \end{equation*} for $n$ ($m$) incoming (outgoing) massive objects with momenta $p_{j\mu}$ ($p'_{j\mu}$) and where $\Vec{k}$ is the null vector pointing from the collision region to infinity. To gain insight, it is again helpful to consider the graviton field in Fourier space, assuming that the stationary approximation holds as at large $r$,
            \begin{equation*}
                h_{\mu \nu}^{TT} (\omega, \Vec{k}) = 4 \pi i \lim_{r \to \infty} r \int \deriv u e^{i \omega u} h_{\mu \nu}^{TT} (u, r \Vec{k}).
            \end{equation*} The field $h^{TT}_{\mu \nu}(u,r \Vec{k})$ is expected to decay at and approach different finite values as $u \to \pm \infty$ (with $r = r_0$ large). 
            Then it must be that 
            \begin{equation*}
                \Delta h_{\mu \nu}^{TT} (\omega, \Vec{k}) = \frac{1}{4 \pi r_0} \lim_{\omega \to 0} \left(- i \omega h_{\mu \nu}^{TT} (\omega, \Vec{k}) \right),
            \end{equation*} Finally, we find an expression for $\omega h_{\mu \nu}^{TT}$ noting that \cite{strominger_gravitational_2016}
            \begin{equation*}
                \begin{aligned}
                    \lim _{\omega \to 0} \omega h_{\mu \nu}^{T T}(\omega, k) \epsilon^{\mu \nu} & =\lim _{\omega \to 0} \frac{\omega \mathcal{A}_{m+n+1}\left(p_1, \ldots p_n ; p_1^{\prime}, \ldots p_m^{\prime},\left(\omega k, \epsilon_{\mu \nu}\right)\right)}{\mathcal{A}_{m+n}\left(p_1, \ldots p_n ; p_1^{\prime}, \ldots p_m^{\prime}\right)} \\ &= \sqrt{8 \pi G} \epsilon^{\mu \nu} \lim _{\omega \to 0} \omega S_{\mu \nu}(\omega k) =\sqrt{8 \pi G} \epsilon^{\mu \nu}\left(\sum_{j=1}^m \frac{p_{j \mu} p_{j \nu}}{k \cdot p_j}-\sum_{j=1}^n \frac{p_{j \mu}^{\prime} p_{j \nu}^{\prime}}{k \cdot p_j^{\prime}}\right)^{T T},
                \end{aligned}
            \end{equation*}  where $\mathcal{A}_{m+n}$ denotes a collision involving $n$ incoming objects with $m$ other massive objects. In short, acting with a Fourier transform on the momentum space formula allows to get to the expression we obtained from Weinberg's soft graviton theorem. This concludes our study of supertranslations and their implications for the infrared structure of gravity.

\newpage
\section{Conclusion and Outlook \label{sec: conclusion}}

    Let us now summarize the review. Starting from simple considerations of gravitational scattering processes between $\cal I^-$ and $\cal I^+$, we constructed the asymptotic symmetry group of asymptotically flat spacetimes, the (global) BMS group. This group turned out to be larger than the Poincaré group of isometries of Minkowski space, showing that general relativity does not reduce to its special counterpart at large distances from an isolated system.
    Recently, Compère et al.~\cite{compere_asymptotic_2023} extended these concepts to $i^\pm$ and $i^0$, showing how individual ingoing and outgoing massive bodies may be ascribed initial or final BMS charges and deriving associated global conservation laws. The BMS analysis has also been carried out in three spacetime dimensions, see \cite{oblak_classical_2017} for details. The framework was extended to higher dimensions as well as of late by Capone and collaborators \cite{capone_phase_2023}.
    
    \subsection{The infrared triangle for BMS supertranslations}

\begin{figure}[h!]
            \centering
            \includegraphics[width=0.6\linewidth]{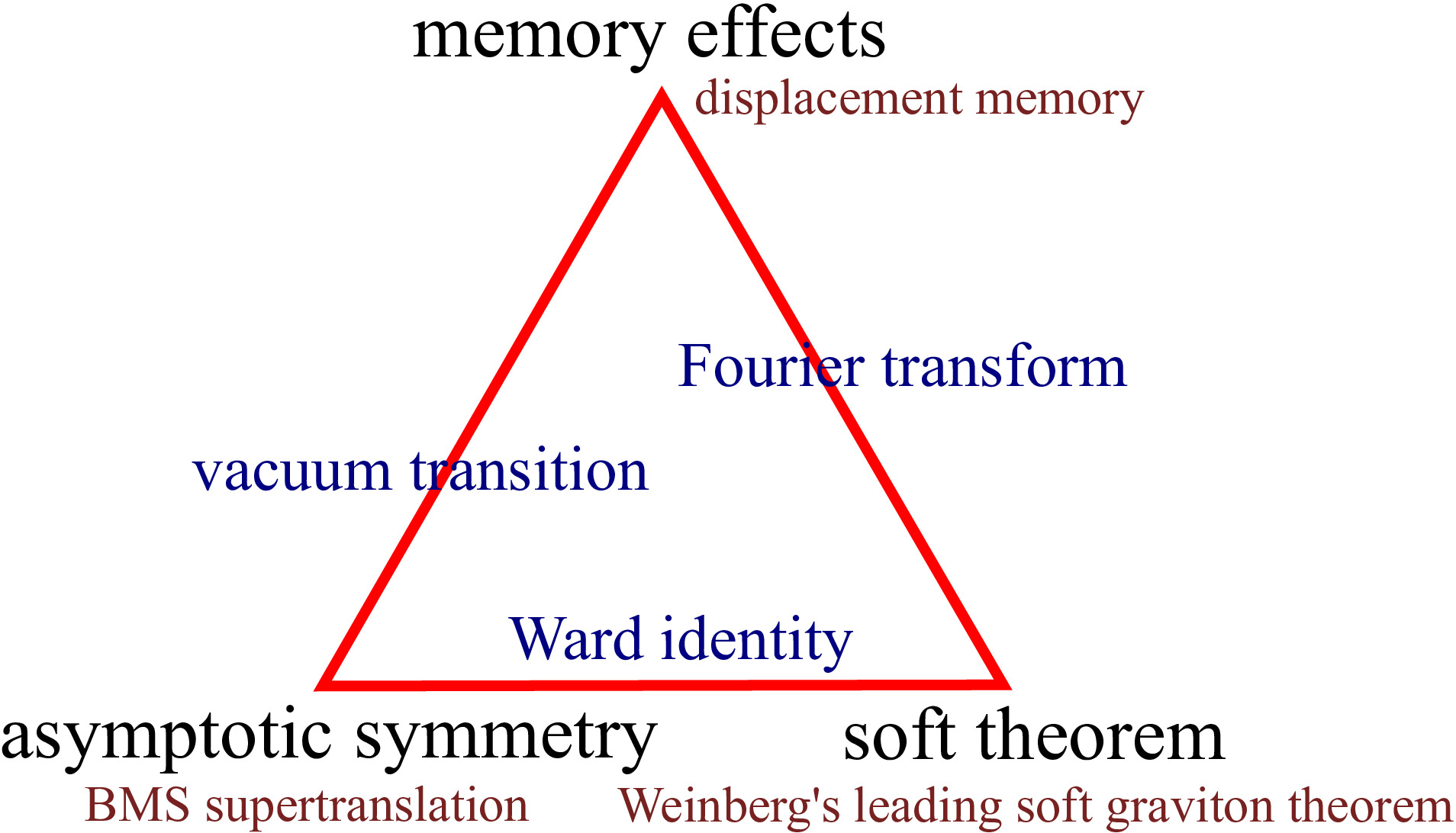}
            \caption{The infrared triangle of  gravity for BMS supertranslations \label{fig:IR triangle supertranslation}}
        \end{figure}

        Building upon our knowledge of the Killing vectors of BMS$_4$, we were able to derive a series of relationships between asymptotic symmetries, soft theorems and memory effects. Figure~\ref{fig:IR triangle supertranslation} summarizes what we learned from this in the case of supertranslations. 
         After having obtained the form of the generators of supertranslations, we were able to put the covariant phase space formalism to use and compute the conserved charges associated with this symmetry. Considering the diagonal subgroup BMS$^0$ of the BMS group then allowed us to show the supertranslation invariance of the $\cal S$-matrix, and subsequently derive a quantum Ward identity for the process. 

        \smallskip \noindent
        By relating the free graviton field to the coefficients $C_{AB}$ of the Bondi metric, we were then able to interpret this Ward identity as a soft theorem relating two matrix elements with or without the insertion of a soft graviton, and recover Weinberg's leading soft graviton theorem.

        \smallskip \noindent
        Meanwhile, the constraint equation on the Bondi mass aspect allowed to view displacement memory effects as the result of a vacuum transition induced by supertranslations. The final edge of the triangle was found by noting that Braginski and Thorne's result for the fluctuations in the graviton field resulting from the collision of large massive objects can be recast as a soft theorem for the scattering of gravitons. Memory effects are key in understanding the propagation of gravitational waves. Their newly found connections to asymptotic symmetries and soft theorems provides a direct mean of studying asymptotic symmetries through gravitational waves \cite{lasky_detecting_2016}, and vice-versa. 

    \subsection{Outlook}
        \subsubsection{Other copies of the infrared triangle \label{sec:conclu superrot}}

        As mentioned in the introduction, the triangular equivalence between asymptotic symmetries, soft theorems and memory effects is not unique to BMS supertranslations. In this section, we first explain how an analogous picture can be obtained in the case of BMS superrotations and then review a few other examples from abelian and non-abelian gauge theories.

        \subsubsection*{Infrared triangle for BMS superrotations}
            Let us give a brief qualitative overview of how the relations introduced for supertranslations also arise when looking at gravity from the viewpoint of BMS superrotation symmetry. We start with superrotation charges, then motivate superrotation invariance of the $\cal S$-matrix and formulate an associated Ward identity. Finally, we also discuss the connection of the latter to soft theorems.
    
            \paragraph{Superrotation charges} It is also possible to construct the charges associated with superrotation symmetry by means of CPS methods. From \eqref{eq: effect BMS LT Czz}, we can deduce the variation of the boundary fields $C$, $N$ and bulk contribution $\hat C_{zz}$ under $\xi_R^+$ \eqref{eq: asymp superrotation scri+} \cite{alessio_note_2019}
            \begin{align*}
                D^2_z \mathcal{L}_\xi C &= \mathcal{L}_R D^2_z C - \frac{1}{2} D \cdot R D_z^2 C + \lim_{u \to \infty} u D_z^2 R^z, \\
                D^2_z \mathcal{L}_\xi N &= \mathcal{L}_R D^2_z N - \frac{1}{2} D \cdot R D_z^2 N - 2 \lim_{u \to \infty} u D_z^3 R^z, \\
                \mathcal{L}_\xi \hat C_{zz} &= \frac{u}{2} D \cdot R N_{zz} + \mathcal{L}_R \hat C_{zz} - \frac{1}{2} D \cdot R \hat C_{zz} - u D_z^3 R^z.
            \end{align*} One immediately notices that the first two are divergent. This is due to the fact that superrotations map the fields outside of the phase space we defined earlier for supertranslations. Conserved charges can nonetheless be constructed from $\cancel{\delta}Q_R = \Omega_{\mathcal{I}^+}[\delta \phi, \delta_R \phi]$ as before for supertranslations. Doing so, one naturally finds that these split into a bulk $\cancel{\delta} \hat Q_R$ and boundary contributions $\cancel{\delta} \tilde Q_R$, and that each of these further splits into integrable and non-integrable parts, $\cancel{\delta} \hat Q_R = \delta \hat Q_R + \hat \Theta_R$, etc. A detailed construction of the charges can be found in \cite{alessio_note_2019}. There are again soft and hard contributions.
            
            \paragraph{Ward identity} As for supertranslations, we seek a subset of the superrotations generators that leave the $\cal S$-matrix invariant. We already introduced the diagonal subgroup BMS$^0$ whose generators commute with $\cal S$, and discussed the form of the supertranslations $\xi_T \in \text{BMS}^0$. Let's now see what $\xi_R \in \text{BMS}^0$ looks like.\footnote{Note that we implicitly extended the definition of BMS$^0$ here from a subgroup of $\text{BMS}^+ \times \text{BMS}^-$ to a subgroup of $\text{eBMS}^+ \times \text{eBMS}^-$ since we do not restrict ourselves to only considering the Lorentz generators as "superrotations".} A careful analysis of the effect of a superrotation on $m|_{\mathcal{I}^+_-}$ and $m^-|_{\mathcal{I}^-_+}$ shows that \cite{strominger_bms_2014}
            \begin{equation*}
                \mathcal{L}_\xi m |_{\mathcal{I}^+_-} = \left( R^z \partial_z + \frac{3D_z R^z}{2} \right) m |_{\mathcal{I}^+_-}, \quad \text{and} \quad \mathcal{L}_\xi m^- |_{\mathcal{I}^-_+} = \left(R^{-z} \partial_z + \frac{3 D_z R^{-z}}{2} \right) m^- |_{\mathcal{I}^-_+}.
            \end{equation*} Adding to the continuity condition \eqref{eq: continuity condition} the matching $m|_{\mathcal{I}^+_-} = m^-|_{\mathcal{I}^-_+}$ allows to single out the generators $\xi_R$ of BMS$^0$ as those superrotations satisfying $R^z = R^{-z}$. It remains to impose the matching of the angular momentum aspect to equate the charges on both sides of $i^0$ as we did for supertranslations, even though this last condition has only been conjectured yet \cite{capone_charge_2023}. Assuming that it holds, one is then in position to show that the $\cal S$-matrix is invariant, with Ward identity \cite{kapec_semiclassical_2014}
            \begin{align*}
                \langle out | Q^+(R) \mathcal{S} - \mathcal{S} Q^-(R) | in \rangle &= 0 \quad \Rightarrow \quad : Q_S(R) \mathcal{S}: \equiv Q_S^+ \mathcal{S} - \mathcal{S}Q_S^-, \quad \text{or equivalently} \\ \langle z_{n+1}, z_{n+2}, ...| : Q_S(R) \mathcal{S} :| z_1, z_2, ... \rangle &= - i \sum_{k=1}^{n+m} \left(R^z(z_k) \partial_{z_k} - \frac{E_k}{2} D_z R^z(z_k) \partial_{E_k}  \right) \langle z_{n+1}, ... |\mathcal{S} |z_1, ... \rangle,
            \end{align*} where $Q_S$ denotes the soft graviton that can be readily obtained from $Q_R$. This is to be compared with what we derived in \eqref{eq:supertranslation ward identity} for supertranslations. Again, this relates a particular $\cal S$-matrix element to the same element with a soft graviton inserted by acting with the charges. 
            
            \paragraph{Subleading soft graviton theorem} To show that this Ward identity bridges the gap between asymptotic symmetry and soft theorems, one needs to recast the above expression in momentum space. This yields \cite{kapec_semiclassical_2014}
            \begin{equation*}
                \lim_{\omega\to 0^+} (1+\omega \partial_\omega) \langle z_{n+1}, ...| a_-(q) \mathcal{S}|z_1, ... \rangle = S^{(1)-} \langle z_{n+1}, ...| \mathcal{S}| z_1, ... \rangle, \quad S^{(1)-}  = -i \sum_k \frac{p_{k\mu} \varepsilon^{- \mu \nu} q^\lambda J_{k \lambda \nu}}{p_k \cdot q}
            \end{equation*} with the universal subleading soft factor $S^{(1)-}$ and $J_{k \lambda \nu} \equiv L_{k \lambda \nu} + S_{k \lambda \nu}$ the total ingoing angular momentum of the $k$-th particle (orbital + spin). Cachazo and Strominger \cite{cachazo_evidence_2014} verified this formula for all tree-level graviton scattering. When the dust settles, one finds that the Ward identity derived from superrotation invariance of the $\cal S$-matrix  indeed relates asymptotic symmetries to Weinberg's soft graviton theorem (subleading this time). 

\begin{figure}
                \centering
            \includegraphics[width=0.6\linewidth]{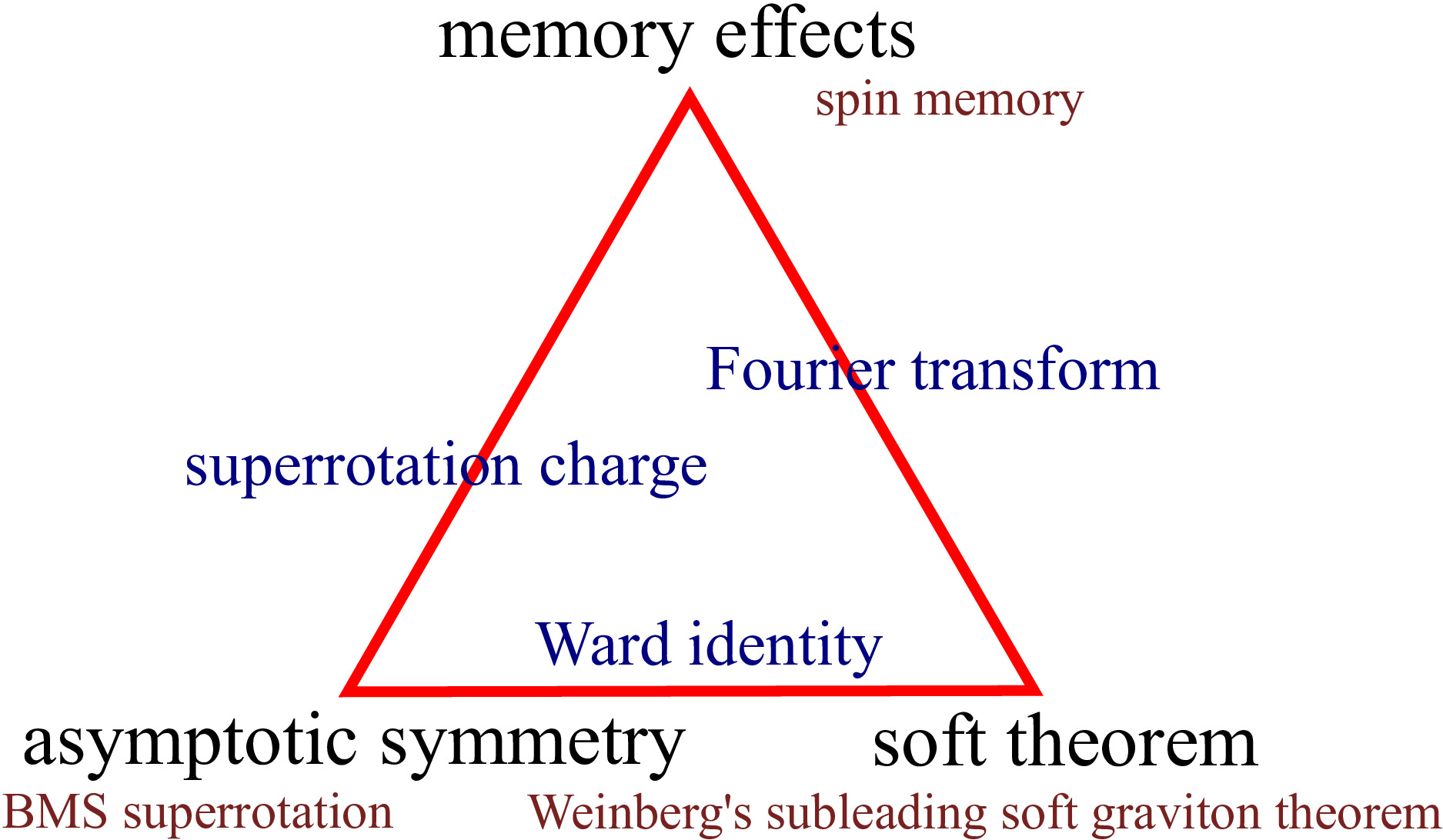}
                \caption{The infrared triangle of  gravity for BMS superrotations \label{fig:IR triangle superrotation}}
            \end{figure}

            \paragraph{Spin memory effect} We argued that supertranslations could be associated to a displacement memory effect through a vacuum transition. The same story repeats in the case of superrotations. Only this time the phenomenon is called \textit{spin memory effect} \cite{pasterski_new_2016}. While displacement memory is sourced by moments of the energy flux through $\cal I$, this one stems from moments of the angular momentum flux. Integrating the conservation law of the angular momentum density \eqref{eq:constraint angular momentum aspect}, one finds that these arise from: a change of angular momentum aspect, angular momentum flux from null matter and angular momentum flux from gravitational waves \cite{compere_advanced_2019}. This indeed corresponds to the action of supertranslation charges on our data, which thus gives us a connection between memory effects and asymptotic symmetries \cite{himwich_note_2019}. As for supertranslations, the spin memory effect is also related to Weinberg's (subleading) soft theorem by a Fourier transform in space. This completes the IR triangle of BMS superrotations, depicted in Figure~\ref{fig:IR triangle superrotation}.  

            \smallskip \noindent We now review analogous results that have been obtained for abelian and non-abelian gauge theories.

        \subsubsection*{Infrared structure of other gauge theories} As mentioned in the introduction, this triangular equivalence relation is in fact common to the infrared dynamics of all physical theories involving massless particles \cite{strominger_lectures_2017}. Analogous triangles have thus been constructed for abelian and non-abelian gauge theories such as massless scalar QED and Yang-Mills, leading to crucial insights into these theories.

        \paragraph{QED} The explicit derivation of this triangle can be seen for instance in Strominger's lecture notes \cite{strominger_lectures_2017} or in the work of Frieswijk  \cite{frieswijk_infrared_2018}. In this story, the Bondi gauge is replaced by a set of gauge-fixing conditions for the gauge field $A_\mu$
        \begin{equation*}
            A_r = 0, \quad A_u|_{\mathcal{I}_+} = 0
        \end{equation*} together with falloff conditions which ensure finite energy configurations,
        \begin{equation*}
            A_u \sim \mathcal{O}(r^{-1}), \quad A_r= 0, \quad A_z \sim \mathcal{O}(1).
        \end{equation*} The equation of motion for the field in coordinates are $\nabla^\mu F_{\mu \nu} = e^2 j_{\nu}$. They are invariant under gauge transformations of the form
        \begin{equation*}
            A_\mu \to A_\mu' = A_\mu + \partial_\mu \lambda.
        \end{equation*} With this at hand, one can identify a set of \textit{large gauge symmetries} of the gauge field which preserve the falloffs mentioned above. These are the equivalent of the asymptotic symmetries we started with in the gravitational case. From a careful matching of the field at $\mathcal{I}_-^+$ and $\mathcal{I}_+^-$
        \begin{equation*}
            A_z(z, \bar z) |_{\mathcal{I}^+_-} = A_z(z, \bar z) |_{\mathcal{I}^-_+} \quad \text{(antipodal matching condition)}
        \end{equation*}one can construct an infinity of conserved charges
        \begin{equation*}
            Q^+_{\epsilon} = \frac{1}{e^2}\int_{\mathcal{I}^+_-} \epsilon \star F, \quad Q^-_{\epsilon} = \frac{1}{e^2}\int_{\mathcal{I}^-_+} \epsilon \star F, \quad \text{such that} \quad Q_\epsilon^+ = Q_\epsilon^-,
        \end{equation*} where $\epsilon = \epsilon(z, \bar z) \equiv \lambda$ denotes an infinitesimal gauge transformation preserving the boundary and falloff conditions. These charges again split in soft and hard contributions. The soft part is the one of interest, as it creates and annihilates incoming and outgoing soft photons upon quantization of the theory. In 2014, He et al.~\cite{he_new_2014} showed that Weinberg’s soft photon theorem \cite{weinberg_infrared_1965} is equivalent to the Ward identity that can be derived from these conserved charges. The third corner of the triangle is given by the memory effect \cite{bieri_electromagnetic_2013,pasterski_asymptotic_2017,susskind_electromagnetic_2019}, which can be experimentally measured as a change in relative phases between two test particles.

        \paragraph{Yang-Mills} A similar story repeats for Yang-Mills theory \cite{fan_soft_2019, pate_color_2017, jokela_memory_2019}, where the infrared structure is put to use to try to find an underlying symmetry that could explain the cancellation of certain terms in the theory, or provide a framework for deriving scattering amplitudes in $\mathcal{N} = 4$ Yang-Mills. More recently,  He and Mitra \cite{he_asymptotic_2023} used the covariant phase space formalism to construct the phase space for non-Abelian gauge theories in $(d+2)$-dimensional Minkowski spacetime for any $d \geq 2$. Upon quantization, they recovered the algebra of the vacuum sector of the Hilbert space and derived a Ward identity that implies the leading soft gluon theorem in $(d+2)$-dimensional spacetime. 
        
        \smallskip \noindent
        Through these last two examples, we learn that soft theorems, memory effects and asymptotic symmetries are three equivalent ways of looking at the behavior of massless gauge theories at large distances, thereby providing a powerful tool for the exploration of new symmetries, memory effects or soft theorems. Recent works show that this is still an active area of research where many questions remain to be answered.
        In the next subsections, we conclude our discussion of the BMS symmetries by motivating the role of the infrared structure of gravity in building flat space holography and tackling the black hole information paradox.

        \subsubsection{Towards flat space holography} 
            The formulation of a complete and coherent theory of quantum gravity has been one of the most active fields of study in theoretical physics in the last decades. In such a quest, the \textit{holographic principle} plays a important role. It posits that all the information contained withing a region of spacetime can be described by the boundary of that region. The prime example of such a duality is usually given by the AdS/CFT correspondence \cite{brown_central_1986, maldacena_large_1999}, which relates a theory of gravity in Anti-de-Sitter (AdS) spacetime of any arbitrary number of dimensions to a conformal field theory (CFT) without gravity living in one dimension less. Such a correspondence emerged from matching the isometries of AdS with the conformal group.
            
            \smallskip
            In this context, it is natural to ask whether BMS symmetries can provide a similar framework for gauge/gravity duality in asymptotically flat spacetimes, especially given the enhancement of the BMS algebra to Virasoro or Diff($\mathbb{S}^2$). This led to the development of \textit{Carollian} and \textit{celestial holography}. In 2017, Kapec et al.~\cite{kapec_2d_2017} used the subleading soft-graviton theorem to construct an operator $T_{zz}$ whose insertion in the four-dimensional tree-level quantum gravity $\cal S$-matrix obeys the Virasoro-Ward identities of the energy momentum tensor of a two-dimensional conformal field theory. Celestial holography builds upon these evidences to posit a duality of the form
            \begin{equation*}
                \text{gravitational scattering in 4D AFS} \quad \leftrightarrow \quad \text{2D CFT on the celestial sphere}.
            \end{equation*} In practice, the gravitational $S$-matrix and its associated soft graviton modes are expressed in terms of so-called \textit{celestial amplitudes} by means of a Mellin transform in energy \cite{pasterski_celestial_2021, raclariu_lectures_2021}. Celestial amplitudes are themselves expressed in boost eigenstates, and while there has been significant progress in understanding the basic principles and applications of celestial holography, the relationship of the former to traditional scattering amplitudes in a basis of plane waves is highly non-trivial.

            \smallskip \noindent
            The Carrollian approach to flat space holography, on the other hand, proposes that the role of the dual theory is played by a conformal \textit{Carrollian CFT} that lives on the codimension-one boundary of spacetime (null infinity $\cal I$). This approach builds upon the isomorphy of the BMS group is isomorphic to the conformal Carroll group \cite{duval_conformal_2014} and has proven to be very successful in the context of three-dimensional gravity, while in 4d it is shown to describe the dynamics of non-radiative spacetime \cite{donnay_bridging_2023}. Little is however known so far about Carollian CFTs, and other technical difficulties remain to be solved. Overall, though promising, both approaches to flat space holography are still recent and active areas of research where many problems are yet to be addressed.

        \subsubsection{Black hole information and soft hairs} 
            The \textit{no-hair theorem} states that any stationary black hole in an AFS can be described by only a small number of parameters, such as its mass, charge and total angular momentum \cite{strominger_lectures_2017}. The discovery in the 1970s by Bekenstein \cite{bekenstein_black_1973} and Hawking \cite{hawking_particle_1975} respectively of black hole thermodynamics and the \textit{Hawking radiation} led to the \textit{black hole information paradox}. Indeed, the loss of information in black hole evaporation conflicts with the principle of unitarity in quantum mechanics, which requires that the evolution of a system be reversible and that the information contained in the initial state be preserved throughout the evolution. 
            
            \smallskip
            The study of soft theorems in the context of gravitational scattering shed a new light on this problem \cite{strominger_lectures_2017}. In 2016, Hawking, Perry and Strominger suggested that BMS transformations could be used to relate the radiation of a black hole to a soft graviton which they named \textit{soft hair} \cite{hawking_soft_2016, hawking_superrotation_2017}. These soft hairs supposedly allow information to be stored in the gravitational field outside the event horizon and can be used to carry information out of the black hole. This circumvents the no-hair theorem but does not solve the black hole information paradox, as one would still need to find methods to extract information from the soft hair for it to be the case. This is still an active area of research and remains a subject of debate within the scientific community.

 \vfill 

    \addcontentsline{toc}{section}{Acknowledgements}
    \section*{Acknowledgements}
        I would like to express my gratitude to Dr.~Prahar Mitra for accepting to set the essay from which this article originates, and then for his time, valuable feedback and advices not only throughout the project but also later when turning my work into a publishable document. 
        Many thanks also to Dr.~Andrea Puhm, who sparked my interest in flat space holography with her talk \textit{Holography and the Celestial Sphere} (26th October 2022), and Simon Heuveline who later helped me decide on an essay topic and incited me to reach out to Dr.~Mitra  in the first place. 
        
        \smallskip \noindent
        Cian Evans-Cowie, Ben Graham, Adam Keyes, Ian Le Meur, Federico Lucas, Cole Meldorf, Venkatesh Srinivasan and everyone in Warkworth 6 and Peterhouse, thank you all for sticking around and making my time in Cambridge so special. Kudos also to everyone at the Cambridge University Hares \& Hounds for the great runs and even greater memories together. 

        \smallskip \noindent
        I am grateful to my parents for their continuous support of my journey as a scientist.
        
        \smallskip \noindent
        Finally, I would also like to acknowledge the generous financial support of the Fondation Suisse d'Etudes (Swiss Study Foundation) and the Colbianco Stiftung of my year in Cambridge.

\newpage
\addcontentsline{toc}{section}{References}
\bibliography{literature.bib}
\bibliographystyle{JHEP}

\appendix

\newpage
\section{Appendix: detailed calculations}

    \subsection{\texorpdfstring{BMS$_4$}{BMS4} equations of motion and constraint equations for \texorpdfstring{$m$}{m} and \texorpdfstring{$N^A$}{NA} \label{app: coefficients metric expansion}}
    We start from \eqref{eq: Bondi metric},
    \begin{equation*}
            g_{\mu \nu} = 
            \begin{pmatrix} 
            -U + \frac{1}{4} g_{AB} U^A U^B & - e^{2 \beta} & \frac{1}{2}g_{2A}U^A & \frac{1}{2}g_{3A} U^A \\
            -e^{2 \beta} & 0 & 0 & 0 \\
            \frac{1}{2}g_{2A}U^A & 0 & g_{22} & g_{23} \\
            \frac{1}{2}g_{3A} U^A & 0 & g_{23} & g_{33}
            \end{pmatrix}_{\mu \nu}
        \end{equation*} and compute the Einstein tensor with the help of \verb!Mathematica!. We have at our diposal the large $r$ expansions \eqref{eq: falloffs metric} for the metric coefficients,
        \begin{equation*}
            \begin{aligned}
                U(u,r,x^A) &= 1 - \frac{2m(u,x^A)}{r} + \frac{U_2(u,x^A)}{r^2} + \mathcal{O}(r^{-3}), \\
                \beta(u,r,x^A) &= \frac{\beta_1(u,x^A)}{r} + \frac{\beta_2(u,x^A)}{r^2} + \frac{\beta_3(u,x^A)}{r^3} + \mathcal{O}(r^{-4}), \\
                U^A(u,r,x^B) &= \frac{U^{A}_{2}(u,x^B)}{r^2} + \frac{U^A_3(u,x^B)}{r^3} + \frac{U^A_4(u,x^B)}{r^4} +\mathcal{O}(r^{-5}), \\
                g_{AB}(u,r,x^C) &= r^2 \gamma_{AB}(x^C) + r C_{AB}(u,x^C) + D_{AB}(u,x^C) + \mathcal{O}(r^{-1}),
            \end{aligned}
        \end{equation*} the fact that $C_{AB}$ is traceless,
        \begin{equation*}
            C_{AB} = \begin{pmatrix}
                C_{zz} & 0 \\ 0 & C_{\bar z \bar z}
            \end{pmatrix},
        \end{equation*} as well as the large $r$ behavior of the stress-energy tensor,
        \begin{gather*}
            T_{uu} = \frac{1}{r^2} \hat T_{uu}(u, x^A) + \mathcal{O}(r^{-3}), \qquad
            T_{rr} = \frac{1}{r^4} \hat T_{rr}(u, x^A) + \frac{1}{r^5} \tilde T_{rr}(u, x^A) + \mathcal{O}(r^{-6}), \\
            T_{uA} = \frac{1}{r^2} \hat T_{uA}(u, x^A) + \mathcal{O}(r^{-3}), \qquad
            T_{rA} = \frac{1}{r^3} \hat T_{rA}(u, x^A) + \mathcal{O}(r^{-4}), \\
            T_{AB} = \frac{1}{r}\hat T(u, x^A) \gamma_{AB} + \mathcal{O}(r^{-2}) \qquad T_{ur} = \mathcal{O}(r^{-4}).
        \end{gather*} We start by imposing the gauge condition \eqref{eq:determinant gauge condition} on the determinant of $g_{AB}$. We get
        \begin{gather*}
            \text{det}(g_{AB}) = - \frac{4 r^4}{(1 + z \bar z)^4} + r^2 \left( - \frac{4 D_{z \bar z}}{(1+z \bar z)^2} + C_{\bar z \bar z} C_{zz} \right) + \mathcal{O}(r),
        \end{gather*} from which we deduce
        \begin{equation*}
             - \frac{4 D_{z \bar z}}{(1+z \bar z)^2} + C_{\bar z \bar z} C_{zz} \overset{!}{=}0 \quad \Leftrightarrow \quad D_{z \bar z} = \frac{1}{2}\gamma^{z \bar z} C_{\bar z \bar z} C_{zz}.
        \end{equation*}This is helpful in simplifying the remaining expressions. We now enforce the Einstein equation
        \begin{equation*}
            G_{\mu \nu} \equiv R_{\mu \nu} - \frac{1}{2}R g_{\mu \nu} = 8 \pi G T_{\mu \nu}.
        \end{equation*} Looking at $G_{rr}$ and $T_{rr}$, we get
        \begin{equation*}
            G_{rr} = - \frac{4 \beta_1}{r^3} + \mathcal{O}(r^{-4}) \overset{!}{=} 0 \quad \Rightarrow \quad \beta_1 \equiv 0.
        \end{equation*} The $\mathcal{O}(r^{-4})$ terms must also vanish when $\hat T = 0$, yielding an expression for $\beta_2$
        \begin{equation*}
            \frac{-\frac{1}{8} (z \bar z +1)^4 C_{\bar z \bar z} C_{zz}-8 \beta_2}{r^4} = 0 \quad \Rightarrow \quad \beta_2  = - \frac{1}{16}\gamma^{z \bar z} \gamma^{z \bar z} C_{\bar z \bar z} C_{zz} = - \frac{1}{16} C_{zz}C^{zz} = - \frac{1}{32} C_{AB} C^{AB}.
        \end{equation*} Now, looking at $G_{rA}$ and cancelling the $\mathcal{O}(r^{-2})$ terms we get
        \begin{equation*}
           \frac{1}{2} U_2^{\bar z} \gamma_{z \bar z} - \gamma^{z \bar z} \frac{\partial_{\bar z}C_{zz}}{2} = 0 \Rightarrow U_2^z = D_z C^{zz} \text{ and similarly } U_2^{\bar z} = D_{\bar z} C^{\bar z \bar z} \text{ so } U_2^A = D_B C^{AB}.
        \end{equation*}
        
        \paragraph{Constraint equations} Looking at $G_{uu}$ and $T_{uu}$ we get at order $\mathcal{O}(r^{-2})$ that
        \begin{gather*}
            \frac{-(1+z \bar z)^4 N_{\bar z \bar z} N_{z z} - 16 \partial_u m }{8} + \frac{1}{8(1+z \bar z)}(-8 \bar z \partial_u U^z_2 - 8 z \partial_u U_2^{\bar z} + 4 \partial_u \partial_{\bar z} U_2^{\bar z} +  4 \partial_u \partial_{z} U_2^{z}) \overset{!}{=} 8 \pi G \hat T_{uu} \\
        \Leftrightarrow - \frac{1}{2}N_{zz}N^{zz} - 2 \partial_u m  + \frac{1}{2}\partial_u\left[ \partial_z U_2^z - \frac{2\bar z}{1 + z \bar z}U_2^z + \partial_{\bar z} U_2^{\bar z} - \frac{2 z}{1+z \bar z} U^{\bar z}_2 \right]  = 8 \pi G \hat T_{uu}.
    \end{gather*} Hence
    \begin{align*}
        \partial_u m &= - 4 \pi G \hat T_{uu} - \frac{1}{4}N_{zz}N^{zz} + \frac{1}{4} \partial_u [D^z U_{2z} + D^{\bar z}U_{2\bar z}] \\ &= - 4 \pi G \hat T_{uu} - \frac{1}{4}N_{zz}N^{zz} + \frac{1}{4} [D^{z2} N_{zz} + D^{\bar z2}N_{\bar z \bar z}] = - T_{uu} + \frac{1}{4} [D^{z2} N_{zz} + D^{\bar z2}N_{\bar z \bar z}],
    \end{align*}with 
    \begin{equation*}
        T_{uu} = 4 \pi G \hat T_{uu} + \frac{1}{4}N_{zz}N^{zz}
    \end{equation*}
    the total stress-energy tensor. Without matter we have $\hat T= 0$ and
    \begin{equation*}
        \partial_u m = \frac{1}{4}[D^2_z N^{zz} + D_{\bar z}^2 N^{\bar z \bar z}] - \frac{1}{4} N_{zz} N^{zz}
    \end{equation*} as in \cite{alessio_note_2019}. Similarly, looking at $G_{uA}$ yields a constraint equation for $N^A$ of the form \cite{strominger_lectures_2017, alessio_asymptotic_2019}
    \begin{equation*}
        \partial_u N_z  = \frac{1}{4} D_z [D_z^2 C^{zz} - D_{\bar z}^2 C^{\bar z \bar z}] - u D_z \partial_u m + \frac{1}{4} D_z (C_{zz} N^{zz}) + \frac{1}{2} C_{zz} D_z N^{zz} - 8 \pi G \hat T_{uz}.
    \end{equation*}

    \subsection{Determination of the \texorpdfstring{BMS$^+$}{BMS+} generators \label{app: determination BMS generators}}
        We want to find the general form of the Killing vector field $\xi$ such that \eqref{eq:gauge preservation condition} and \eqref{eq: falloff preservation condition} hold. Recall that the change of the metric under a diffeomorphism is \eqref{eq: Lie derivative metric}
        \begin{equation}
             \mathcal{L}_\xi g_{\mu \nu} \equiv \delta g_{\mu \nu} = \xi^\rho \partial_\rho g_{\mu \nu} + g_{\mu \rho} \partial_\nu \xi^\rho + g_{\nu \rho} \partial_\mu \xi^\rho, \label{eq: Killing contravariant}
        \end{equation} which writes in a general basis
        \begin{equation}
            (\mathcal{L}_\xi g)_{ab} \equiv \delta g_{ab} = 0 \Leftrightarrow \nabla_a \xi_b + \nabla_b \xi_a = 0. \label{eq: Killing covariant}
        \end{equation} Both are equivalent statements of the \textit{Killing equation}. Working with the first one allows to determine the contravariant components of $\xi$, while the second one refers to covariant terms. Of course, both are related by $\xi^a = g^{ab}\xi_b$. To solve for $\xi$, we supplement the Killing equation with the conditions given by the Bondi gauge \eqref{eq:gauge preservation condition}
        \begin{equation}
            \delta g_{rr} = 0, \quad \delta g_{rA} = 0, \quad g^{AB} \delta g_{AB} = 0 \label{eq: change metric gauge}
        \end{equation}and AFS falloffs \eqref{eq: falloff preservation condition}
        \begin{equation} 
            \delta g_{uu} = \mathcal{O}(r^{-1}), \quad \delta g_{uA} = \mathcal{O}(1), \quad \delta g_{ur} = \mathcal{O}(r^{-2}), \quad \delta g_{AB} = \mathcal{O}(r). \label{eq: change metric falloffs}
        \end{equation} In the following, we work with \eqref{eq: Killing contravariant}. The second approach can be seen e.g.~in \cite{alessio_asymptotic_2019} but is more tedious as it involves an explicit manipulation of the Christoffel symbols for the metric \eqref{eq: Bondi metric}.
        
        \paragraph{Applying the gauge constraints} We follow the steps of Strominger \cite{strominger_lectures_2017} and Zheng Liang \cite{zheng_liang_bms_2017}. Starting from $\delta g_{rr} = 0$ in \eqref{eq: change metric gauge} and using the Bondi gauge \eqref{eq: bondi gauge} and \eqref{eq: Killing contravariant} we have
        \begin{equation*}
            \xi^\rho \partial_\rho \cancel{g_{rr}} + 2 g_{r \mu} \partial_r \xi^\rho = 2(g_{ur} \partial_r \xi^u + \cancel{g_{rr}} \partial_r \xi^r + \cancel{g_{rA}} \partial_r \xi^A) = 0.
        \end{equation*} Our expression for $g_{ur}$ in \eqref{eq: Bondi metric} gives $- 2 e^{2 \beta} \partial_r \xi^u = 0$, which is solved by
        \begin{equation}
            \xi^u \equiv f(u, x^A), \quad \text{for $f$ suitably differentiable.} \label{eq: xi u up}
        \end{equation} Next, $\delta g_{rA} = 0$ in \eqref{eq: change metric gauge} gives with \eqref{eq: Killing contravariant}
        \begin{multline*} 
            \xi^\rho \partial_\rho \cancel{g_{rA}} + g_{\mu A} \partial_r \xi^\mu + g_{r\mu} \partial_A \xi^\mu = g_{uA} \cancel{\partial_r \xi^u} + \cancel{g_{rA}} \partial_r \xi^r \\ + g_{AB} \partial_r \xi^B + g_{ur} \partial_A \xi^u + \cancel{g_{rr}} \partial_A \xi^r + \cancel{g_{Br}} \partial_A \xi^B
            = g_{AB} \partial_r \xi^B - e^{2\beta} \partial_A \xi^u.
        \end{multline*} Multiplying on both sides by $g^{AC}$, we arrive at 
        \begin{equation}
            \partial_r \xi^C = g^{AC} (\partial_A f) e^{2 \beta}
            \Rightarrow \xi^B = R^B(u, x^A) + \partial_A f \int_r^\infty \deriv r' g^{AB} e^{2 \beta}, \label{eq: xi A up}
        \end{equation} for the integration function $R^B(u, x^A)$ a suitably differentiable function of its coordinates. Finally, let's look at $g^{AB} \delta g_{AB} = 0$ in \eqref{eq: change metric gauge}. We take $g_{AB} = r^2 \gamma_{AB} + \mathcal{O}(r)$ so $\partial_u g_{AB} \approx 0$, and
        \begin{equation*} \begin{aligned}
            \delta g_{AB} &= \xi^u \cancel{\partial_u g_{AB}} + g_{uB} \partial_A \xi^u + \xi^r \partial_r g_{AB} + \xi^C \partial_C g_{AB} + g_{CB} \partial_A \xi^C + g_{Au} \partial_B \xi^u + g_{AC}  \partial_B \xi^C \\
            &= \frac{g_{BC}}{2}U^C  \partial_A \xi^u + \xi^r \partial_r (r^2 \gamma_{AB} + ...) + \xi^C \partial_C g_{AB} + g_{CB} \partial_A \xi^C + \frac{g_{AC}}{2}  U^C \partial_B \xi^u + g_{AC} \partial_B \xi^C  \\
            &\approx \frac{g_{BC}}{2}U^C  \partial_A \xi^u + \xi^r 2r \gamma_{AB} + \xi^C \partial_C g_{AB} + g_{CB} \partial_A \xi^C + g_{AC} \partial_B \xi^C + \frac{g_{AC}}{2}  U^C \partial_B \xi^u 
        \end{aligned} \end{equation*} having used $g_{uA}  = \frac{1}{2}g_{AB}U^B$ from \eqref{eq: Bondi metric}. Now, consider the covariant derivative $D_A$ with respect to $g$. On the one hand,
        \begin{equation}\begin{aligned}
            D_C(g_{AB}) = 0 = \partial_C g_{AB} - \Gamma^D_{AC}g_{DB} - \Gamma^D_{BC}g_{AD} \Rightarrow \partial_C g_{AB} = \Gamma^D_{AC}g_{DB} + \Gamma^D_{BC}g_{AD}. \label{eq: covariant derivative}
        \end{aligned}\end{equation} On the other hand,
        \begin{equation} \begin{aligned}
            D_A \xi_B + D_B \xi_A &= D_A (g_{BC} \xi^C) + D_B (g_{AC} \xi^C) = g_{BC} \partial_A \xi^C + g_{AC} \partial_B \xi^C + (\partial_C g_{AB}) \xi^C, \label{eq: CKV 1}
        \end{aligned}\end{equation} where we relabelled dummy indices and used \eqref{eq: covariant derivative}. We thus rewrite $\delta g_{AB}$ as 
        \begin{equation}
            \delta g_{AB} = \xi^r 2r \gamma_{AB} + D_B \xi_A + D_A \xi_B + \frac{g_{BC}}{2}U^C  \partial_A \xi^u + \frac{g_{AC}}{2}  U^C \partial_B \xi^u \label{eq: change gAB covar}
        \end{equation} and evaluate $g^{AB} \delta g_{AB}$ using the fact that $g^{AB}g_{AB} = 2$, \eqref{eq: xi u up} and rewriting $\gamma_{AB} \approx g_{AB}/r^2$,
        \begin{equation} \begin{aligned}
            g^{AB} \delta g_{AB} &\approx g^{AB}\left(\xi^r 2r \frac{g_{AB}}{r^2} + D_B \xi_A + D_A \xi_B + \frac{g_{BC}}{2}U^C  \partial_A \xi^u + \frac{g_{AC}}{2}  U^C \partial_B \xi^u\right) \\
            &= \frac{4}{r} \xi^r + 2 D_A \xi^A + U^C\partial_C f \overset{!}{=}0 \quad \Rightarrow \xi^r = - \frac{r}{4}(2 D_A \xi^A + U^C \partial_C f). \label{eq: xi r up}
        \end{aligned}\end{equation} We have obtained an expression for the four components of $\xi$,
        \begin{equation}
            \xi^u = f, \quad \xi^B = R^B + \partial_A f \int_r^\infty \deriv r' g^{AB} e^{2 \beta}, \quad \text{and} \quad \xi^r = - \frac{r}{4}(2 D_A \xi^A + U^C \partial_C f) \label{eq: components xi up}
        \end{equation} in terms of the integration functions $f(u, x^A)$ and $R^B(u, x^A)$. We now apply the constraints \eqref{eq: change metric falloffs} from the falloffs, to gain insight into the form of these three functions.

        \paragraph{Applying the falloffs} Starting with $\delta g_{ur} = \mathcal{O}(r^{-2})$  in \eqref{eq: change metric falloffs} and using our previous results \eqref{eq: components xi up} as well as \eqref{eq: Bondi metric}, we get
        \begin{equation*}\begin{aligned}
            \delta g_{ur} &= \xi^r \partial_r g_{ur} + g_{ur} \partial_u f + g_{ur} \partial_r \xi^r + g_{uu} \cancel{\partial_r \xi^u} + g_{uA} \partial_r \xi^A \\
            &\approx \xi^r \partial_r (-e^{2\beta}) - e^{2\beta} \partial_u f - e^{2 \beta} \partial_r \xi^r + \frac{g_{BC}}{2}U^C \partial_r \xi^A \overset{!}{=} \mathcal{O}(r^{-2}).
        \end{aligned}\end{equation*} Inserting our results, and keeping in mind that $\beta = \mathcal{O}(r^{-2})$, $U^A = \mathcal{O}(r^{-3})$ from \eqref{eq: falloffs metric}
        \begin{multline*}
            \delta g_{ur} = - \frac{r}{4}(2D_A \xi^A + U^C \partial_C f)\cancel{(-2 \partial_r \beta)} e^{2 \beta} - e^{2\beta} \partial_u f - e^{2\beta}\partial_r \left( -\frac{r}{4} (2 D_A \xi^A + \cancel{U^C} \partial_C f) \right) \\
            + \frac{g_{BC}}{2} \cancel{U^C}\partial_r \xi^A = - e^{2\beta}\partial_u f + \frac{e^{2\beta}}{2}D_A \xi^A \approx -e^{2 \beta} \partial_u f + \frac{e^{2\beta}}{2}D_A R^A \overset{!}{=} 0,
        \end{multline*} implying that $f$ and $R^A$ are related by
        \begin{equation}
            \partial_u f = \frac{1}{2} D_A R^A. \label{eq: partial u f}
        \end{equation} Looking at $\delta g_{uA} = \mathcal{O}(1)$ in \eqref{eq: change metric falloffs} gives us
        \begin{multline*}
            \delta g_{uA} = \xi^u \cancel{\partial_u g_{uA}} + \xi^r \partial_r g_{uA} + \xi^C \partial_C g_{uA} + g_{uu} \partial_A \xi^u + g_{ur} \partial_A \xi^r + g_{uB} \partial_A \xi^B + g_{Au} \partial_u \xi^u + \cancel{g_{Ar}} \partial_u \xi^r + g_{AB} \partial_u \xi^B \\ = \xi^r \partial_r \left(\frac{g_{AC}}{2} U^C\right) + \xi^C \partial_C \left(\frac{g_{AC}}{2} U^C\right) + \left(-U + \frac{1}{4} g_{CB}U^C U^B\right) \partial_A \xi^u - e^{2\beta} \partial_A \xi^r \\ + \frac{g_{BC}}{2} U^C \partial_A \xi^B + \frac{g_{AC}}{2}U^C \partial_u \xi^u + g_{AB} \partial_u \xi^B   \overset{!}{=} \mathcal{O}(1).
        \end{multline*} The first term in the first line vanishes since $g_{uA} = g_{AB}U^B/2$ but $U^B = \mathcal{O}(r^{-2})$ so $\partial_u g_{uA} = 0$ at $\mathcal{O}(1)$. Since $g_{AB} = \mathcal{O}(r^2)$ and $U^B = \mathcal{O}(r^{-2})$, the only term at $\mathcal{O}(r^2)$ is $g_{AB}\partial_u \xi^B$, which we set to $0$. Using \eqref{eq: xi A up}, one arrives at
        \begin{equation}
            \partial_u R^B = 0 \Rightarrow R^B \equiv R^B(x^A). \label{eq: partial u R^A}
        \end{equation} Looking at $\delta g_{AB} = \mathcal{O}(r)$ in \eqref{eq: change metric falloffs}, we see that it is already satisfied from $g^{AB} \delta g_{AB} = 0$ in \eqref{eq: change metric gauge}. Since $g_{uA} = \mathcal{O}(r^{-3})$, we rewrite \eqref{eq: change gAB covar} as
        \begin{equation*}
            \delta g_{AB} = \xi^r 2r \gamma_{AB} + r^2 (\xi^C \partial_C \gamma_{AB} + \gamma_{CB} \partial_A \xi^C + \gamma_{AC} \partial_B \xi^C) = \mathcal{O}(r).
        \end{equation*} Using and \eqref{eq: xi A up} and \eqref{eq: xi r up}, we have
        \begin{equation*}
            - \frac{r}{4}(2 D_A \xi^A + ...) 2r \gamma_{AB} + r^2 \left[\left(R^C + ...\right) \partial_C \gamma_{AB} + \gamma_{CB} \partial_A \left(R^C + ... \right) + \gamma_{AC} \partial_B \left(R^C+ ... \right) \right] \overset{!}{=} 0
        \end{equation*} such that the $r^2$ terms give
        \begin{equation}
            D_C R^C \gamma_{AB} = R^C \partial_C \gamma_{AB} + \gamma_{CB} \partial_A R^C + \gamma_{AC} \partial_B R^C. \label{Conformal killing final}
        \end{equation} This is \eqref{eq: CKV 1}, but for the round metric on the unit two-sphere, $\gamma$. We thus learn that $R^A$ are conformal Killing vectors on the two-sphere, cf. \eqref{eq:conformal Killing equation}! This is confirmed by \eqref{eq: partial u R^A}, which tells us that $R^A$ does not depend on $u$. Thus \eqref{eq: xi A up} becomes
        \begin{equation*}
            \xi^B = R^B(x^A) + \partial_A f \int^\infty_r \deriv r' g^{AB} e^{2 \beta}.
        \end{equation*} Integrating \eqref{eq: partial u f} with respect to $u$, one obtains
        \begin{equation}
            f = T(x^A) + \frac{u}{2} D_C R^C, \label{eq: definition f bis}
        \end{equation} for $T$ an unconstrained and suitably differentiable function of $x^A$. $\xi^r$ remains unchanged through \eqref{eq: xi r up}. Combining all these results, we have for the asymptotic Killing vector field
        \begin{multline}
            \xi_{T,R} = \xi^u \partial_u + \xi^A \partial_A + \xi^r \partial_r
            = f \partial_u + \left[R^A + \partial_B f \int^\infty_r \deriv r' g^{BA} e^{2 \beta} \right]\partial_A \\ + \left[- \frac{r}{4}\left(2 D_A \left( R^A + \partial_B f \int^\infty_r \deriv r' g^{BA} e^{2 \beta}\right) + U^C \partial_C f\right)\right] \partial_r. \label{eq: asymp Killing temp}
        \end{multline} Now, considering the expansion \eqref{eq: falloffs metric} for $g_{AB}$ to next order, $g_{AB} = r^2 \gamma_{AB} + C_{AB}r + \mathcal{O}(1)$, we have
        \begin{equation*}
            g^{AB} = \frac{\gamma^{AB}}{r^2} - \frac{C^{AB}}{r^3} + \mathcal{O}(r^{-4}),
        \end{equation*} such that the integral in \eqref{eq: asymp Killing temp} can be computed, with $\beta \ll 1$:
        \begin{equation*}
            \int_r^\infty \deriv r' g^{AB} e^{2 \beta} \approx \int_r^\infty \deriv r' \left( \frac{\gamma^{AB}}{r'^2} - \frac{C^{AB}}{r'^3} \right) = -\frac{\gamma^{AB}}{r} + \frac{C^{AB}}{2r^2}
        \end{equation*} to give
        \begin{equation*}
            \xi^A = R^A + (\partial_B f) \left(-\frac{\gamma^{AB}}{r} + \frac{C^{AB}}{2r^2} \right) + \mathcal{O}(r^{-3}) = R^A - \frac{\partial^A f}{r} + \frac{(\partial_B f) C^{AB}}{2r^2} + \mathcal{O}(r^{-3})
        \end{equation*} and
        \begin{equation*}\begin{aligned}
            \xi^r &= - \frac{r}{4}\left(2 D_A \left( R^A - \frac{\partial^A f}{r} + \frac{(\partial_B f) C^{AB}}{2r^2} + \mathcal{O}(r^{-3}) \right) + U^C \partial_C f\right) \\
            &= - \frac{rD_A R^A}{2} + \frac{D_A D^A f}{2r} - \frac{2 (D_A C^{AB}) D_B f + C^{AB} D_A D_B f}{4r} + \mathcal{O}(r^{-2}),
        \end{aligned}\end{equation*} where we used \eqref{eq: U2A} to approximate $U^C = D_B C^{BC}/r^2 + \mathcal{O}(r^{-3})$ and Leibniz rule as well as $D_A f = \partial_A f$. Putting everything together, we obtain
        \begin{multline}
            \xi_{T,R} = f \partial_u + \left[R^A - \frac{D^A f}{r} + \frac{(D_B f) C^{AB}}{2r^2} + \mathcal{O}(r^{-3}) \right]\partial_A \\ + \left[- \frac{rD_A R^A}{2} + \frac{D_A D^A f}{2r} - \frac{2 (D_A C^{AB}) D_B f + C^{AB} D_A D_B f}{4r} + \mathcal{O}(r^{-2}) \right] \partial_r. \label{eq: asymp killing final}
        \end{multline} Note also that from \eqref{Conformal killing final}, $D_A$ denotes the covariant derivative with respect to $\gamma$ and not $g$. Our asymptotic Killing vector field \eqref{eq: asymp killing final} matches \eqref{eq: BMS generators}, while we also recovered \eqref{eq: definition f from T and R} in \eqref{eq: definition f bis}. In the limit where $r \to \infty$, we ignore the $r$ component of \eqref{eq: asymp killing final} which blows up and recover our exact Killing vector field at $\cal I^+$ \eqref{eq: BMS generators scri+},
        \begin{equation*}
            \xi_{T,R}|_{\mathcal{I}^+} = \left[T(x^A) + \frac{u}{2} D_C R^C(x^A) \right]\partial_u + R^B(x^A) \partial_B .
        \end{equation*} This concludes the construction of the BMS generators at $\cal I^+$.

    \subsection{Effect of a supertranslation \label{app: effect supertranslation}}
    Let us start with $\cal I^+$ and denote the variation under $\xi_T^+(f)$ by $\mathcal{L}_f \equiv \mathcal{L}_{\xi_f^+}$ with $f \equiv T$. Since 
    \begin{equation*}
        g_{AB} = r^2 \gamma_{AB} + r C_{AB} + \mathcal{O}(1),
    \end{equation*}the change in $\mathcal{L}_f C_{AB}$ in $C_{AB}$ will be given by the $\mathcal{O}(r)$ terms in $\mathcal{L}_f g_{AB}$. To make it easier to keep track of the order in $1/r$ in our manipulations, it is useful to rewrite \eqref{eq: BMS generators} with $R=0$ as
        \begin{equation}
            \xi_T^+(f) = f \partial_u + \sum_{n=0}^\infty \frac{\xi^{r(n)}}{r^n} \partial_r + \sum_{n=1}^\infty \frac{\xi^{A(n)}}{r^n} \partial_A. \label{eq: expansion translation 1/r}
        \end{equation} One can readily read off the first few coefficients from \eqref{eq: BMS generators} with $R=0$
        \begin{align}
            \xi^{r(0)} &= \frac{D^2 f}{2}, \quad \text{and} \quad
            \xi^{r(1)} = - \frac{1}{2}D_A C^{AB} D_B f - \frac{1}{4} C^{AB} D_A D_B f \label{eq: xi r expansion}\\
            \xi^{A(1)} &= - D^A f, \quad \text{and} \quad \xi^{A(2)} = \frac{D_B f C^{AB}}{2}. \label{eq: xi A expansion}
        \end{align} Then we have for $\mathcal{L}_f g_{AB}$, using  the equations \eqref{eq: U2A} and \eqref{eq: expansion translation 1/r} (note that $\partial_u g_{AB} \neq 0$ this time since we consider $g_{AB} = r^2 \gamma_{AB} + r C_{AB}$ and $\partial_u C_{AB} \neq 0$)
        \begin{equation*}\begin{aligned}
            \mathcal{L}_f g_{AB} &= \xi^u \partial_u (r^2 \gamma_{AB} + r C_{AB}) + \xi^r \partial_r g_{AB} + \xi^C \partial_C g_{AB} + g_{Au} \partial_B \xi^u + \cancel{g_{Ar}}\partial_B \xi^r + g_{AC} \partial_B \xi^C \\ &\qquad + g_{Bu} \partial_A \xi^u + \cancel{g_{Br}}\partial_A \xi^r + g_{BC} \partial_A \xi^C \\
            &= f \partial_u r C_{AB} + \left( \xi^{r(0)} + \frac{\xi^{r(1)}}{r} \right)(2r \gamma_{AB}  + C_{AB})  + \left( \frac{\xi^{C(1)}}{r} + \frac{\xi^{C(2}}{r^2}\right) \partial_C(r^2 \gamma_{AB} + r C_{AB})\\
            &\qquad + \frac{g_{AC}}{2}U^C \partial_B f + (r^2 \gamma_{AC} +r C_{AB}) \partial_B \left( \frac{\xi^{C(1)}}{r} + \frac{\xi^{C(2)}}{r^2}\right) \\ &\qquad + \frac{g_{BC}}{2}U^C \partial_A f + (r^2 \gamma_{BC} + r C_{BC}) \partial_A \left( \frac{\xi^{C(1)}}{r} + \frac{\xi^{C(2)}}{r^2}\right). 
        \end{aligned}\end{equation*} Keeping only terms to $\mathcal{O}(r)$ above (again, $g_{BC}U^B$ is of order $\mathcal{O}(1)$ at most) and using \eqref{eq: CKV 1},
        \begin{equation*}\begin{aligned}
            \mathcal{L}_f g_{AB} &= r \left(f \partial_u C_{AB} + 2 r \gamma_{AB} \xi^{r(0)} + D_A \xi_B^{(1)} + D_B \xi^{(1)}_A \right) + \mathcal{O}(1) \\
            \Rightarrow \mathcal{L}_f C_{AB} &= f \partial_u C_{AB} + 2  \gamma_{AB} \xi^{r(0)} + D_A \xi_B^{(1)} + D_B \xi^{(1)}_A \\
            &= f \partial_u C_{AB} +  \gamma_{AB} D^2 f - 2 D_A D_B f \quad \text{using \eqref{eq: xi r expansion} and \eqref{eq: xi A expansion}.} 
        \end{aligned}\end{equation*} We can easily check that $\gamma^{AB} \mathcal{L}_f C_{AB} = 0$, meaning that the traceless property of $C_{AB}$ is preserved under supertranslations. As $f$ is independent of $u$, the effect of supertranslations on the Bondi news \eqref{eq: bondi news tensor} is straightforwardly obtained from the previous result as
        \begin{equation*}
            \mathcal{L}_f N_{AB} = \mathcal{L}_f \partial_u C_{AB} = \partial_u \mathcal{L}_f C_{AB} = f \partial_u N_{AB}. 
        \end{equation*} Now, we look at $\mathcal{L}_f m$. From \eqref{eq: guu asymptotic} we see that
        \begin{equation*}
            \mathcal{L}_f m = \frac{1}{2} \mathcal{L}_f g_{uu} \quad \text{at order } \mathcal{O}(r^{-1}).
        \end{equation*} Applying the same procedure as before, we use \eqref{eq: Killing covariant} to write
        \begin{equation*}
            \mathcal{L}_f g_{uu} = \xi^u \partial_u g_{uu} + \xi^r \partial_r g_{uu} + \xi^A \partial_A g_{uu} + 2( g_{uu} \partial_u \xi^u + g_{ur} \partial_u \xi^r + g_{uA} \partial_u \xi^A)
        \end{equation*} and insert our asymptotic expansions to get
        \begin{multline*}
             \mathcal{L}_f g_{uu} = f \partial_u \left( -1 + \frac{2m}{r} + ... \right) +  \left(\xi^{r(0)} + \frac{\xi^{r(1)}}{r} + ...\right) \partial_r \left( -1 + \frac{2m}{r}+...\right) \\ + \left(\frac{\xi^{A(1)}}{r} + \frac{\xi^{A(2)}}{r^2} + ... \right) \partial_A \left( -1 + \frac{2m}{r}+...\right) + 2 \left[\left( -1 + \frac{2m}{r}+...\right) \cancel{\partial_u f} \right. \\ \left. + (-1 + ...) \partial_u \left(\xi^{r(0)} + \frac{\xi^{r(1)}}{r} + ...\right) + \left(\frac{1}{2} D^B C_{BA}+...\right) \partial_u  \left(\frac{\xi^{A(1)}}{r} + \frac{\xi^{A(2)}}{r^2} + ... \right)\right],
        \end{multline*} where we combined \eqref{eq: guz guzbar} and \eqref{eq: U2A} to write
        \begin{equation*}\begin{aligned}
            g_{uA} &= \frac{1}{2} g_{AB}U^B = \frac{1}{2}(r^2 g_{AB} + r C_{AB} + ...) \left( \frac{U_{2}^B}{r^2} + ...\right) \\
            &= \frac{1}{2}(r^2 g_{AB} + r C_{AB} + ...)\left( \frac{D_C C^{CB}}{r^2} + ...\right) = \frac{1}{2} D^B C_{BA} + \mathcal{O}(r^{-1}).
        \end{aligned}\end{equation*} Expanding, we get to $\mathcal{O}(r^{-1})$
        \begin{equation*}
            \mathcal{L}_f g_{uu} = - 2 \partial_u \xi^{r(0)} + \frac{-2 \partial_u \xi^{r(1)} + 2 f\partial_u m + \partial_u \xi^{A(1)} D^B C_{BA}}{r} + \mathcal{O}(r^{-2})
        \end{equation*} and hence using \eqref{eq: xi r expansion} and \eqref{eq: xi A expansion}
        \begin{equation*}\begin{aligned}
            \mathcal{L}_f m &= \frac{1}{2} \left[ -2 \partial_u \left( - \frac{1}{2}D_A C^{AB} D_B f - \frac{1}{4} C^{AB} D_A D_B f \right) + 2 f \partial_u m + \partial_u \left( - D^A f \right) D^B C_{BA} \right] \\
            &= f \partial_u m + \frac{1}{4} \left( N^{AB} D_A D_B f + 2 D_A f D_B N^{AB} \right).
        \end{aligned}\end{equation*} Finally, recall that $U_{2z} = D^z C_{zz}$ \eqref{eq: U2A}. Hence, $\mathcal{L}_f U_{2z} = D^z \mathcal{L}_f C_{zz}$,
        \begin{equation*}
            \mathcal{L}_f U_{2z} = D^z(f \partial_u C_{zz} - 2 D_z^2 f) = D^z f \partial_u C_{zz} + f \partial_u U_{2z} - 2 D^z D^2_z f.
        \end{equation*}To sum up our results, we have at $\cal I^+$
        \begin{align*}
            \mathcal{L}_f C_{AB} &= f \partial_u C_{AB} + \gamma_{AB} D^2 f - 2 D_A D_B f, \\
            \mathcal{L}_f N_{AB} &= f \partial_u N_{AB}, \\
            \mathcal{L}_f m &= f \partial_u m + \frac{1}{4} \left( N^{AB} D_A D_B f + 2 D_A f D_B N^{AB}\right), \\
            \mathcal{L}_f U_{2z} &= f \partial_u U_{2z} + D^z f \partial_u C_{zz} - 2 D^z D^2_z f.
        \end{align*} while at $\cal I^-$ we have ($f \equiv f^-$ here for notational simplicity)
        \begin{align*}
            \mathcal{L}_f M_{zz} &= f \partial_v M_{zz}, \\
            \mathcal{L}_{f} D_{zz} &= f \partial_v D_{zz} + 2 D^2_z f, \\
            \mathcal{L}_f V_{2z} &= - D^z f \partial_v D_{zz} + f \partial_v V_{2z} - 2 D^z D^2_z f.
        \end{align*}

\end{document}